\documentstyle[11pt,epsf]{article}

\begin{document}

\begin{center}

{\bf Los Alamos Electronic Archives: physics/9906066}\\

\bigskip

{\large 
{\bf IFUG}\\

\bigskip

{\it La verdad os har\'a libres}\\

\bigskip

{\bf CURSO:}  MEC\'ANICA CL\'ASICA\\

\bigskip

{\bf EDITOR:}  HARET C. ROSU\\
rosu@ifug3.ugto.mx

\bigskip
\bigskip

$\;$\\
$\;$\\

\vskip 1ex
\centerline{
\epsfxsize=180pt
\epsfbox{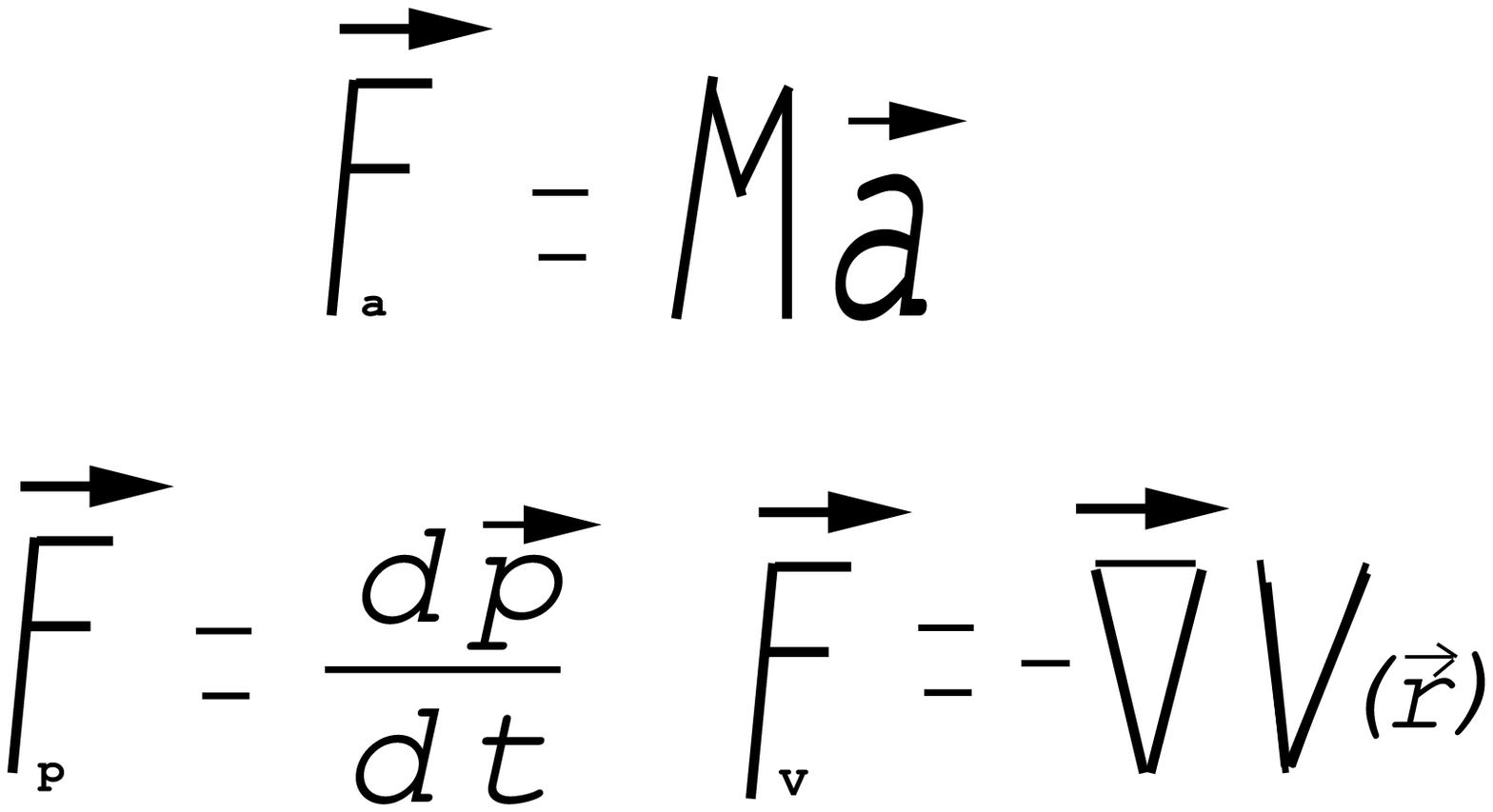}}
\vskip 2ex

\bigskip
\bigskip

{\bf curso de maestr\'{\i}a}\\

({\bf graduate course})\\

\bigskip


\bigskip


$\;$\\
$\;$\\
$\;$\\
$\;$\\
$\;$\\
$\;$\\
$\;$\\
$\;$\\
$\;$\\

Copyright \copyright $\;$ 1999 H.C. Rosu\\
Le\'on, Guanajuato, M\'exico\\ 
v1: Junio de 1999.}

\end{center}

\newpage

\begin{center}

\underline{INDICE DE CONTENIDO}

\end{center}

\bigskip

{\large

1. LOS PRINCIPIOS DE MINIMO ... 3.

\bigskip

2. MOVIMIENTO BAJO FUERZAS CENTRALES ... 20.

\bigskip

3. CUERPO RIGIDO ... 34.

\bigskip

4. OSCILACIONES PEQUE\~NAS ... 55.

\bigskip

5. TRANSFORMACIONES CANONICAS ... 73.

\bigskip

6. PARENTESIS DE POISSON ... 83.

\bigskip

7. LAS ECUACIONES DE HAMILTON-JACOBI ... 86.

\bigskip

8. VARIABLES ACCION-ANGULO ... 94.

\bigskip

9. TEORIA CANONICA DE PERTURBACIONES ... 101.

\bigskip

10. INVARIANTES ADIABATICOS ... 116.

\bigskip

11. MECANICA DE SISTEMAS CONTINUOS ... 120.}

\bigskip
\bigskip

Estudiantes colaboradores:

\bigskip

Eri Mena (1)\\

Julio L\'opez (2,3)\\

Alberto Ju\'arez (4)\\

Mario Ranfer\'{\i} Guti\'errez (5-8)\\

Zaida Urrutia (9,10)\\

M\'onica Beltr\'an (11)\\

{\it Gran parte de la responsabilidad del idioma pertenece a los estudiantes}.


\newpage

\newtheorem{theorem}{Theorem}
\newtheorem{acknowledgement}[theorem]{Acknowledgement}
\newtheorem{algorithm}[theorem]{Algorithm}
\newtheorem{axiom}[theorem]{Axiom}
\newtheorem{claim}[theorem]{Claim}
\newtheorem{conclusion}[theorem]{Conclusion}
\newtheorem{condition}[theorem]{Condition}
\newtheorem{conjecture}[theorem]{Conjecture}
\newtheorem{corollary}[theorem]{Corollary}
\newtheorem{criterion}[theorem]{Criterion}
\newtheorem{definition}[theorem]{Definition}
\newtheorem{example}[theorem]{Example}
\newtheorem{exercise}[theorem]{Exercise}
\newtheorem{lemma}[theorem]{Lemma}
\newtheorem{notation}[theorem]{Notation}
\newtheorem{problem}[theorem]{Problem}
\newtheorem{proposition}[theorem]{Proposition}
\newtheorem{remark}[theorem]{Remark}
\newtheorem{solution}[theorem]{Solution}
\newtheorem{summary}[theorem]{Summary}




\newlength{\defaultparindent}
\setlength{\defaultparindent}{\parindent}
\newenvironment{Default Paragraph Font}{}{}


\centerline{\Large 1. LOS PRINCIPIOS DE MINIMO}

\bigskip
\bigskip

\noindent
{\bf Pr\'ologo}:
La historia de los principios de ``m\'{\i}nimo" en la f\'{\i}sica es
larga e interesante. La investigaci\'{o}n de tales principios se argumenta
sobre la idea de que la Naturaleza act\'{u}a siempre de tal forma que
determinadas cantidades de importancia resultan ser siempre minimizadas cuando
tiene lugar un proceso f\'{\i}sico. La base matematica de estos principios
es el calculo variacional.

\bigskip
\bigskip

{\bf CONTENIDO}:

1. Introducci\'on

2. Principio de m\'{\i}nima acci\'on

3. Principio de D'Alembert

4. Espacio f\'asico

5. Espacio de configuraciones

6. Ligaduras

7. Ecuaciones de movimiento de Hamilton

8. Leyes de conservaci\'on

9. Aplicaciones del principio de acci\'on

\newpage

{\bf 1. Introducci\'on}\\

\noindent
La experiencia ha demostrado que, cuando sea posible despreciar los efectos
relativistas, el movimiento de una part\'{\i}cula dentro de un sistema de
referencia inercial queda correctamente descrito mediante la ecuaci\'{o}n de
Newton $\vec{F} = d\vec{p}/dt$. Cuando suceda que la part\'{\i}cula no haya de
ejecutar un
movimiento complicado y se utilicen coordenadas rectangulares para
describirlo, generalmente las ecuaciones de movimiento ser\'{a}n
relativamente sencillas; ahora bien si no se verifica ninguna de estas
condiciones, las ecuaciones pueden hacerse bastante complicadas y
dif\'{\i}ciles de manejar.

\noindent
Cuando una part\'{\i}cula est\'{a} limitada a moverse sobre una superficie
dada, deben existir ciertas fuerzas (llamadas fuerzas de ligadura) que
mantengan a la part\'{\i}cula en contacto con dicha superficie. Con el fin
de facilitar algunos problemas de \'{\i}ndole pr\'{a}ctico que aparecen al
aplicar las f\'{o}rmulas de Newton a ciertos problemas, pueden desarrollarse
otros procedimientos. Esencialmente, todos estos procedimientos de abordar
los problemas son a posteriori, puesto que sabemos de antemano que hemos de
obtener resultados equivalentes a las f\'{o}rmulas de Newton. Entonces, no
es necesario formular una nueva teor\'{\i}a de la mec\'{a}nica --la
teor\'{\i}a de Newton es suficientemente correcta- para efectuar una
simplificaci\'{o}n, sino que basta con idear otro m\'{e}todo que nos permita
abordar problemas complicados de forma general. El principio de Hamilton
contiene un m\'{e}todo de este car\'{a}cter y las ecuaciones de movimiento
que resultan de la aplicaci\'{o}n del mismo se llaman ecuaciones de Lagrange.

\noindent
Si las ecuaciones de Lagrange han de constituir una descripci\'{o}n adecuada
de la din\'{a}mica de las part\'{\i}culas, deber\'{a}n ser equivalentes a
las ecuaciones que resulten de las f\'{o}rmulas de Newton. Por otra parte,
el principio de Hamilton es de aplicaci\'{o}n a una amplia gama de
fen\'{o}menos f\'{\i}sicos (en especial los relativos a campos) con los
que generalmente no se relacionan las ecs. de Newton. Es seguro que cada una
de las consecuencias que pueden extraerse del principio de Hamilton fue
deducida primero, al igual que las ecs. de Newton, relacionando entre
s\'{\i} hechos experimentales. El principio de Hamilton no nos proporciona
teor\'{\i}a f\'{\i}sica nueva alguna, pero nos ha permitido unificar
satisfactoriamente muchas teor\'{\i}as separadas, partiendo de un postulado
fundamental sencillo. Ello no constituye un ejercicio de habilidad, puesto
que el objetivo de la f\'{\i}sica no es \'{u}nicamente dar una
formulaci\'{o}n matem\'{a}tica precisa para los fen\'{o}menos observados,
sino tambi\'{e}n describir sus efectos con ahorro de postulados
fundamentales y de la manera m\'{a}s unificada posible.

\bigskip 

\bigskip 

\bigskip 




\noindent
El primero de los principios de m\'{\i}nimo se desarrollo en el campo de la
\'{o}ptica por Her\'{o}n de Alejandr\'{\i}a hace casi dos mil a\~nos. 
Encontr\'{o} que la ley que
rige la reflexi\'{o}n de la luz puede obtenerse admitiendo que un rayo
luminoso, que viaje de un punto a otro reflej\'{a}ndose en un espejo plano,
recorre siempre el camino m\'{a}s corto. No obstante, el principio del
camino m\'{a}s corto de Her\'{o}n no puede proporcionar una expresi\'{o}n
correcta de la ley de la refracci\'{o}n. En 1657, Fermat formul\'{o}
nuevamente el principio postulando que los rayos luminosos viajan siempre de
un punto a otro de un medio siguiendo el camino que requiera el menor
tiempo. Este principio del tiempo m\'{\i}nimo de Fermat conduce
inmediatamente, no s\'{o}lo a la ley correcta de la reflexi\'{o}n, sino 
tambi\'{e}n a la ley de la refracci\'{o}n de Snell.

\noindent
Los estudios acerca de los principios de m\'{\i}nimo continuaron y, en la
\'{u}ltima parte del siglo XVII, Newton, Leibniz y los hermanos Bernoulli
iniciaron el desarrollo del c\'{a}lculo variacional. En a\~nos posteriores
este principio recibi\'{o} de Lagrange una base matem\'{a}tica s\'{o}lida
(1760). En 1828, Gauss desarrollo un m\'{e}todo para estudiar la mec\'{a}nica
mediante su principio de ligadura m\'{\i}nima. En sendos trabajos,
publicados en 1834 y 1835, Hamilton expuso el principio din\'{a}mico
de la m\'{\i}nima acci\'on sobre
el cu\'{a}l es posible formular toda la mec\'{a}nica y a decir verdad, la
mayor parte de la f\'{\i}sica. 

\noindent
Acci\'{o}n es una magnitud de dimensiones longitud por \'{\i}mpetu o bien
energ\'{\i}a por tiempo.

\bigskip


{\bf 2.} {\bf Principio de m\'{\i}nima acci\'{o}n}

\noindent
La formulaci\'{o}n m\'{a}s general de la ley del movimiento de los
sistemas mec\'{a}nicos es el {\it principio de m\'{\i}nima acci\'{o}n (o de
Hamilton). }Seg\'{u}n este principio, todo sistema mec\'{a}nico est\'{a}
caracterizado por una funci\'{o}n definida:

\[
L\left( q_{1},q_{2},..., q_{s},\stackrel{\cdot }{q_{1}},%
\stackrel{\cdot}{q_{2}},\stackrel{\cdot }{q_{s}}, t\right) , 
\]

\noindent
o m\'{a}s brevemente $L\left(q,\stackrel{\cdot}{q},
t\right) $, y el movimiento del sistema satisface la siguiente
condici\'{o}n: Supongamos que en los instantes $t=t_{1}$ y $t=t_{2}$ el
sistema ocupa posiciones dadas, caracterizadas por los dos conjuntos de
valores de las coordenadas $q^{\left( 1\right) }$ y $q^{\left( 2\right) };$
el sistema se mueve entre estas posiciones de manera que la integral

\begin{equation} \label{e1}
S=\int_{t_{1}}^{t_{2}}L\left( q,\stackrel{\cdot }{q},
t\right) dt  
\end{equation}

\noindent
tome el menor valor posible. La funci\'{o}n $L$ se llama {\it Lagrangiana }%
del sistema, y la integral (1) la {\it acci\'{o}n}. La funci\'{o}n de
Lagrange no contiene m\'{a}s que $q$ y $\stackrel{\cdot}{q}$, y no las
derivadas superiores, eso es debido al hecho que el estado mec\'{a}nico de
un sistema est\'{a} completamente definido por sus coordenadas y sus
velocidades.

\noindent
Establezcamos ahora las ecuaciones diferenciales que determinan el
m\'{\i}nimo de la integral (1). Por simplicidad empecemos suponiendo que el
sistema no tiene m\'{a}s que un s\'{o}lo grado de libertad, de manera que
hace falta determinar una sola funci\'{o}n $q\left( t\right)$.
Sea precisamente $q=q\left( t\right)$ la funci\'{o}n para la cual $S$ es un
m\'{\i}nimo. Esto significa que $S$ crece cuando se sustituye $q\left(
t\right) $ por una funci\'{o}n cualquiera

\begin{equation} \label{e2}
q\left( t\right) +\delta q\left( t\right) ,  
\end{equation}

\noindent
donde $\delta q\left( t\right) $ es una funci\'{o}n que es peque\~{n}a en
todo el intervalo de $t_{1}$a $t_{2}$ [se le llama variaci\'{o}n de la
funci\'{o}n $q\left( t\right) $]. Puesto que para $t=t_{1}$ y $t=t_{2}$
todas las funciones (2) deben tomar los mismos valores $q^{\left( 1\right) }$
y $q^{\left( 2\right) }$, se tiene:

\begin{equation} \label{e3}
\delta q\left( t_{1}\right) =\delta q\left( t_{2}\right) =0. 
\end{equation}

\noindent
Lo que var\'{\i}a $S$ cuando se reemplaza $q$ por $q+\delta q$ est\'{a} dado
por:

\[
\int_{t_{1}}^{t_{2}}L\left( q+\delta q,\stackrel{\cdot }{q}+\delta 
\stackrel{\cdot }{q},t\right) dt-\int_{t_{1}}^{t_{2}}L\left(q,
\stackrel{\cdot }{q},t\right) dt. 
\]

\noindent
El desarrollo en serie de esta diferencia en potencias de $\delta q$ y $%
\delta \stackrel{\cdot}{q}$ comienza por t\'{e}rminos de primer orden. La
condici\'{o}n necesaria de m\'{\i}nimo (en general extremal) de $S$ es que
el conjunto de estos t\'{e}rminos se anule; As\'{\i} el principio de
m\'{\i}nima acci\'{o}n puede escribirse:

\begin{equation}\label{e4}
\delta S=\delta \int_{t_{1}}^{t_{2}}L\left( q,\stackrel{\cdot }{q},
t\right) dt=0,  
\end{equation}

\noindent
o, efectuando la variaci\'{o}n:

\[
\int_{t_{2}}^{t_{1}}\left( \frac{\partial L}{\partial q}\delta q+\frac{%
\partial L}{\partial \stackrel{\cdot }{q}}\delta \stackrel{\cdot }{q}\right)
dt=0~.
\]

\noindent
Teniendo en cuenta que $\delta \stackrel{\cdot}{q}=d/dt\left( \delta
q\right) $, integramos el sugundo t\'{e}rmino por partes y se obtiene:

\begin{equation}
\delta S=\left[ \frac{\partial L}{\partial \stackrel{\cdot }{q}}\delta q%
\right] _{t_{1}}^{t_{2}}+\int_{t_{2}}^{t_{1}}\left( \frac{\partial L}{%
\partial q}-\frac{d}{dt}\frac{\partial L}{\partial \stackrel{\cdot }{q}}%
\right) \delta qdt=0~.  \label{5}
\end{equation}

\noindent
En virtud de las condiciones (3), el primer t\'{e}rmino de esta
expresi\'{o}n desaparece. Queda una integral, la cual debe anularse para
todo valor de $\delta q$. Esto es solamente posible si el integrando es
id\'{e}nticamente nulo, y consecuentemente se obtiene la ecuaci\'{o}n:

\[
\frac{\partial L}{\partial q}-\frac{d}{dt}\frac{\partial L}{\partial 
\stackrel{\cdot }{q}}=0~.
\]

\noindent
Si hay varios grados de libertad, las $s$ funciones diferentes $q_{i}(t)$
deben variar independientemente. Es evidente que entonces obtenemos $s$
ecuaciones de la forma:

\begin{equation}
\frac{d}{dt}\left( \frac{\partial L}{\partial \stackrel{\cdot }{q_{i}}}%
\right) -\frac{\partial L}{\partial q_{i}}=0\qquad
\qquad\left( i=1,2,...,s\right)  \label{6}
\end{equation}

\noindent
Estas son las ecuaciones diferenciales buscadas; en Mec\'{a}nica se les
llama {\it ecuaciones de Lagrange. }Si se conoce la lagrangiana de un
sistema mec\'{a}nico dado, entonces las ecuaciones (6) establecen la
relacion entre las aceleraciones, las velocidades y las coordenadas, es
decir, son las ecuaciones del movimiento del sistema. Desde un punto de
vista matem\'{a}tico, las ecuaciones (6) forman un sistema de $s$ ecuaciones
diferenciales de segundo orden con $s$ funciones desconocidas $q_{i}(t).$ La
soluci\'{o}n general del sistema contiene $2s$ constantes arbitrarias. Para
determinarlas y, por lo tanto, para definir completamente el movimiento del
sistema mec\'{a}nico, es necesario conocer las condiciones iniciales que
caractericen el estado del sistema en un instante dado, por ejemplo los
valores iniciales de las coordenadas y de las velocidades.

{\bf 3.} {\bf Principio de D'Alembert}

\noindent
Desplazamiento virtual (infinitesimal) de un sistema es el cambio de
configuraci\'{o}n de \'{e}ste a consecuencia de una variaci\'{o}n
infinitesiamal arbitraria de las coordenadas $\delta {\bf r}_{i},${\it %
compatible con las fuerzas y ligaduras impuestas al sistema en el instante
dado t.} Se llama virtual al desplazamiento para distinguirlo del
desplazamiento real del sistema que tiene lugar en un intervalo de tiempo $%
dt $, durante el cual puede variar las fuerzas y ligaduras.

\noindent
Las ligaduras (o restricciones) introducen 2 tipos de dificultades en la
soluci\'{o}n de problemas mec\'{a}nicos:

(1) No todas las coordenadas son independientes.

(2) Las fuerzas de ligadura por lo general no se conocen a priori, son parte
de las inc\'{o}gnitas del problema y han de obtenerse a partir de la
soluci\'{o}n buscada.

\noindent
En el caso de ligaduras holon\'{o}micas la dificultad (1) se salva
introduciendo coordenadas independientes ($q_{1,} q_{2,...,} q_{m}$,
donde m es el numero de grados de libertad). Esto es, si hay $m$ ecuaciones
de ligadura y $3N$ coordenadas $(x_{1},...,x_{3N})$, podemos eliminar esas n
ecs. introduciendo las variables independientes $(q_{1},q_{2},..,,q_{n})$
mediante una transformaci\'{o}n de la forma
$$
x_{1}=f_{1}(q_{1},...,q_{m},t)
$$
\begin{eqnarray}
\vdots  \nonumber
\end{eqnarray}
$$
x_{3N}=f_{3N}(q_{1},...,q_{n},t)~,%
$$

\noindent
donde $\ n=3N-m$.

\noindent
Para librarnos del problema (2) necesitamos formular la mec\'{a}nica de modo
que las fuerzas de ligadura NO APAREZCAN en la soluci\'{o}n de problemas.
Esta labor constituye la esencia de lo que llamaremos
``{\it El Principio de Trabajos Virtuales}".

\noindent
{\bf Trabajo Virtual:} Supongamos que un sistema de N part\'{\i}culas se
describe por $3N$ coordenadas $(x_{1},x_{2},...,x_{3N})$ y sean $%
F_{1,}F_{2,...,}F_{3N}$ las componentes de las fuerzas que
act\'{u}an sobre cada uno. Si las part\'{\i}culas del sistema experimentan
desplazamientos infinit\'{e}simales e instant\'{a}neos $\delta x_{1},\delta
x_{2},...,\delta x_{3N}$ debido a dichas fuerzas, entonces el trabajo
realizado es:

\begin{equation}
\delta W=\sum_{j=1}^{3N}F_{j}\delta x_{j}~.  \label{7}
\end{equation}

\noindent
Dichos desplazamientos se llaman {\it desplazamientos virtuales} y $\delta W$
se llama {\it trabajo virtual}; (7{\it )} tambien puede escribirse como:

\begin{equation}
\delta W=\sum_{\alpha =1}^{N}{\bf F}_{\alpha }\cdot \delta {\bf r}~. 
\label{8}
\end{equation}

\noindent
{\bf Fuerzas de ligadura o de restricci\'{o}n}: adem\'{a}s las fuerzas
externas ${\bf F}_{\alpha }^{\left( e\right) }$ las part\'{\i}culas pueden
estar sujetas a fuerzas de ligadura ${\bf F}_{\alpha }.$

\noindent
{\bf Principio de trabajo virtual:} Sea ${\bf F}_{\alpha }$ la fuerza que
act\'{u}a sobre la $\alpha $-\'{e}sima part\'{\i}cula, si
separamos ${\bf F}_{\alpha }$ en la contribuci\'{o}n de origen
externo ${\bf F}_{\alpha}^{\left( e\right) }$ y la
ligadura ${\bf R}_{\alpha}$

\begin{equation}
{\bf F}_{\alpha }={\bf F}_{\alpha }^{\left( e\right) }+{\bf R}_{\alpha}~.
\label{9}
\end{equation}
\noindent
Si el sistema est\'{a} en equilibrio entonces

\begin{equation}
{\bf F}_{\alpha }={\bf F}_{\alpha }^{\left( e\right) }+{\bf R}_{\alpha }=0~.
\label{10}
\end{equation}

\noindent
As\'{\i} que el trabajo virtual debido a todas las posibles
fuerzas ${\bf F}_{\alpha}$ es:

\begin{equation}
W=\sum_{\alpha =1}^{N}{\bf F}_{\alpha }\cdot \delta {\bf r}_{\alpha
}=\sum_{\alpha =1}^{N}\left( {\bf F}_{\alpha }^{\left( e\right) }+{\bf R}%
_{\alpha }\right) \cdot \delta {\bf r}_{\alpha }=0~.  \label{11}
\end{equation}

\noindent
Si el sistema es tal que sus fuerzas de ligadura no producen trabajo virtual
entonces de (11) concluimos que:

\begin{equation}
\sum_{\alpha =1}^{N}{\bf F}_{\alpha }^{\left( e\right) }\cdot \delta {\bf r}%
_{\alpha }=0~.  \label{12}
\end{equation}

\noindent
Ya hechan las definiciones anteriores podemos llegar a lo que es el principio
de D'Alembert. La ecuaci\'{o}n del movimiento es seg\'{u}n Newton:

\[
{\bf F}_{\alpha }=\stackrel{\cdot }{{\bf p}}_{\alpha } 
\]

\noindent
y puede escribirse de la forma

\[
{\bf F}_{\alpha }-\stackrel{\cdot }{{\bf p}}_{\alpha }=0~,
\]

\noindent
que dice que las part\'{\i}culas del sistema estar\'{a}n en equilibrio bajo
una fuerza igual a la real m\'{a}s una fuerza invertida
$-\stackrel{\cdot }{{\bf p}}_{i}.$ En vez de (12) podemos escribir
inmediatamente

\begin{equation}
\sum_{\alpha =1}^{N}\left( {\bf F}_{\alpha }-\stackrel{\cdot }{{\bf p}}%
_{\alpha }\right) \cdot \delta {\bf r}_{\alpha }=0  \label{13}
\end{equation}

\noindent
y haciendo la misma descomposici\'{o}n en fuerzas aplicadas y fuerzas de
ligadura $\left( {\bf f}_{\alpha }\right) $, resulta:

\[
\sum_{\alpha =1}^{N}\left( {\bf F}_{\alpha }^{\left( e\right) }-\stackrel{
\cdot }{{\bf p}}_{\alpha }\right) \cdot \delta {\bf r}_{\alpha
}+\sum_{\alpha =1}^{N}{\bf f}_{\alpha }\cdot \delta {\bf r}_{\alpha }=0~.
\]

\noindent
Limit\'{e}monos de nuevo a sistemas para los cuales el trabajo virtual de
las fuerzas de ligadura sea nulo y obtendremos

\begin{equation}
\sum_{\alpha =1}^{N}\left( {\bf F}_{\alpha }^{\left( e\right) }-\stackrel{%
\cdot }{{\bf p}}_{\alpha }\right) \cdot \delta {\bf r}_{\alpha }=0~,
\label{14}
\end{equation}

\noindent
que constituye el {\it principio de D'Alembert}. Ahora esta
ecuaci\'{o}n anterior a\'{u}n no tiene forma \'{u}til
para proporcionar las ecuaciones de movimiento del sistema, por lo que
debemos transfomar el principio en una expresi\'{o}n que contenga
desplazamientos virtuales de las coordenadas generalizadas, las cuales son
entonces independientes entre si esto implica que se podran hacer cero los
coeficientes de $\delta {\bf q}_{\alpha }$ y la velocidad en t\'{e}rmino de
las coordenadas generalizadas es:

\[
{\bf v}_{\alpha }=\frac{d{\bf r}_{\alpha }}{dt}=\sum_{k}\frac{\partial {\bf r
}_{\alpha }}{\partial q_{k}}\stackrel{\cdot }{q_{k}}+\frac{\partial {\bf r}
_{\alpha }}{\partial t}\ \ \ \ \ \  {\rm donde}
\qquad {\bf r}_{\alpha }={\bf r}%
_{\alpha }\left( q_{1},q_{2},...,q_{n},t\right)~. 
\]

\noindent
An\'{a}logamente, el desplazamiento virtual arbitrario $\delta {\bf r}
_{\alpha }$ se puede relacionar con los desplazamientos virtuales $\delta 
{\bf q}_{j}$ mediante

\[
\delta {\bf r}_{\alpha }=\sum_{j}\frac{\partial {\bf r}_{\alpha }}{\partial
q_{j}}\delta q_{j}~.
\]

\noindent
Entonces en funci\'{o}n de las coordenadas generalizadas, el trabajo virtual
de las ${\bf F}_{\alpha }$ \ ser\'{a}:

\begin{equation}
\sum_{\alpha =1}^{N}{\bf F}_{\alpha }\cdot \delta {\bf r}_{\alpha
}=\sum_{j,\alpha }{\bf F}_{\alpha }\cdot \frac{\partial {\bf r}_{\alpha }}{
\partial q_{j}}\delta q_{j}=\sum_{j}Q_{j}\delta q_{j}~,  \label{15}
\end{equation}

\noindent
donde las $Q_{j}$ son las llamadas componentes de la fuerza generalizada,
las cuales se definen en la forma

\[
Q_{j}=\sum_{\alpha }{\bf F}_{\alpha }\cdot \frac{\partial {\bf r}_{\alpha }}{
\partial q_{j}}~.
\]

\noindent
Ahora si vemos a la ec. (14) como:

\begin{equation}
\sum_{\alpha }\stackrel{\cdot }{{\bf p}}\cdot \delta {\bf r}_{\alpha
}=\sum_{\alpha }m_{\alpha }\stackrel{\cdot \cdot }{{\bf r}_{\alpha }}\cdot 
\delta {\bf r}_{\alpha }  \label{16}
\end{equation}

\noindent
y sustituyendo en esta ultima ec. los resultados anteriores podemos ver que
(16):

\begin{equation}
\sum_{\alpha }\left\{ \frac{d}{dt}\left( m_{\alpha }{\bf v}_{\alpha }\cdot 
\frac{\partial {\bf v}_{\alpha }}{\partial \stackrel{\cdot }{q_{j}}}\right)
-m_{\alpha }{\bf v}_{\alpha }\cdot \frac{\partial {\bf v}_{\alpha }}{%
\partial q_{j}}\right\} =\sum_{j}\left[ \left\{ \frac{d}{dt}\left( \frac{
\partial T}{\partial \stackrel{\cdot }{q_{j}}}\right) -\frac{\partial T}{
\partial q_{j}}\right\} -Q_{j}\right] \delta q_{j}=0~.  \label{17}
\end{equation}

\noindent
Las variables $q_{j}$ pueden ser un sistema cualquiera de coordenadas para
describir el movimiento del sistema. Sin embargo, si las ligaduras son
holonomas, ser\'{a} posible encontrar sistemas de coordenadas $q_{j}$
independientes que contengan impl\'{\i}citamente las condiciones de ligadura
en las ecuaciones de transformaci\'{o}n $x_{i}=f_{i}$ es que se anulen por 
separado los coeficientes:

\begin{equation}
\frac{d}{dt}\left( \frac{\partial T}{\partial \stackrel{\cdot }{q}}\right) -
\frac{\partial T}{\partial q_{\alpha }}=Q_{j}~.  \label{18}
\end{equation}

\noindent
En total hay $m$ ecuaciones.
Las ecuaciones (18) suelen llam\'{a}rseles ecuaciones de Lagrange, si bien
esta designaci\'{o}n se reserva frecuentemente para la forma que toman estas
ecuaciones, cuando las fuerzas se derivan de un potencial escalar V

\[
{\bf F}_{\alpha }=-\nabla _{i}V. 
\]

\noindent
Entonces $Q_{j}$ puede escribirse como:

\[
Q_{j}=-\frac{\partial V}{\partial q_{j}}~.
\]

\noindent
Las ecuaciones (18) pueden escribirse tambi\'{e}n en la forma:

\begin{equation}
\frac{d}{dt}\left( \frac{\partial T}{\partial \stackrel{\cdot }{q_{j}}}%
\right) -\frac{\partial (T-V)}{\partial q_{j}}=0  \label{19}
\end{equation}

\noindent
y definiendo la funci\'{o}n {\it Lagrangiana} $L$, en la forma $L=T-V$
se obtiene
\begin{equation}
\frac{d}{dt}\left( \frac{\partial L}{\partial \stackrel{\cdot }{q_{j}}}%
\right) -\frac{\partial L}{\partial q_{j}}=0~.  \label{20}
\end{equation}

\noindent
Estas son las {\bf Ecuaciones de Lagrange}.

\bigskip

{\bf 4. - Espacio F\'{a}sico}

\noindent
En la interpretaci\'{o}n geom\'{e}trica de los fen\'{o}menos mec\'{a}nicos
se hace frecuentemente uso del concepto de {\it espacio f\'{a}sico}, es un
espacio de $2s$ dimensiones cuyos ejes coordenados corresponden a los $s$
coordenadas generalizadas y a los $s$ \'{\i}mpetus del sistema mec\'{a}nico
considerado. Cada punto en este espacio corresponde a un estado definido del
sistema. Cuando el sistema se mueve, el punto representativo en el espacio
f\'{a}sico describe una l\'{\i}nea denominada {\it trayectoria f\'{a}sica.}

{\bf 5. - Espacio de Configuraci\'{o}nes}

\noindent
El estado de un sistema compuesto de $n$ part\'{\i}culas y sometido a $m$
ligaduras que relacionen entre s\'{\i} a algunas de las $3n$ coordenadas
rectangulares queda especificado por completo mediante $s=3n-m$ coordenadas
generalizadas. Es posible, pues, representar el estado de tal sistema por un
punto de un espacio de $s$ dimensiones que llamamos {\it espacio de
configuraciones}, correspondiendo cada una de las diemensiones de este
espacio a una de las $q_{j}$. La historia, o evoluci\'{o}n a trav\'{e}s del
tiempo, del sistema, estar\'{a} representada por una curva del espacio de
confuguraciones, cada uno de cuyos puntos represaentar\'{a} la
configuraci\'{o}n del sistema en un instante determinado.

{\bf 6. - Ligaduras}

\noindent
Es necesario tener en cuenta las {\it ligaduras} que limitan el movimiento
del sistema. Las ligaduras pueden clasificarse de divesas maneras. En el
caso general en que la ecuaciones de ligadura puedan expresarse como:

\[
\sum_{i}c_{\alpha i}\stackrel{\cdot }{q_{i}}=0~, 
\]

\noindent
donde las $c_{\alpha i}$ son funciones de las coordenadas solamente (el
\'{\i}ndice $\alpha$ numera las ecuaciones de ligadura). Si los primeros
miembros de estas ecuaciones no son derivadas totales con respecto al tiempo
de funciones de las coordenadas, estas ecuaciones no pueden ser integradas.
En otras palabras, no pueden reducirse a relaciones entre las coordenadas
solamente, que podr\'{\i}an utilizarse para expresar la posici\'{o}n e los
cuerpos por un n\'{u}mero menor de coordenadas, correspondiente al
n\'{u}mero real de grados de libertad. Tales ligaduras se llaman {\it no
holon\'{o}micas} (en oposici\'{o}n a las ligaduras anteriores son las
llamadas {\it holon\'{o}micas }que relacionan solamente las coordenadas del
sistema{\it )}.

{\bf 7.} {\bf Ecuaciones de movimiento de Hamilton}

\noindent
La formulaci\'{o}n de las leyes de la Mec\'{a}nica con la ayuda de la
Lagrangiana, presupone que el estado mec\'{a}nico del sistema est\'{a}
determinado dando sus coordenadas y velocidades generalizadas. Sin embargo,
\'{e}ste no es el \'{u}nico m\'{e}todo posible; la descripci\'{o}n del
estado de un sistema en funci\'{o}n de sus coordenadas e \'{\i}mpetus
generalizados presenta un cierto n\'{u}mero de ventajas.

\noindent
El paso de un conjuto de variables independientes a otro puede realizarse
mediante lo que se llama en matem\'{a}ticas {\it tranformaci\'{o}n de
Legendre.} En este caso la transformaci\'{o}n toma la siguinte forma, donde
la diferencial total de la Lagrangiana como funci\'{o}n de las coordenadas y
de las velocidades es:

\[
dL=\sum_{i}\frac{\partial L}{\partial q_{i}}dq_{i}+\sum_{i}\frac{\partial L}{%
\partial \stackrel{\cdot }{q_{i}}}d\stackrel{\cdot }{q}_{i}~,
\]

\noindent
la cual puede escribirse como:

\begin{equation}
dL=\sum_{i}\stackrel{\cdot }{p_{i}}dq_{i}+\sum_{i}p_{i}d\stackrel{\cdot }{q}%
_{i}~,  \label{21}
\end{equation}

\noindent
donde ya sabemos que las derivadas $\partial L/\partial \stackrel{\cdot}{
q_{i}}$, son por definici\'{o}n, los \'{\i}mpetus
generalizados y $\partial L$ $/$ $\partial q_{i}=\stackrel{\cdot }{p_{i}}$
por las ecuaciones de Lagrange. El segundo t\'{e}rmino de
la ec. (21) puede escribirse en la forma

\[
\sum_{i}p_{i}d\stackrel{\cdot }{q}_{i}=d\left( \sum p_{i}\stackrel{\cdot }{q}%
_{_{i}}\right) -\sum \stackrel{\cdot }{q_{i}}dq_{i}~. 
\]

\noindent
Llevando la diferencial total $d\left( \sum p_{i}\stackrel{\cdot }{q}
_{_{i}}\right) $ al primer miembro, y cambiando los signos, se obtiene de
(21):

\begin{equation}
d\left( \sum p_{i}\stackrel{\cdot }{q}_{_{i}}-L\right) =-\sum \stackrel{%
\cdot }{p}_{_{i}}dq_{i}+\sum p_{i}\stackrel{\cdot }{q}_{_{i}}~.  \label{22}
\end{equation}

\noindent
La cantidad bajo el signo de la diferencial es la energ\'{\i}a del sistema
expresada en funci\'{o}n de las coordenadas y de los \'{\i}mpetus y se llama
{\it funci\'{o}n de Hamilton o Hamiltoniana} del sistema:

\begin{equation}
H\left( p,q,t\right) =\sum_{i}p_{i}\stackrel{\cdot }{q}_{_{i}}-L~.
\label{23}
\end{equation}

\noindent
Entonces de la ec. (22)

\[
dH=-\sum \stackrel{\cdot }{p}_{_{i}}dq_{i}+\sum p_{i}\stackrel{\cdot }{q}%
_{i} 
\]

\noindent
en lo cual las variables independientes son las coordenadas y los
\'{\i}mpetus, se obtienen las ecuaciones

\begin{equation}
\stackrel{\cdot }{q}_{i}=\frac{\partial H}{\partial p_{i}}\ \ \ \ \ \
\ \ \ \ \ \ \ \ \stackrel{\cdot }{p}_{_{i}}=-\frac{\partial H}{\partial
q_{i}}~.  \label{24}
\end{equation}

\noindent
Estas son las ecuaciones de movimiento en las variables $q$ y $p$ y
se llaman {\it ecuaciones de Hamilton.}

{\bf 8.} {\bf Leyes de conservaci\'{o}n}

\noindent
{\bf 8.1  Energ\'{\i}a}

\noindent
Consideremos primero el teorema de conservaci\'{o}n que resulta de la 
{\it homogeneidad del tiempo}. En virtud de esta homogeneidad, la
Lagrangiana de un sistema cerrado no depende expl\'{\i}citamente del tiempo.
Entonces la derivada total respecto al tiempo de la Lagrangiana (no
dependiente expl\'{\i}citamente del tiempo) puede escribirse como:

\[
\frac{dL}{dt}=\sum_{i}\frac{\partial L}{\partial q_{i}}\stackrel{\cdot }{q}%
_{i}+\sum_{i}\frac{\partial L}{\partial \stackrel{\cdot }{q}_{i}}\stackrel{%
\cdot \cdot }{q}_{i} 
\]

\noindent
y de acuerdo a las ecs. de Lagrange podemos reescribir la ec. anterior como:

\[
\frac{dL}{dt}=\sum_{i}\stackrel{\cdot }{q}_{i}\frac{d}{dt}\left( \frac{%
\partial L}{\partial \stackrel{\cdot }{q}_{i}}\right) +\sum_{i}\frac{%
\partial L}{\partial \stackrel{\cdot }{q}_{i}}\stackrel{\cdot \cdot }{q}%
_{i}=\sum_{i}\frac{d}{dt}\left( \stackrel{\cdot }{q}_{i}\frac{\partial L}{%
\partial \stackrel{\cdot }{q}_{i}}\right)~, 
\]

\noindent
\'{o}

\[
\sum_{i}\frac{d}{dt}\left( \stackrel{\cdot }{q}_{i}\frac{\partial L}{%
\partial \stackrel{\cdot }{q}_{i}}-L\right) =0~.
\]

\noindent
De donde se deduce que la magnitud

\begin{equation}
E\equiv \sum_{i}\stackrel{\cdot }{q}_{i}\frac{\partial L}{\partial \stackrel{%
\cdot }{q}_{i}}-L  \label{25}
\end{equation}

\noindent
permanece constante durante el movimiento de un sistema cerrado, es decir es
una integral del movimiento. A esta magnitud se le llama {\it energ\'{\i}a} E
del sistema.

\noindent
{\bf 8.2 Impetu}

\noindent
De la {\it homogeneidad del espacio} se deduce otro teorema de
conservaci\'{o}n. En virtud de dicha homogeneidad, las propiedades
mec\'{a}nicas de un sistema cerrado no var\'{\i}an por un desplazamiento
paralelo de todo el sistema en el espacio. Consideremos un desplazamiento
infinitesimal $\epsilon$ (i.e., los vectores de posici\'{o}n {\bf r}$
_{\alpha }$ se convierten en {\bf r}$_{a}+\epsilon $) y busquemos la
condici\'{o}n para que la lagrangiana no var\'{\i}e. La variaci\'{o}n de la
funci\'{o}n L, consecuencia de un cambio infinitesimal en las coordenadas
(permaneciendo constantes las velocidades de las part\'{\i}culas), est\'{a}
dado por:

\[
\delta L=\sum_{a}\frac{\partial L}{\partial {\bf r}_{a}}\cdot \delta {\bf r}
_{a}=\epsilon \cdot \sum_{a}\frac{\partial L}{\partial {\bf r}_{a}}~,
\]

\noindent
extendiendo la suma a todas las part\'{\i}culas del sistema. Como $\epsilon$
es arbitrario, la condici\'{o}n $\delta L=0$ es equivalente a

\begin{equation}
\sum_{a}\frac{\partial L}{\partial {\bf r}_{a}}=0  \label{26}
\end{equation}

\noindent
y en virtud de las ecs. de Lagrange ya mencionadas

\[
\sum_{a}\frac{d}{dt}\left( \frac{\partial L}{\partial {\bf v}_{a}}\right) =%
\frac{d}{dt}\sum_{a}\frac{\partial L}{\partial {\bf v}_{a}}=0~.
\]

\noindent
Llegamos as\'{\i} a la conclusi\'{o}n de que en un sistema mec\'{a}nico
cerrado, la magnitud vectorial (llamada {\it \'{\i}mpetu})

\[
{\bf P\equiv }\sum_{a}\frac{\partial L}{\partial {\bf v}_{a}} 
\]

\noindent
permanece constante durante el movimiento.

\noindent
{\bf 8.3 Momento angular \'{o} cin\'{e}tico}

\noindent
Estudiemos ahora el teorema de conservaci\'{o}n que infiere de la {\it %
isotrop\'{\i}a del espacio}. Consideremos una rotaci\'{o}n infinitesimal del
sistema, y busquemos la condici\'{o}n para que la Lagrangiana no var\'{\i}e.

\noindent
Llamaremos vector de rotaci\'{o}n infinitesimal $\delta {\bf \phi }$ al
vector cuyo m\'{o}dulo es igual al \'{a}ngulo de rotaci\'{o}n $\delta \phi $
y cuya direcci\'{o}n coincide con el eje de rotaci\'{o}n. Consideremos
primero el incremento en el vector de posici\'{o}n correspondiente a una
part\'{\i}cula del sistema, tomando un origen de coordenadas situado en el
eje de rotaci\'{o}n. El desplazamiento lineal extremo del vector de
posici\'{o}n en funci\'{o}n de \'{a}ngulo es

\[
\left| \delta {\bf r}\right| =r\sin \theta \delta \phi~, 
\]

\noindent
(ver fig.). La direcci\'{o}n del vector $\delta {\bf r}$ es perpendicular
al plano definido por ${\bf r}$ y $\delta {\bf \phi}$, y por tanto,

\begin{equation}
\delta {\bf r=}\delta {\bf \phi \times r}~.  \label{27}
\end{equation}

\vskip 1ex
\centerline{
\epsfxsize=80pt
\epsfbox{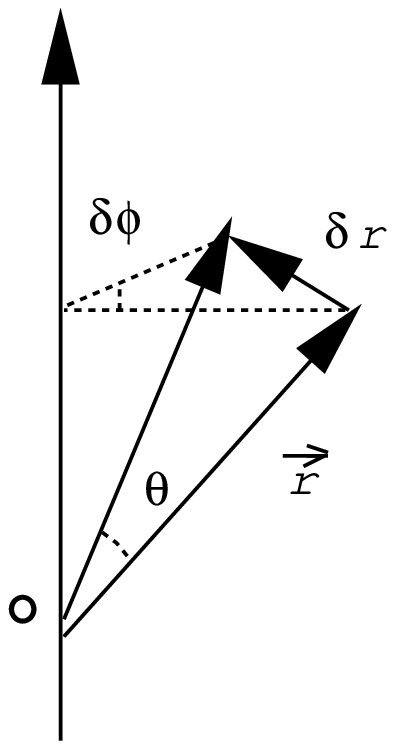}}
\vskip 2ex


\noindent
La rotaci\'{o}n del sistema no solamente modifica la direcci\'{o}n de los
vectores de posici\'{o}n sino tambi\'{e}n las velocidades de las
part\'{\i}culas, transform\'{a}ndose en todos los vectores seg\'{u}n la misma
ley. El incremento de velocidad con respecto a un sistema fijo de
coordenadas ser\'{a};

\[
\delta {\bf v}=\delta {\bf \phi \times v}~.
\]

\noindent
Llevemos estas expresiones a la condici\'{o}n de que la Lagrangiana no
var\'{\i}a por la rotaci\'{o}n:

\[
\delta L=\sum_{a}\left( \frac{\partial L}{\partial {\bf r}_{a}}\cdot \delta 
{\bf r}_{a}+\frac{\partial L}{\partial {\bf v}_{a}}\cdot \delta {\bf v}%
_{a}\right) =0 
\]

\noindent
y sustituyendo, por definici\'{o}n las derivadas $\partial L/\partial
{\bf v}_{a}$ por ${\bf p}_{a}$ y las derivadas $\partial L/\partial
{\bf r}_{a}$ de acuerdo con las ecs. de Lagrange,
por $\stackrel{\cdot }{{\bf p}}_{a}$; obtenemos

\[
\sum_{a}\left( \stackrel{\cdot }{{\bf p}}_{a}\cdot \delta {\bf \phi
\times r}_{a}+{\bf p}_{a}\cdot \delta {\bf \phi \times v}_{a}\right) =0~,
\]

\noindent
o permutando circularmente los factores y sacando $\delta {\bf \phi }$ fuera
del signo suma:

\[
\delta {\bf \phi }\sum_{a}\left( {\bf r}_{a}{\bf \times }\stackrel{\cdot }{
{\bf p}}_{a}+{\bf v}_{a}{\bf \times p}_{a}\right) =\delta {\bf \phi }\cdot 
\frac{d}{dt}\sum_{a}{\bf r}_{a}{\bf \times p}_{a}=0~,
\]

\noindent
puesto que $\delta {\bf \phi }$ es arbitrario, resulta

\[
\frac{d}{dt}\sum_{a}{\bf r}_{a}{\bf \times p}_{a}=0 
\]

\noindent
y se concluye que en el movimiento de un sistema cerrado se conserva la
magnitud vectorial (llamada {\it momento angular \'{o} momento cin\'{e}tico})

\[
M\equiv \sum_{a}{\bf r}_{a}{\bf \times p}_{a}~.
\]

\bigskip

{\bf 9.- Aplicaciones del principio de Acci\'{o}n}

\noindent
{\bf a) Ecuaciones de movimiento}

\noindent
Hallar las ecuaciones del movimiento para una masa pendular suspendida de un
resorte, por aplicaci\'{o}n directa del principio de Hamilton

\vskip 2ex
\centerline{
\epsfxsize=200pt
\epsfbox{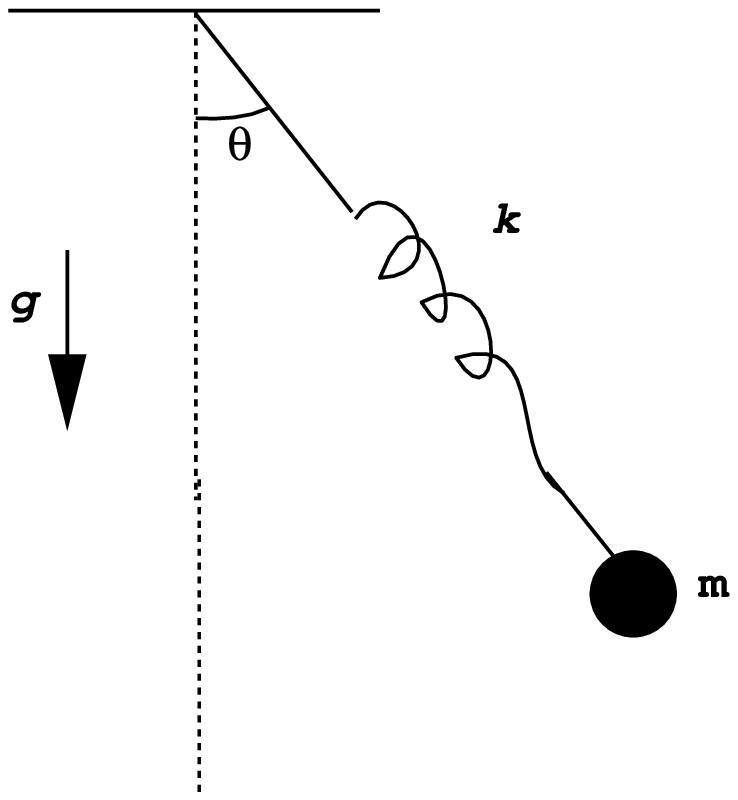}}
\vskip 4ex


\noindent
Para el p\'{e}ndulo de la figura la Lagrangiana es de la forma

\[
L=\frac{1}{2}m(\stackrel{\cdot}{r}^{2}+r^{2}\stackrel{\cdot}{\theta}^{2}
)+mgr\cos \theta -\frac{1}{2}k(r-r_{o})^{2}~, 
\]

\noindent
por lo tanto

\[
\int_{t_{1}}^{t_{2}}\delta Ldt=\int_{t_{1}}^{t_{2}}\left[
 m\left( 
\stackrel{\cdot }{r}\delta \stackrel{\cdot }{r}+r\stackrel{\cdot}{\theta}
^{2}+r^{2}\stackrel{\cdot }{\theta }\delta \stackrel{\cdot }{\theta }\right)
+mg\delta r\cos \theta -mgr\delta \theta \sin \theta -k(r-r_{o})\delta r%
\right] dt 
\]

\[
m\stackrel{\cdot }{r}\delta \stackrel{\cdot }{r}dt=m\stackrel{\cdot }{r}%
d(\delta r)=d\left( m\stackrel{\cdot }{r}\delta r\right) -m\delta r\stackrel{%
\cdot \cdot }{r}dt~.
\]

\noindent
Igualmente

\[
mr^{2}\theta ^{2}\delta \stackrel{\cdot }{\theta }dt=d\left( mr^{2}\stackrel{%
\cdot }{\theta }\delta \stackrel{\cdot }{\theta }\right) -\delta \theta 
\frac{d\left( mr^{2}\stackrel{\cdot }{\theta }\right) }{dt}dt 
\]

\[
=d\left( mr^{2}\stackrel{\cdot }{\theta }\delta \stackrel{\cdot }{\theta }%
\right) -\delta \theta \left( mr^{2}\stackrel{\cdot \cdot }{\theta }+2mr%
\stackrel{\cdot }{r}\stackrel{\cdot }{\theta }\right) dt~.
\]

\noindent
Por tanto la integral anterior se escribe como

\[
\int_{t_{1}}^{t_{2}}\left[ \left\{ m\stackrel{\cdot \cdot }{r}-mr\stackrel{%
\cdot }{\theta }^{2}-mg\cos \theta +k\left( r-r_{o}\right) \right\} +\left\{
mr^{2}\stackrel{\cdot \cdot }{\theta }+2mr\stackrel{\cdot }{r}\stackrel{%
\cdot }{\theta }+mgr\sin \theta \right\} \delta \theta \right] dt 
\]

\[
-\int_{t_{1}}^{t_{2}}\left[ d\left( m\stackrel{\cdot }{r}\delta r\right)
+d\left( mr^{2}\theta ^{2}\stackrel{\cdot }{\theta }\delta \theta \right) %
\right] =0~.
\]

\noindent
Suponiendo que $\delta r$ y $\delta \theta $ son ambas iguales a cero en $%
t_{1}$ y $t_{2}$ la segunda integral es evidentemente nula. Como $%
\delta r$ y $\delta \theta $ son completamente independientes, la primera
integral puede ser cero solamente si

\[
m\stackrel{\cdot \cdot }{r}-mr\stackrel{\cdot }{\theta }^{2}-mg\cos \theta
+k(r-r_{o})=0 
\]

\noindent
y

\[
mr^{2}\stackrel{\cdot \cdot }{\theta }+2mr\stackrel{\cdot }{r}\stackrel{
\cdot }{\theta }+mgr\sin \theta =0~,
\]

\noindent
pero estas son las ecuaciones de movimiento del sistema.

\noindent
{\bf b) Ejemplo de c\'{a}lculo de minimo}

\noindent
Se trata de demostrar que la l\'{\i}nea m\'{a}s corta entre dos puntos
cualesquiera $p_{1}$ y $C$ sobre un cilindro es una h\'{e}lice.

\noindent
La longitud $S$ de una l\'{\i}nea cualquiera sobre un cilindro
entre $p_{1}$ y $p_{2}$ est\'{a} dada por

\[
S=\int_{p_{1}}^{p_{2}}\left[ 1+r^{2}\left( \frac{d\theta }{dz}%
\right) ^{2}\right] ^{1/2}dz~,
\]

\noindent
donde r, $\theta $ y z son las coordenadas cil\'{\i}ndricas usuales
con $r=cte.$ Puede determinarse una relaci\'{o}n entre $\theta$ y z que le
d\'{e} a esta integral un valor extremo, por medio de

\[
\frac{d}{dz}\left( \frac{\partial \phi }{\partial \theta ^{^{\prime }}}%
\right) -\frac{\partial \phi }{\partial \theta }=0~,
\]

\noindent
donde $\phi =\left[ 1+r^{2}\theta ^{\prime 2}\right] ^{1/2}$ y $\theta
^{\prime }=\frac{d\theta }{dz}$, pero como $\partial \phi /\partial \theta
=0$,

\[
\frac{\partial \phi }{\partial \theta ^{\prime }}=\left( 1+r^{2}\theta
^{\prime 2}\right) ^{-1/2}r^{2}\theta ^{\prime }=c_{1}=cte.~,
\]

\noindent
por esto $r\theta ^{\prime }=c_{2}.$ Y as\'{\i}, $r\theta =c_{2}z+c_{3}$ que
es la ecuaci\'{o}n de una h\'{e}lice. Supongamos que en $p_{1}$ se
tiene $\theta =0$ y $z=0$, entonces $c_{3}=0$.
En $p_{2}$, h\'{a}gase $\theta
=\theta _{2}$ y $z=z_{2}$ por tanto $c_{2}=r\theta _{2}/z_{2}$, y $r\theta
=\left( r\theta _{2}/z_{2}\right) z$ es la ecuaci\'{o}n final.

\begin{center}  {\bf Bibliografia}  \end{center}

L. D. Landau y E. M Lifshitz, {\it Mec\'anica}, F\'{\i}sica Te\'{o}rica,
 vol I, editorial 

Revert\'{e}, S.A. (1969)

H. Goldstein, {\it Mec\'{a}nica Cl\'{a}sica}, editorial Revert\'{e}, S.A.
(1992)


\newpage



\centerline{\Large 2. MOVIMIENTO BAJO FUERZAS CENTRALES}

\bigskip

\noindent
{\bf Pr\'ologo}:
Por motivos astronomicos este fue el movimiento m\'as estudiado del punto
de vista experimental en los primeros dos siglos de f\'{\i}sica moderna
y representa un ejemplo principal para cualquier tipo de formalismo matematico.
En su variante relativista, el problema de Kepler sigue siendo de gran
interes.

\bigskip
\bigskip

{\bf  CONTENIDO:}

\bigskip

2.1 Problema de dos cuerpos: reducci\'on al problema de un cuerpo.

\bigskip

2.2 Ecuaciones de movimiento.

\bigskip

2.3 Ecuaci\'on diferencial de la \'orbita.

\bigskip

2.4 Problema de Kepler.

\bigskip

2.5 Dispersion por un centro de fuerzas (con ejemplo).

\bigskip

\newpage


\noindent {\bf
2.1 Problema de 2 cuerpos: Reducci\'{o}n al problema de un cuerpo.}

\noindent
Consideremos un sistema mon\'{o}geno de dos puntos materiales de masas $%
m_{1} $ y $m_{2}$, en el cual las unicas fuerzas son las debidas a un
potencial de interaccion $V$. Supondremos que $V$ es funci\'{o}n de
cualquier vector entre $m_{1}$ y $m_{2}$, ${\bf r}_{2}-{\bf r}_{1}$, o de
sus velocidades relativas $\stackrel{\cdot }{{\bf r}}_{2}-\stackrel{\cdot }{%
{\bf r}}_{1}$, o de derivadas superiores de ${\bf r}_{2}-{\bf r}_{1}$. Este
sistema tiene 6 grados de libertad y por lo tanto 6 coordenadas
generalizadas independientes.

\noindent
Consideremos que estas son las coordenadas del vector de posici\'{o}n del
centro de masa, ${\bf R}$, mas las tres componentes del vector diferencia \
${\bf r}={\bf r}_{2}-{\bf r}_{1}$. La Lagrangiana del sistema tendra entonces
la forma:
\setcounter{equation} {0}\\
\begin{equation}
L=T({\bf \dot{R}},{\bf \dot{r}})-V({\bf r},{\bf \dot{r}},{\bf \ddot{r}}%
,.....).  \label{eq1}
\end{equation}

\noindent
La energ\'{\i}a cin\'{e}tica $T$ es la suma de la energ\'{\i}a cin\'{e}tica
del movimiento del centro de masa mas la energ\'{\i}a cin\'{e}tica del
movimiento en torno al centro de masa, $T{\acute{}}$:
\[
T=\frac{1}{2}(m_{1}+m_{2}){\bf \dot{R}}^{2}+T{\acute{}}~,
\]
siendo 
\[
T{\acute{}}=\frac{1}{2}m_{1}{\bf \dot{r}}_{1}^{2}{\acute{}}
+\frac{1}{2}m_{2}{\bf \dot{r}}_{2}^{2}
{\acute{}}.
\]

\noindent
Aqu\'{\i} ${\bf r}_{1}{\acute{}}$ y ${\bf r}_{2}{\acute{}}$ son los vectores
de posici\'{o}n de las dos part\'{\i}culas relativas al
centro de masa y estan relacionadas con ${\bf r}$ a travez de 
\begin{equation}
{\bf r}_{1}{\acute{}}
=-\frac{m_{2}}{m_{1}+m_{2}}{\bf r},\;{\bf r}_{2}{\acute{}}
=\frac{m_{1}}{m_{1}+m_{2}}{\bf r}~,  \label{eq2}
\end{equation}
entonces $T
{\acute{}}$ toma la forma
\[
T{\acute{}}
=\frac{1}{2}\frac{m_{1}m_{2}}{m_{1}+m_{2}}{\bf \dot{r}}^{2} 
\]
y la Lagrangiana total dada por la ecuaci\'{o}n (1) es:
\begin{equation}
L=\frac{1}{2}(m_{1}+m_{2}){\bf \dot{R}}^{2}+\frac{1}{2}\frac{m_{1}m_{2}}{%
m_{1}+m_{2}}{\bf \dot{r}}^{2}-V({\bf r},{\bf \dot{r}},{\bf \ddot{r}},.....)~,
\label{eq3}
\end{equation}
de donde definimos la masa reducida como 
\[
\mu =\frac{m_{1}m_{2}}{m_{1}+m_{2}}\;\;\;o{\acute{}}
\;\;\;\frac{1}{\mu }=\frac{1}{m_{1}}+\frac{1}{m_{2}}~. 
\]
Entonces nuestra ecuacion (\ref{eq3}) se puede escribir como 
\[
L=\frac{1}{2}(m_{1}+m_{2}){\bf \dot{R}}^{2}+\frac{1}{2}\mu {\bf \dot{r}}%
^{2}-V({\bf r},{\bf \dot{r}},{\bf \ddot{r}},.....). 
\]
De esta ecuaci\'{o}n vemos que las coordenadas de ${\bf \dot{R}}$ son
c\'{\i}clicas por lo que el centro de masa estara fijo o se movera
uniformemente.

\noindent
Ahora, ninguna de las ecuaciones de movimiento para ${\bf r}$ contendra
t\'{e}rminos donde aparesca ${\bf R}$ o ${\bf \dot{R}}$, este t\'{e}rmino de
la ecuaci\'{o}n es exactamente lo que tendriamos si tuvieramos un centro de
fuerzas fijo (en el centro de masa) con una part\'{\i}cula a una distancia $%
{\bf r}$ de \'{e}l con masa $\mu$ (masa reducida).

\noindent
As\'{\i} pues, el movimiento de dos cuerpos en torno a su centro de masa 
debido a
una fuerza central se puede reducir siempre a un problema equivalente de un
cuerpo.

\bigskip

\noindent {\bf 2.2 Ecuaciones de movimiento.}

\noindent
Ahora nos limitaremos a fuerzas centrales conservativas, en donde el
potencial es funci\'{o}n solo de $r$, $V(r)$, por lo que la fuerza solo
estara dirijida a lo largo de{\bf \ }${\bf r}$. Ya vimos que para resolver
el problema solo necesitamos considerar el problema de una part\'{\i}cula de
masa $m$ que se mueva en torno a un centro de fuerzas fijo, el cual sera el
origen del sistema de coordenadas. Como el potencial solo depende de $r$, el
problema tiene simetr\'{\i}a esf\'{e}rica, es decir, cualquier rotaci\'{o}n,
en torno a cualquier eje fijo, puede no tener efecto sobre la soluci\'{o}n.
Por tanto, una coordenada angular que represente rotaci\'{o}n alrededor de
un eje fijo debe de ser c\'{\i}clica, lo cual da una simplificaci\'{o}n
considerable al problema. Debido a la simetr\'{\i}a esf\'{e}rica el vector
de momento angular total 
\[
{\bf L}={\bf r}\times{\bf p} 
\]
se conserva. Se deduce, por tanto, que ${\bf r}$ es perpendicular a la
direcci\'{o}n fija de ${\bf L}$. Ahora si ${\bf L}=0$ el movimiento debe ser
a lo largo de una recta que pase por el centro de fuerzas, ya que para ${\bf %
L}=0$ ${\bf r}$ y ${\bf \dot{r}}$ son paralelas, cosa que solo se cumple en
el movimiento rectilineo, por tanto el movimiento bajo fuerza central es
siempre un movimiento plano.

\noindent
Ahora bien, tomando el eje $z$ en direcci\'{o}n de ${\bf L}$, el movimiento
tendra simpre lugar en un plano normal al eje $z$. La coordenada
esf\'{e}rica $\phi$ tendra entonces el valor constante $\pi/2$ y podemos
prescindir de ella en el estudio que sigue. La conservaci\'{o}n del momento
cin\'{e}tico proporciona 3 constantes de movimiento independientes. De
hecho, dos de ellas, que expresan la {\it direcci\'{o}n} constante del
momento cin\'{e}tico, se han utilizado para reducir el problema de 3 grados
de libertad a dos. La tercera corresponde a la conservaci\'{o}n del modulo
de ${\bf L}$.

\noindent
En coordenadas polares la Lagrangiana es
\begin{equation}
L=\frac{1}{2}m(\dot{r}^{2}+r^{2}\dot{\theta}^{2})-V(r)~.  \label{eq4}
\end{equation}
como vimos $\theta $ es coordenada c\'{\i}clica cuya cantidad de momento
canonico es el momento cin\'{e}tico 
\[
p_{\theta }=\frac{\partial L}{\partial \dot{\theta}}=mr^{2}\dot{\theta}~, 
\]
entonces una de las dos ecuaciones de movimiento sera 
\begin{equation}
\dot{p}_{\theta }=\frac{d}{dt}(mr^{2}\dot{\theta})=0~,  \label{eq5}
\end{equation}
lo que nos conduce a 
\begin{equation}
mr^{2}\dot{\theta}=l=cte~,  \label{eq6}
\end{equation}
donde $l$ es la magnitud constante del momento cin\'{e}tico. De la
ecuaci\'{o}n (5) se deduce tambien que 
\begin{equation}
\frac{d}{dt}
{r^{2}\dot{\theta} \overwithdelims() 2}
=0.  \label{eq7}
\end{equation}

\noindent
Se introduce el termino $1/2$ por la raz\'{o}n de
que $(r^{2}\dot{\theta})/2$
es la {\it velocidad areolar} (\'{a}rea barrida por el vector de
posici\'{o}n por unidad de tiempo).

\noindent
La conservaci\'{o}n del momento cin\'{e}tico es equivalente a decir que la
velocidad areolar es constante. Tenemos aqui la demostraci\'{o}n de la
segunda ley de Kepler del movimiento planetario: el radio vector barre
areas iguales en tiempos iguales. Sin embargo debemos recalcar que la
constancia de la velocidad areolar es una propiedad de movimiento debido a
una fuerza central y no esta limitada a una ley de fuerza inversamente
proporcional al cuadrado de la distancia.

\noindent
La ecuaci\'{o}n de Lagrange restante, para la coordenada $r$ es 
\begin{equation}
\frac{d}{dt}(m\dot{r})-mr\dot{\theta}^{2}+\frac{\partial V}{\partial r}=0~.
\label{eq8}
\end{equation}
Designando por $f(r)$ el valor de la fuerza, podemos escribir la
ecuaci\'{o}n en la forma 
\begin{equation}
m\ddot{r}-mr\dot{\theta}^{2}=f(r)~.  \label{eq9}
\end{equation}
Utilizando la ecuaci\'{o}n (6), esta ecuaci\'{o}n se puede reescribir como 
\begin{equation}
m\ddot{r}-\frac{l^{2}}{mr^{3}}=f(r).  \label{eq10}
\end{equation}

\noindent
Basandonos en el teorema de la conservaci\'{o}n de la energ\'{\i}a 
\begin{equation}
E=T+V=\frac{1}{2}m(\dot{r}^{2}+r^{2}\dot{\theta}^{2})+V(r)~.  \label{eq11}
\end{equation}
$E$ es una constante de movimiento. Esto lo podemos deducir de las
ecuaci\'{o}nes de movimiento. La ecuaci\'{o}n (10) la podemos escribir en la
forma 
\begin{equation}
m\ddot{r}=-\frac{d}{dr}\left[ V(r)+\frac{1}{2}\frac{l^{2}}{mr^{2}}\right]~.
\label{eq12}
\end{equation}
Ahora multipliquemos por $\dot{r}$ ambos lados de la ecuaci\'{o}n 
\[
m\ddot{r}\dot{r}=\frac{d}{dt}(\frac{1}{2}m\dot{r})=-\frac{d}{dt}\left[ V(r)+%
\frac{1}{2}\frac{l^{2}}{mr^{2}}\right]~, 
\]
o bien 
\[
\frac{d}{dt}\left[ \frac{1}{2}m\dot{r}^{2}+V(r)+\frac{1}{2}\frac{l^{2}}{%
mr^{2}}\right] =0~. 
\]
Por lo tanto 
\begin{equation}
\frac{1}{2}m\dot{r}^{2}+V(r)+\frac{1}{2}\frac{l^{2}}{mr^{2}}=cte
\label{eq13}
\end{equation}
y ya que $(l^{2}/2mr^{2})=(mr^{2}\dot{\theta}/2)$, la ecuaci\'{o}n (13) se
reduce a (11).

\noindent
Ahora resolvamos las ecuaciones de movimiento para $r$ y $\theta $.
Despejando $\dot{r}$ de la ecuaci\'{o}n (\ref{eq13}), tenemos 
\begin{equation}
\dot{r}=\sqrt[2]{\frac{2}{m}(E-V-\frac{l^{2}}{2mr^{2}})}~,  \label{eq14}
\end{equation}
o bien 
\begin{equation}
dt=\frac{dr}{\sqrt[2]{\frac{2}{m}(E-V-\frac{l^{2}}{2mr^{2}})}}~.  \label{eq15}
\end{equation}

\noindent
Sea $r_{0}$ el valor de $r$ al timepo $t=0$. La integral de los 2 miembros
de la ecuaci\'{o}n toma la forma 
\begin{equation}
t=\int\nolimits_{r_{0}}^{r}\frac{dr}{\sqrt[2]{\frac{2}{m}(E-V-\frac{l^{2}}{%
2mr^{2}})}}.  \label{eq16}
\end{equation}

\noindent
Esta ecuaci\'{o}n nos da $t$ en funci\'{o}n de $r$ y de las constantes de
integraci\'{o}n $E$, $l$ y $r_{0}$. No obstante se puede invertir, al menos
formalmente, para dar $r$ en funci\'{o}n de $t$ y de las constantes. Una vez
hallada $r$, se deduce inmediatamente $\theta $ a partir de la ecuaci\'{o}n
(6), que se puede escribir 
\begin{equation}
d\theta =\frac{ldt}{mr^{2}}~.  \label{eq17}
\end{equation}

\noindent
Si $\theta _{0}$ es el valor inicial de $\theta $, entonces (17) sera 
\begin{equation}
\theta =l\int\nolimits_{0}^{t}\frac{dt}{mr^{2}(t)}+\theta _{0}.  \label{eq18}
\end{equation}

\noindent
Asi pues hemos ya obtenido las ecuaciones de movimiento para las
variables $r$ y $\theta$.\bigskip

\noindent {\bf 2.3 Ecuaci\'{o}n diferencial de la \'{o}rbita.}

\noindent
Al tratar detalles concretos de problemas de fuerzas centrales reales
conviene efectuar un cambio en la orientaci\'{o}n de nuestro tratamiento.
Hasta ahora, resolver el problema significa hallar $r$ y $\theta $ en
funci\'{o}n del tiempo siendo $E$, $l$, etc. constantes de integraci\'{o}n.
Pero muy a menudo, lo que realmente buscamos es la ecuaci\'{o}n de la
\'{o}rbita, es decir, la dependencia entre $r$ y $\theta $, eliminando el
par\'{a}metro $t$. En el caso de problemas de fuerzas centrales, esta
eliminaci\'{o}n es particularmente sencilla ya que $t$ solo figura en las
ecuaciones de movimiento en forma de variable respecto a la cual se deriva.
En verdad, una ecuaci\'{o}n de movimiento, (\ref{eq6}), no hace sino darnos
la una relaci\'{o}n definida entre una variaci\'{o}n infinitesimal $dt$ y la
variaci\'{o}n $d\theta $ correspondiente 
\begin{equation}
ldt=mr^{2}d\theta .  \label{eq19}
\end{equation}

\noindent
La relaci\'{o}n correspondiente entre sus derivadas respecto a $t$ y $\theta 
$ es 
\begin{equation}
\frac{d}{dt}=\frac{l}{mr^{2}}\frac{d}{d\theta }.  \label{eq20}
\end{equation}

\noindent
Estas relaciones se pueden usar para convertir la ecuaci\'{o}n (\ref{eq10})
en una ecuaci\'{o}n diferente para la \'{o}rbita. Tambien se pueden
solucionar las ecuaciones de movimiento formalmente y llegar a la
ecuaci\'{o}n de la \'{o}rbita. De momento continuemos con la primera
p\'{o}sibilidad.

\noindent
A partir de la ecuaci\'{o}n (\ref{eq20}) podemos escribir la segunda
derivada con respecto a $t$%
\[
\frac{d^{2}}{dt^{2}}=\frac{d}{d\theta }\frac{l}{mr^{2}}\left( \frac{d}{%
d\theta }\frac{l}{mr^{2}}\right) 
\]
y la ecuaci\'{o}n de Lagrange para $r$, (\ref{eq10}), queda en la forma 
\begin{equation}
\frac{l}{r^{2}}\frac{d}{d\theta }\left( \frac{l}{mr^{2}}\frac{dr}{d\theta }%
\right) -\frac{l}{mr^{3}}=f(r)~.  \label{eq21}
\end{equation}
Pero 
\[
\frac{1}{r^{2}}\frac{dr}{d\theta }=-\frac{d(1/r)}{d\theta }~. 
\]
Haciendo el cambio de variable $u=1/r$, tenemos 
\begin{equation}
\frac{l^{2}u^{2}}{m}\left( \frac{d^{2}u}{d\theta ^{2}}+u\right) =-f\left( 
\frac{1}{u}\right)~.  \label{eq22}
\end{equation}
Como 
\[
\frac{d}{du}=\frac{dr}{d\theta }\frac{d}{dr}=-\frac{1}{u^{2}}\frac{d}{dr}~, 
\]
la ecuaci\'{o}n (\ref{eq22}) puede escribirse en la forma 
\begin{equation}
\frac{d^{2}u}{d\theta ^{2}}+u=-\frac{m}{l^{2}}\frac{d}{du}V\left( \frac{1}{u}%
\right) .  \label{eq23}
\end{equation}

\noindent
Cualquiera de las dos ecuaciones (\ref{eq22}) o (\ref{eq23}) es la ecuacion
diferencial de la \'{o}rbita si se conoce la fuerza $f$ o el potencial $V$.
Inversamente si conocemos la ecuaci\'{o}n de la \'{o}rbita podemos seguir
los pasos inversos y obtener $f$ o $V$.

\noindent
Para una ley de fuerza particular cualquiera, la ecuaci\'{o}n de la
\'{o}rbita debe obtenerse integrando la ecuaci\'{o}n (\ref{eq22}) en una u
otra forma. Puesto que ya se ha realizado la mayor parte del trabajo al
resolver la ecuaci\'{o}n (\ref{eq10}), solo queda eliminar $t$ de la
soluci\'{o}n (\ref{eq15}) por medio de (\ref{eq19}), 
\begin{equation}
d\theta =\frac{ldr}{mr^{2}\cdot \sqrt[2]{\frac{2}{m}\left[ E-V(r)-\frac{l^{2}%
}{2mr^{2}}\right] }}~,  \label{eq24}
\end{equation}
o 
\begin{equation}
\theta =\int_{r_{0}}^{r}\frac{dr}{r^{2}\cdot \sqrt[2]{\frac{2mE}{l^{2}}-%
\frac{2mU}{l^{2}}-\frac{1}{r^{2}}}}+\theta _{0}~.  \label{eq25}
\end{equation}
Haciendo el cambio de variable $u=1/r$, 
\begin{equation}
\theta =\theta _{0}-\int_{u_{0}}^{u}\frac{du}{\sqrt[2]{\frac{2mE}{l^{2}}-%
\frac{2mU}{l^{2}}-u^{2}}}~,  \label{eq26}
\end{equation}
lo que es la soluci\'{o}n formal para la ecuaci\'{o}n de la \'{o}rbita.

\bigskip

\noindent {\bf 2.4 Problema de Kepler: Fuerza inversamente proporcional al
cuadrado de la distancia}

\noindent
La ley inversamente proporcional al cuadrado de la distancia es la mas
importante de todas las leyes de fuerzas centrales por lo que le daremos un
tratamiento detallado. En este caso la fuerza y el potencial son: 
\begin{equation}
f=-\frac{k}{r^{2}}\;\;\;\;\;{\rm y}\;\;\;\;\;V=-\frac{k}{r}~.  \label{eq27}
\end{equation}

\noindent
Para integrar la ecuaci\'{o}n de la \'{o}rbita sustituyamos (23) en (22), 
\begin{equation}
\frac{d^{2}u}{d\theta ^{2}}+u=-\frac{mf(1/u)}{l^{2}u^{2}}=\frac{mk}{l^{2}}~.
\label{eq28}
\end{equation}
Hacemos el cambio de variable $y=u-\frac{mk}{l^{2}}$~, para que la
ecuaci\'{o}n diferencial quede en la forma 
\[
\frac{d^{2}y}{d\theta ^{2}}+y=0~, 
\]
cuya soluci\'{o}n es 
\[
y=B\cos (\theta -\theta {\acute{}}~,
) 
\]
siendo $B$ y $\theta {\acute{}}$ las correspondientes constantes de
integraci\'{o}n. La solucin en
fuenci\'{o}n de $r$ es 
\begin{equation}
\frac{1}{r}=\frac{mk}{l^{2}}\left[ 1+e\cos (\theta -\theta {\acute{}}
)\right] ,  \label{eq29}
\end{equation}
donde 
\[
e=B\frac{l^{2}}{mk}~. 
\]

\noindent
Podemos obtener la ecuaci\'{o}n de la \'{o}rbita a partir de la soluci\'{o}n
formal (\ref{eq26}). A pesar de que este procedimiento es mas largo que
resolver la ecuaci\'{o}n (\ref{eq28}), resulta ilustrativo hacerlo ya que la
constante de integraci\'{o}n $e$ se evalua automaticamente en funci\'{o}n de
$E$ y $l$.

\noindent
Escribamos (\ref{eq26}) en la forma 
\begin{equation}
\theta =\theta {\acute{}}
-\int \frac{du}{\sqrt[2]{\frac{2mE}{l^{2}}-\frac{2mU}{l^{2}}-u^{2}}}~,
\label{eq30}
\end{equation}
donde ahora se trata de una integral indefinida. La
cantidad $\theta {\acute{}}$ que aparece en (\ref{eq30}) es una constante de
integraci\'{o}n
determinada por las condiciones iniciales y no tiene por que ser el angulo
inicial $\theta _{0}$ al tiempo $t=0$. La soluci\'{o}n a este tipo de
integrales es 
\begin{equation}
\int \frac{dx}{\sqrt[2]{\alpha +\beta x+\gamma x^{2}}}=\frac{1}{\sqrt[2]{%
-\gamma }}\arccos \left[ -\frac{\beta +2\gamma x}{\sqrt[2]{q}}\right]~,
\label{eq31}
\end{equation}
donde 
\[
q=\beta ^{2}-4\alpha \gamma . 
\]

\noindent
Para aplicar este tipo de soluciones a la ecuaci\'{o}n (\ref{eq30}) debemos
hacer 
\[
\alpha =\frac{2mE}{l^{2}},\;\;\;\;\;\beta =\frac{2mk}{l^{2}}%
,\;\;\;\;\;\gamma =-1, 
\]
y el discriminante $q$ sera por lo tanto 
\[
q=\left( \frac{2mk}{l^{2}}\right) ^{2}\left( 1+\frac{2El^{2}}{mk^{2}}\right)
. 
\]

\noindent
Con estas sustitucion (\ref{eq30}) queda en la forma 
\[
\theta =\theta {\acute{}}
-\arccos \left[ \frac{\frac{l^{2}u}{mk}-1}{\sqrt[2]{1+\frac{2El^{2}}{mk^{2}}}
}\right]~. 
\]
Despejando $u\equiv 1/r$, la ecuaci\'{o}n de la \'{o}rbita resulta ser 
\begin{equation}
\frac{1}{r}=\frac{mk}{l^{2}}\left[ 1+\sqrt[2]{1+\frac{2El^{2}}{mk^{2}}}\cos
(\theta -\theta {\acute{}})\right] .  \label{eq32}
\end{equation}

\noindent
Comparando (\ref{eq32}) con la ecuaci\'{o}n (\ref{eq29}) observamos que el
valor de $e$ es: 
\begin{equation}
e=\sqrt[2]{1+\frac{2El^{2}}{mk^{2}}}~.  \label{eq33}
\end{equation}

\noindent
La naturaleza de la \'{o}rbita depende del valor de $e$ seg\'{u}n el esquema
siguiente:

\begin{center}
\begin{tabular}{lll}
$e>1,$ & $E>0:$ & hip\'{e}rbola, \\ 
$e=1,$ & $E=0:$ & par\'{a}bola, \\ 
$e<1,$ & $E<0:$ & elipse, \\ 
$e=0$ & $E=-\frac{mk^{2}}{2l^{2}}:$ & circunferencia.
\end{tabular}

\bigskip
\end{center}

\noindent {\bf 2.5 Dispersi\'{o}n por un centro de fuerzas.}

\noindent
Desde un punto de vista hist\'{o}rico, el inter\'{e}s acerca de las fuerzas
centrales surgio en los problemas astron\'{o}micos del movimiento
planetario. Sin embargo, no hay raz\'{o}n alguna para que s\'{o}lo las
consideremos en este tipo de problemas. Otra cuesti\'{o}n que podemos
estudiar mediante la Mec\'{a}nica Cl\'{a}sica es la {\it dispersi\'{o}n} de
part\'{\i}culas por campos de fuerzas centrales. Desde luego, si el
tama\~{n}o de las part\'{\i}culas es del orden at\'{o}mico, debemos esperar
que los resultados espec\'{\i}ficos de un tratamiento cl\'{a}sico sean a
menudo incorrectos desde un punto de vista f\'{\i}sico, ya en que tales
regiones suelen ser importantes los efectos cu\'{a}nticos. A pesar de todo
hay predicciones cl\'{a}sicas que siguen siendo v\'{a}lidas con buena
aproximaci\'{o}n. M\'{a}s importante a\'{u}n, los procedimientos de {\it %
descripci\'{o}n} de los fen\'{o}menos de dispersi\'{o}n son los mismos en la
Mec\'{a}nica cl\'{a}sica que en la cu\'{a}ntica; podemos aprender a hablar
el lenguaje igualmente bien bas\'{a}ndonos en la Mec\'{a}nica cl\'{a}sica.

\noindent
En su formulaci\'{o}n para un cuerpo, el problema de la dispersi\'{o}n se
ocupa de la desviaci\'{o}n de part\'{\i}culas por un {\it centro de fuerzas}.
Consideremos un haz uniforme de part\'{\i}culas -da igual que sean
electrones, protones o planetas- todas de igual masa y energ\'{\i}a que
inciden sobre un centro de fuerzas. Podemos suponer que la fuerza disminuye
tendiendo a cero a grandes distancias. El haz incidente se caracteriza
especificando su {\it intensidad }$I$ (tambi\'{e}n llamada densidad de
flujo), la cual da el n\'{u}mero de part\'{\i}culas que atraviesan en unidad
de tiempo la unidad de superficie colocada normalmente al haz. Al acercarse
una part\'{\i}cula al centro de fuerzas ser\'{a} atra\'{\i}da o repelida y
su \'{o}rbita se desviar\'{a} de la trayectoria rectil\'{\i}nea incidente.
Despu\'{e}s de haber pasado el centro de fuerzas, la fuerza que se ejerce
sobre la part\'{\i}cula ir\'{a} disminuyendo de manera que la \'{o}rbita
tender\'{a} de nuevo a tener forma rectil\'{\i}nea. En general, la
direcci\'{o}n final del movimiento no coincide con la incidente y diremos
que la part\'{\i}cula se ha desviado o dispersado. Por definici\'{o}n la 
{\it secci\'{o}n eficaz}, $\sigma (\Omega )$, {\it de dispersi\'{o}n en una
direcci\'{o}n dada} es 
\begin{equation}
\sigma (\Omega )d\Omega =\frac{dN}{I},
\label{eq34}
\end{equation}
donde $dN$ es el n\'{u}mero de part\'{\i}culas dispersadas por unidad de
tiempo en un elemento de \'{a}ngulo s\'{o}lido $d\Omega$ en la direcci\'{o}n
$\Omega$. A menudo, a $\sigma (\Omega )$ se le llama tambi\'{e}n {\it %
secci\'{o}n eficaz diferencial de dispersi\'{o}n}. En el caso de fuerzas
centrales debe haber una simetr\'{\i}a total en torno al eje del haz
incidente, por lo que el elemento de \'{a}ngulo s\'{o}lido podr\'{a}
escribirse en la forma 
\begin{equation}
d\Omega =2\pi \sin \Theta d\Theta ,  \label{eq35}
\end{equation}
donde $\Theta $ es el \'{a}ngulo que forman las direcciones desviadas e
incidentes, al cual se le da el nombre de {\it \'{a}ngulo de dispersi\'{o}n}.

\noindent
Para una part\'{\i}cula dada cualquiera, las constantes de la \'{o}rbita y
por lo tanto la magnitud de la dispersi\'{o}n, est\'{a}n determinadas por su
energ\'{\i}a y su momento cin\'{e}tico. Conviene expresar el momento
cin\'{e}tico en funci\'{o}n de la energ\'{\i}a \ y de una cantidad $s$
llamada {\it par\'{a}metro de impacto} que es, por definici\'{o}n, la
distancia del centro de fuerzas a la recta soporte de la velocidad
incidente. Si $u_{0}$ es la velocidad incidente de la part\'{\i}cula,
tendremos 
\begin{equation}
l=mu_{0}s=s\cdot \sqrt[2]{2mE}.  \label{eq36}
\end{equation}

\noindent
Una vez fijadas $E$ y $s$, queda determinado un\'{\i}vocamente el \'{a}ngulo
de dispersi\'{o}n $\Theta $. De momento supondremos que valores diferentes
de $s$ no pueden llevar un mismo \'{a}ngulo de dispersi\'{o}n. Por tanto, el
n\'{u}mero de part\'{\i}culas dispersadas por un \'{a}ngulo s\'{o}lido $%
d\Omega $ comprendido entre $\Theta $ y $\Theta +d\Theta $ deber\'{a} ser
igual al n\'{u}mero de part\'{\i}culas incidentes cuyo par\'{a}metro de
impacto est\'{e} comprendido entre los valores correspondientes $s$ y $s+ds$%
: 
\begin{equation}
2\pi Is\left| ds\right| =2\pi \sigma (\Theta )I\sin \Theta \left| d\Theta
\right| .  \label{eq37}
\end{equation}

\noindent
En la ecuaci\'{o}n (\ref{eq37}) se han introducido los valores absolutos por
que los n\'{u}meros de part\'{\i}culas tiene que ser siempre positivos,
mientras que $s$ y $\Theta $ var\'{\i}an a menudo en sentidos opuestos. Si
consideramos $s$ funci\'{o}n de la energ\'{\i}a y del \'{a}ngulo de
dispersi\'{o}n correspondiente, 
\[
s=s(\Theta ,E), 
\]
la dependencia entre la secci\'{o}n eficaz diferencial y $\Theta $
vendr\'{a} dada por 
\begin{equation}
\sigma (\Theta )=\frac{s}{\sin \Theta }\left| \frac{ds}{d\Theta }\right| .
\label{eq38}
\end{equation}

\noindent
A partir de la ecuaci\'{o}n de la \'{o}rbita (\ref{eq25}) se puede obtener
directamente una expresi\'{o}n formal del \'{a}ngulo de dispersi\'{o}n.
Tambi\'{e}n ahora, para mayor sencillez, consideraremos el caso de una
dispersi\'{o}n puramente repulsiva. Como la \'{o}rbita debe ser
sim\'{e}trica respecto a la direcci\'{o}n del peri\'{a}pside, el \'{a}ngulo
de dispersi\'{o}n vendr\'{a} dado por 
\begin{equation}
\Theta =\pi -2\Psi~,  \label{eq39}
\end{equation}
donde $\Psi $ es el \'{a}ngulo que forma la direcci\'{o}n de la
as\'{\i}ntota incidente con la direcci\'{o}n del peri\'{a}pside. A su vez, $%
\Psi $ puede obtenerse de la ecuaci\'{o}n (\ref{eq25}) haciendo $%
r_{0}=\infty $ cuando $\theta _{0}=\pi $ (direcci\'{o}n incidente), por
consiguiente $\theta =\pi -\Psi $ cuando $r=r_{m}$, distancia de mayor
acercamiento. F\'{a}cilmente se llega a 
\begin{equation}
\Psi =\int\nolimits_{r_{m}}^{\infty }\frac{dr}{r^{2}\cdot \sqrt[2]{\frac{2mE%
}{l^{2}}-\frac{2mV}{l^{2}}-\frac{1}{r^{2}}}}~.  \label{eq40}
\end{equation}

\noindent
Expresando $l$ en funci\'{o}n del par\'{a}metro de impacto $s$ (ec. (\ref
{eq36})), resulta 
\begin{equation}
\Theta =\pi -2\int\nolimits_{r_{m}}^{\infty }\frac{sdr}{r\cdot \sqrt[2]{r^{2}%
\left[ 1-\frac{V(r)}{E}\right] -s^{2}}}~,  \label{eq41}
\end{equation}
o bien 
\begin{equation}
\Theta =\pi -2\int\nolimits_{0}^{u_{m}}\frac{sdu}{\sqrt[2]{1-\frac{v(u)}{E}%
-s^{2}u^{2}}}~.  \label{eq42}
\end{equation}

\noindent
Las ecuaciones (\ref{eq41}) y (\ref{eq42}) rara vez se utilizan, a no ser en
el c\'{a}lculo num\'{e}rico directo del \'{a}ngulo de dispersi\'{o}n. No
obstante, cuando se disponga de una expresi\'{o}n anal\'{\i}tica para las
\'{o}rbitas, se puede a menudo obtener una relaci\'{o}n entre $\Theta $ y $s$
casi por simple inspecci\'{o}n.

\bigskip

\noindent \underline{{\it EJEMPLO}}:

\noindent
Este ejemplo es hist\'{o}ricamente muy importante. Se trata de la 
dispersi\'{o}n repulsiva de part\'{\i}culas
cargadas por causa de un campo coulombiano. El campo de fuerzas dispersor es
el creado por una carga fija $-Ze$ al ejercerse sobre part\'{\i}culas
incidentes que tienen carga $-Z%
{\acute{}}%
e$; por tanto, la fuerza se puede escribir en la forma 
\[
f=\frac{ZZ{\acute{}}e^{2}}{r^{2}}~,
\]
es decir, se trata de una fuerza repulsiva inversamente proporcional al
cuadrado de la distancia. Escribamos la constante 
\begin{equation}
k=-ZZ%
{\acute{}}%
e^{2}.  \label{eq43}
\end{equation}

\noindent
La energ\'{\i}a $E$ es mayor que cero y la \'{o}rbita ser\'{a} una
hip\'{e}rbola de excentricidad dada por 
\begin{equation}
\epsilon =\sqrt[2]{1+\frac{2El^{2}}{m(ZZ{\acute{}}
e^{2})^{2}}}=\sqrt[2]{1+\left( \frac{2Es}{ZZ{\acute{}}
e^{2}}\right) ^{2}},  \label{eq44}
\end{equation}
donde hemos tenido en cuenta la ecuaci\'{o}n (\ref{eq36}). Si se toma igual
a $\pi $ el \'{a}ngulo $\theta {\acute{}}$ de la ecuaci\'{o}n (\ref{eq29}),
el peri\'{a}pside corresponder\'{a} a $\theta =0$ y la ecuaci\'{o}n de
la \'{o}rbita queda en la forma
\begin{equation}
\frac{1}{r}=\frac{mZZ{\acute{}}
e^{2}}{l^{2}}\left[ \epsilon \cos \theta -1\right] .  \label{eq45}
\end{equation}

\noindent
La direcci\'{o}n $\Psi $ de la as\'{\i}ntota de incidencia queda entonces
determinada por la condici\'{o}n $r\rightarrow \infty $: 
\[
\cos \Psi =\frac{1}{\epsilon}~, 
\]
o sea, seg\'{u}n la ecuaci\'{o}n (\ref{eq39}), 
\[
\sin \frac{\Theta }{2}=\frac{1}{\epsilon }~. 
\]

\noindent
Luego 
\[
\cot ^{2}\frac{\Theta }{2}=\epsilon ^{2}-1, 
\]
y utilizando la ecuaci\'{o}n (\ref{eq44}) 
\[
\cot \frac{\Theta }{2}=\frac{2Es}{ZZ{\acute{}}e^{2}}~.
\]

\noindent
La relaci\'{o}n funcional buscada entre el par\'{a}metro de impacto y el
\'{a}ngulo de dispersi\'{o}n ser\'{a} p\'{u}es, 
\begin{equation}
s=\frac{ZZ{\acute{}}
e^{2}}{2E}\cot \frac{\Theta }{2},  \label{eq46}
\end{equation}
de manera que efectuando la transformaci\'{o}n que exige la ecuaci\'{o}n
(\ref{eq38}), encontramos que $\sigma (\Theta )$ viene dada por
\begin{equation}
\sigma (\Theta )=\frac{1}{4}\left( \frac{ZZ{\acute{}}
e^{2}}{2E}\right) ^{2}\csc ^{4}\frac{\Theta }{2}.  \label{eq47}
\end{equation}

\noindent
La ecuaci\'{o}n (\ref{eq47}) da la famosa secci\'{o}n eficaz de
dispersi\'{o}n de Rutherford, quien la dedujo para la dispersi\'{o}n de
part\'{\i}culas $\alpha $ por los n\'{u}cleos at\'{o}micos. La mec\'{a}nica
cu\'{a}ntica da, en el limite no relativista, una secci\'{o}n eficaz
coincidente con este resultado cl\'{a}sico.

\noindent
En f\'{\i}sica at\'{o}mica tiene mucha importancia el concepto
de {\it secci\'{o}n eficaz total de dispersi\'{o}n} $\sigma _{T}$ cuya
definic\'{o}n es
\[
\sigma _{T}=\int\nolimits_{4\pi }\sigma (\Omega )d\Omega =2\pi
\int\nolimits_{0}^{\pi }\sigma (\Theta )d\Theta~. 
\]
No obstante, si intentamos calcular la secci\'{o}n eficaz total para
dispersi\'{o}n coulombiana sustituyendo la ecuaci\'{o}n (\ref{eq47}) en esta
definici\'{o}n obtenemos un resultado infinito. La raz\'{o}n f\'{\i}sica de
esto es f\'{a}cil de ver. Seg\'{u}n su definici\'{o}n, la secci\'{o}n eficaz
total es el n\'{u}mero de part\'{\i}culas que, por unidad de intensidad
incidente, se dispersan en todas direcciones. Ahora bien, el campo
coulombiano constituye un ejemplo de fuerza de $\ll $largo alcance$\gg $;
sus efectos se extienden hasta el infinito. Las desviaciones muy
peque\~{n}as solo tienen lugar en el caso de part\'{\i}culas de
par\'{a}metro de impacto muy grande. Por tanto, todas las part\'{\i}culas de
un haz incidente de extensi\'{o}n lateral infinita se desviar\'{\i}an
m\'{a}s o menos y deben de incluirse en la secci\'{o}n eficaz total de
dispersi\'{o}n. Queda claro, pues, que el valor infinito de $\sigma _{T}$ no
es peculiar del campo coulombiano, tiene lugar en Mec\'{a}nica cl\'{a}sica
siempre que el campo dispersor sea diferente de cero a todas las distancias
independientemente de lo grande que sean.

\bigskip
\bigskip

\begin{center}  BIBLIOGRAFIA COMPLEMENTARIA  \end{center}

\bigskip

\noindent
L.S. Brown, {\it Forces giving no orbit precession}, Am. J. Phys. 46, 930
(1978)

\bigskip

\noindent
H. Goldstein, {\it More on the prehistory of the Laplace-Runge-Lenz vector},
Am. J. Phys. 44, 1123 (1976)



\newpage



\centerline{{\Large 3. CUERPO R\'{I}GIDO}}

\bigskip
\bigskip

\noindent
{\bf Pr\'ologo}:
Por las particularidades de su movimiento, el estudio del cuerpo r\'{\i}gido
ha generado nuevas tecnicas y procedimientos matematicos interesantes.

\bigskip
\bigskip

{\bf CONTENIDO:}

\bigskip

3.1 Definici\'on.

\bigskip

3.2 Grados de libertad.

\bigskip

3.3 Tensor de inercia (con ejemplo).

\bigskip

3.4 Momento angular.

\bigskip

3.5 Ejes principales de inercia (con ejemplo).

\bigskip

3.6 El teorema de los ejes paraleles (con 2 ejemplos).

\bigskip

3.7 Dinamica del cuerpo r\'{\i}gido (con ejemplo).

\bigskip

3.8 Trompo sim\'etrico libre de torcas.

\bigskip

3.9 Angulos de Euler.

\bigskip

3.10 Trompo sim\'etrico con un punto fijo.

\bigskip

\newpage

\noindent {\bf 3.1 Definici\'{o}n}.

\noindent
Un cuerpo r\'{\i}gido se define como un
sistema de part\'{\i}culas cuyas distancias relativas est\'{a}n obligadas a
permanecer absolutamente fijas.

\bigskip

\noindent{\bf 3.2 Grados de libertad}.

\noindent
Para describir el movimiento general de un s\'{o}lido r\'{\i}gido en el
espacio tridimensional s\'{o}lo requerimos de 6 cantidades, por ejemplo: las
3 coordenadas del centro de masa medidas desde un sistema inercial y 3
\'{a}ngulos para especificar la orientaci\'{o}n del s\'{o}lido (o de un
sistema fijo en el s\'{o}lido con origen en el centro de masa) decimos que
un cuerpo r\'{\i}gido en el espacio tiene 6 grados de libertad.

\noindent
El n\'{u}mero de grados de libertad puede ser menor en los casos en que el
s\'{o}lido est\'{a} sujeto a restricciones, por ejemplo:

\begin{itemize}
\item  Si el s\'{o}lido s\'{o}lo gira alrededor de un eje m\'{o}vil es de un
grado de libertad (basta con un angulo).

\item  Si el s\'{o}lido se mueve en el plano, su movimiento es mas general
requiere de 5 cantidades (2 grados de libertad traslacional y 3 grados de
libertad rotacional).
\end{itemize}

\bigskip

\noindent{\bf 3.3 Tensor de inercia}.

\noindent
Consideremos que el cuerpo constituido por $N$ part\'{\i}culas de masas $%
m_{\alpha }$, $\alpha =1,2,3...,N$. Si el cuerpo rota con velocidad angular $%
{\bf \omega }$ alrededor de un punto fijo del cuerpo, y este punto a su vez
se mueve a velocidad ${\bf v}$ respecto al sistema fijo (inercial), entonces
la velocidad de la $\alpha$ -\'esima part\'{\i}cula respecto al
sistema inercial est\'{a} dada por 
\setcounter{equation} {0}\\
\begin{equation}
{\bf v}_{\alpha }={\bf v+\omega \times r}_{\alpha }.  \label{eq1a}
\end{equation}

\noindent
La energ\'{\i}a cin\'{e}tica de la $\alpha$ -\'esima part\'{\i}cula es
\begin{equation}
T_{\alpha }=\frac{1}{2}m_{\alpha }{\bf v}_{\alpha }^{2}~,  \label{eq2a}
\end{equation}
donde 
\begin{equation}
{\bf v}_{\alpha }^{2}={\bf v}_{\alpha }\cdot {\bf v}_{\alpha }=({\bf %
v+\omega \times r}_{\alpha })\cdot ({\bf v+\omega \times r}_{\alpha }) 
\nonumber
\end{equation}
\[
={\bf v}\cdot {\bf v}+2{\bf v}\cdot ({\bf \omega \times r}_{\alpha })+({\bf %
\omega \times r}_{\alpha })\cdot ({\bf \omega \times r}_{\alpha }) 
\]
\begin{equation}
={\bf v}^{2}+2{\bf v(\omega \times r}_{\alpha })+({\bf \omega \times r}%
_{\alpha })^{2}.  \label{eq3a}
\end{equation}

\noindent
Entonces la energ\'{\i}a total es 
\begin{eqnarray}
T& =\sum_{\alpha }T_{\alpha }=\sum_{\alpha }\frac{1}{2}m_{\alpha }{\bf v}
^{2}+\sum_{\alpha }m_{\alpha }\left[ {\bf v\cdot }\left( {\bf \omega \times r
}_{\alpha }\right) \right] + \nonumber \\
& +\frac{1}{2}\sum_{\alpha }m_{\alpha }({\bf \omega \times r}_{\alpha })^{2}~;
\nonumber \\
T& =\frac{1}{2}M{\bf v}^{2}+{\bf v\cdot }\left[ {\bf \omega \times }%
\sum_{\alpha }m_{\alpha }{\bf r}_{\alpha }\right] +\frac{1}{2}\sum_{\alpha}
m_{\alpha }\left( {\bf \omega \times r}_{\alpha }\right) ^{2}. \nonumber
\end{eqnarray}

\noindent
Si el origen esta fijo al s\'{o}lido lo elegimos en el centro de masas,
entonces 
\[
{\bf R=}\frac{\sum_{\alpha }m_{\alpha }{\bf r}_{\alpha }}{M}=0, 
\]
por lo que 
\begin{equation}
T=\frac{1}{2}M{\bf v}^{2}+\frac{1}{2}\sum_{\alpha }m_{\alpha }\left( {\bf %
\omega \times r}_{\alpha }\right) ^{2}  \label{eq4a}
\end{equation}
\begin{equation}
T=T_{trans}+T_{rot}  \label{eq5a}
\end{equation}
donde 
\begin{equation}
T_{trans}=\frac{1}{2}\sum_{\alpha }m_{\alpha }{\bf v}^{2}=\frac{1}{2}
M{\bf v}^{2}  \label{eq6a}
\end{equation}
\begin{equation}
T_{rot}=\frac{1}{2}\sum_{\alpha }m_{\alpha }\left( {\bf \omega \times r}%
_{\alpha }\right) ^{2}.  \label{eq7a}
\end{equation}

\noindent
Ahora usaremos en la ecuaci\'{o}n (\ref{eq7a}) la identidad vectorial 
\begin{equation}
({\bf A}\times {\bf B})^{2}={\bf A}^{2}{\bf B}^{2}-({\bf A\cdot B})^{2}
\label{eq8a}
\end{equation}
entonces la ecuaci\'{o}n nos queda en la forma 
\[
T_{rot}=\frac{1}{2}\sum_{\alpha}m_{\alpha}\left[ {\bf \omega }^{2}{\bf r}%
^{2}-({\bf \omega \cdot r}_{\alpha })^{2}\right] 
\]
que en t\'{e}rminos de las componentes de ${\bf \omega }$ y ${\bf r}$%
\[
{\bf \omega }={\bf (}\omega _{1},\omega _{2},\omega _{3})\;\;\;{\rm y}\;\;\;%
{\bf r}_{\alpha }=(x_{\alpha 1},x_{\alpha 2},x_{\alpha 3}) 
\]
\[
T_{rot}=\frac{1}{2}\sum_{\alpha }m_{\alpha }\left\{ \left( 
\mathop{\textstyle\sum}
_{i}\omega _{i}^{2}\right) \left( 
\mathop{\textstyle\sum}
_{k}x_{\alpha k}^{2}\right) -\left( 
\mathop{\textstyle\sum}
_{i}\omega _{i}x_{\alpha i}\right) \left( 
\mathop{\textstyle\sum}
_{j}\omega _{j}x_{\alpha j}\right) \right\} . 
\]

\noindent
Ahora introducimos 
\[
\omega _{i}=%
\mathop{\textstyle\sum}%
_{j}\delta _{ij}\omega _{j} 
\]
\begin{equation}
T_{rot}=\frac{1}{2}\sum_{\alpha }\sum_{ij}m_{\alpha }\left\{ \omega
_{i}\omega _{j}\delta _{ij}\left( 
\mathop{\textstyle\sum}%
_{k}x_{\alpha k}^{2}\right) -\omega _{i}\omega _{j}x_{\alpha i}x_{\alpha
j}\right\}  \nonumber
\end{equation}
\begin{equation}
T_{rot}=\frac{1}{2}\sum_{ij}\omega _{i}\omega _{j}\sum_{\alpha }m_{\alpha }%
\left[ \delta _{ij}%
\mathop{\textstyle\sum}%
_{k}x_{\alpha k}^{2}-x_{\alpha i}x_{\alpha j}\right] .  \label{eq9a}
\end{equation}

\noindent
Podemos escribir $T_{rot}$ como 
\begin{equation}
T_{rot}=\frac{1}{2}\sum_{ij}I_{ij}\omega _{i}\omega _{j}  \label{eq10a}
\end{equation}
donde 
\begin{equation}
I_{ij}=\sum_{\alpha }m_{\alpha }\left[ \delta _{ij}
\mathop{\textstyle\sum}
_{k}x_{\alpha k}^{2}-x_{\alpha i}x_{\alpha j}\right] .  \label{eq11a}
\end{equation}

\noindent
Las 9 cantidades de $I_{ij}$ constituyen las componentes de de una nueva
cantidad matem\'{a}tica que denotamos por $\left\{ I_{ij}\right\} $ y se
llama {\it Tensor de Inercia}, $\left\{ I_{ij}\right\} $ se puede escribir
convenientemente mediante un arreglo matricial de ($3\times 3$) 
\[
\left\{ I_{ij}\right\} =\left( 
\begin{array}{ccc}
I_{11} & I_{12} & I_{13} \\ 
I_{21} & I_{22} & I_{23} \\ 
I_{31} & I_{32} & I_{33}
\end{array}
\right) 
\]
\begin{equation}
=\left( 
\begin{array}{ccc}
\sum_{\alpha }m_{\alpha }(x_{\alpha 2}^{2}+x_{\alpha 3}^{2}) & -\sum_{\alpha
}m_{\alpha }x_{\alpha 1}x_{\alpha 2} & -\sum_{\alpha }m_{\alpha }x_{\alpha
1}x_{\alpha 3} \\ 
-\sum_{\alpha }m_{\alpha }x_{\alpha 2}x_{\alpha 1} & \sum_{\alpha }m_{\alpha
}(x_{\alpha 1}^{2}+x_{\alpha 3}^{2}) & -\sum_{\alpha }m_{\alpha }x_{\alpha
2}x_{\alpha 3} \\ 
-\sum_{\alpha }m_{\alpha }x_{\alpha 3}x_{\alpha 1} & -\sum_{\alpha
}m_{\alpha }x_{\alpha 3}x_{\alpha 2} & \sum_{\alpha }m_{\alpha }(x_{\alpha
1}^{2}+x_{\alpha 2}^{2})
\end{array}
\right) .  \label{eq12a}
\end{equation}

\noindent
Podemos notar que $I_{ij}=I_{ji}$ por lo tanto $\left\{ I_{ij}\right\} $ es
un {\it tensor sim\'{e}trico}, por lo tanto solo hay 6 t\'{e}rminos
independientes. Los elementos diagonales de $\left\{ I_{ij}\right\} $ se
llaman {\it momentos de inercia} con respecto a los ejes de coordenadas, los
negativos de los elementos no diagonales se llaman
{\it productos de inercia}. Para una distribuci\'{o}n continua de masa de
densidad $\rho ({\bf r)}$, $\left\{ I_{ij}\right\}$ se escribe en lugar
de (\ref{eq11a}) como
\begin{equation}
I_{ij}=\int_{V}\rho ({\bf r)}\left[ \delta _{ij}%
\mathop{\textstyle\sum}%
_{k}x_{k}^{2}-x_{i}x_{j}\right] dV.  \label{eq13a}
\end{equation}

\bigskip

\noindent \underline{{\it EJEMPLO}}:

\noindent
Calcular los elementos $I_{ij}$ del tensor de
inercia $\left\{ I_{ij}\right\}$ para un cubo uniforme de lado $b$,
masa $M$, una esquina esta en el origen.
$$
I_{11} =\int\limits_{V}\rho \left[ x_{1}^{2}+x_{2}^{2}+x_{3}^{2}-x_{1}x_{1}
\right] dx_{1}dx_{2}dx_{3}
 =\rho
\int\limits_{0}^{b}\int\limits_{0}^{b}\int
\limits_{0}^{b}(x_{2}^{2}+x_{3}^{2})dx_{1}dx_{2}dx_{3}~.
$$

\noindent
El resultado de la integral tres dimencional es $I_{11}=\frac{2}{3}
(\rho b^{3})^{2}=\frac{2}{3}Mb^{2}$.
$$
I_{12} =\int\limits_{V}\rho (-x_{1}x_{2})dV 
 =-\rho
\int\limits_{0}^{b}\int\limits_{0}^{b}\int
\limits_{0}^{b}(x_{1}x_{2})dx_{1}dx_{2}dx_{3} 
 =-\frac{1}{4}\rho b^{5}=-\frac{1}{4}Mb^{2}~.
$$

\noindent
Vemos que las demas integrales son las mismas por lo que 
\[
I_{11}=I_{22}=I_{33}=\frac{2}{3}Mb^{2} 
\]
\[
\mathrel{\mathop{I_{ij}}\limits_{i\neq j}}%
=-\frac{1}{4}Mb^{2}~, 
\]
entonces la matriz queda de la forma 
\[
\left\{ I_{ij}\right\} =\left( 
\begin{array}{ccc}
\frac{2}{3}Mb^{2} & -\frac{1}{4}Mb^{2} & -\frac{1}{4}Mb^{2} \\ 
-\frac{1}{4}Mb^{2} & \frac{2}{3}Mb^{2} & -\frac{1}{4}Mb^{2} \\ 
-\frac{1}{4}Mb^{2} & -\frac{1}{4}Mb^{2} & \frac{2}{3}Mb^{2}
\end{array}
\right) . 
\]

\bigskip

\noindent{\bf 3.4 Momento angular.}

\noindent
El momento angular para el s\'{o}lido r\'{\i}gido constituido por $N$
part\'{\i}culas $m_{\alpha }$ esta dado por 
\begin{equation}
{\bf L=}\sum_{\alpha }{\bf r}_{\alpha }\times {\bf p}_{\alpha }~,
\label{eq14a}
\end{equation}
donde 
\begin{equation}
{\bf p}_{\alpha }=m_{\alpha }{\bf v}_{\alpha }=m_{\alpha }({\bf \omega }%
\times {\bf r}_{\alpha })~.  \label{eq15a}
\end{equation}
Sustituyendo (\ref{eq15a}) en la ecuaci\'{o}n (\ref{eq14a}), tenemos que 
\[
{\bf L=}\sum_{\alpha }m_{\alpha }{\bf r}_{\alpha }\times ({\bf \omega }%
\times {\bf r}_{\alpha })~. 
\]
Utilizando la identidad vectorial 
\[
{\bf A}\times ({\bf B}\times {\bf A)=(A}\cdot {\bf A}){\bf B}-({\bf A}\cdot 
{\bf B}){\bf A}={\bf A}^{2}{\bf B}-({\bf A}\cdot {\bf B}){\bf A}~, 
\]
tenemos 
\[
{\bf L=}\sum_{\alpha }m_{\alpha }({\bf r}_{\alpha }^{2}{\bf \omega -r}%
_{\alpha }({\bf \omega \cdot r}_{\alpha }). 
\]

\noindent
Tomando la $i$-\'esima componente del vector ${\bf L}$
\[
L_{i}=\sum_{\alpha }m_{\alpha }\left( \omega _{i}%
\mathop{\textstyle\sum}%
_{k}x_{\alpha k}^{2}\right) {\bf -}x_{\alpha i}\left( 
\mathop{\textstyle\sum}%
_{j}x_{\alpha j}\omega _{j}\right)~, 
\]
introduciendo la ecuaci\'{o}n 
\[
\omega _{i}=%
\mathop{\textstyle\sum}%
_{j}\omega _{j}\delta _{ij}~, 
\]
obtenemos 
\begin{eqnarray}
L_{i}& =\sum_{\alpha }m_{\alpha }\left( 
\mathop{\textstyle\sum}%
_{j}\delta _{ij}\omega _{j}%
\mathop{\textstyle\sum}%
_{k}x_{\alpha k}^{2}\right) {\bf -}\left( 
\mathop{\textstyle\sum}%
_{j}x_{\alpha j}x_{\alpha j}\omega _{j}\right) \\
& =\sum_{\alpha }m_{\alpha }%
\mathop{\textstyle\sum}%
_{j}\omega _{j}\delta _{ij}\left( 
\mathop{\textstyle\sum}%
_{k}x_{\alpha k}^{2}{\bf -}x_{\alpha i}x_{\alpha j}\right) \\
& =\sum_{j}\omega _{j}%
\mathop{\textstyle\sum}%
_{\alpha }m_{\alpha }\left( \delta _{ij}%
\mathop{\textstyle\sum}%
_{k}x_{\alpha k}^{2}{\bf -}x_{\alpha i}x_{\alpha j}\right)~.
\end{eqnarray}
Comparando con la ecuaci\'{o}n (\ref{eq11a}) 
\begin{equation}
L_{i}=\sum_{j}I_{ij}\omega _{j}~.  \label{eq16a}
\end{equation}
Esta ecuaci\'{o}n tambein se puede escribir en la forma 
\begin{equation}
{\bf L=}\left\{ I_{ij}\right\} {\bf \omega }~,  \label{eq17a}
\end{equation}
o 
\begin{equation}
\left( 
\begin{array}{c}
L_{1} \\ 
L_{2} \\ 
L_{3}
\end{array}
\right) =\left( 
\begin{array}{ccc}
I_{11} & I_{12} & I_{13} \\ 
I_{21} & I_{22} & I_{23} \\ 
I_{31} & I_{32} & I_{33}
\end{array}
\right) \left( 
\begin{array}{c}
\omega _{1} \\ 
\omega _{2} \\ 
\omega _{3}
\end{array}
\right) .  \label{eq18a}
\end{equation}

\noindent
La energ\'{\i}a cin\'{e}tica rotacional, $T_{rot}$, se puede relacionar con
el momento angular de la siguiente manera: multipliquemos la ecuaci\'{o}n (
\ref{eq16a}) por $\frac{1}{2}\omega _{i}$
\begin{equation}
\omega _{i}\frac{1}{2}L_{i}=\frac{1}{2}\omega _{i}\sum_{j}I_{ij}\omega _{j}~,
\label{eq19a}
\end{equation}
sumando sobre todas las $i$ da
\[
\sum_{i}\frac{1}{2}L_{i}\omega _{i}=\frac{1}{2}\sum_{ij}I_{ij}\omega
_{i}\omega _{j}~. 
\]
Al comparar esta ecuaci\'{o}n con (\ref{eq10a}), vemos que el segundo
t\'{e}rmino no es mas que $T_{rot}$, por lo tanto 
\begin{equation}
T_{rot}=\sum_{I}\frac{1}{2}L_{i}\omega _{i}=\frac{1}{2}{\bf L\cdot \omega}~ .
\label{eq20a}
\end{equation}

\noindent
Ahora, sustituimos (\ref{eq17a}) en la ecuaci\'{o}n (\ref{eq20a}), obtenemos
una relaci\'{o}n entre la $T_{rot}$ y el tensor de inercia 
\begin{equation}
T_{rot}=\frac{1}{2}{\bf \omega \cdot }\left\{ I_{ij}\right\} {\bf \cdot
\omega .}  \label{eq21a}
\end{equation}

\bigskip

\noindent{\bf 3.5 Ejes principales de inercia}.

\noindent
Consideremos que el tensor de inercia $\left\{ I_{ij}\right\} $ es diagonal,
es decir $I_{ij}=I_{i}\delta _{ij}$, la energ\'{\i}a cin\'{e}tica rotacional
y el momento angular quedarian expresadas en la siguiente forma 
\[
T_{rot}=\frac{1}{2}\sum_{ij}I_{ij}\omega _{i}\omega _{j} 
\]
\[
=\frac{1}{2}\sum_{ij}\delta _{ij}I_{i}\omega _{i}\omega _{j} 
\]
\begin{equation}
T_{rot}=\frac{1}{2}\sum_{i}I_{i}\omega _{i}^{2}  \label{eq22a}
\end{equation}
y el momento angular 
\[
L_{i}=\sum_{j}I_{ij}\omega _{j} 
\]
\[
=\sum_{j}\delta _{ij}I_{i}\omega _{j}=I_{i}\omega _{i} 
\]
\begin{equation}
{\bf L}={\bf I\omega .}  \label{eq23a}
\end{equation}

\noindent
Encontrar una expresi\'{o}n diagonal para $\left\{ I_{ij}\right\} $ equivale
a encontrar un nuevo sistema de 3 ejes, en los cuales la energ\'{\i}a
cin\'{e}tica y el momento angular se reducen a las expresiones (\ref{eq22a})
y (\ref{eq23a}), tales ejes se les llama {\it Ejes Principales de Inercia},
es decir dado un cierto sistema inicial de coordenadas en el cuerpo, podemos
pasar de \'{e}l a los ejes principales mediante una transformaci\'{o}n
ortogonal particular que, en consecuencia, se llama {\it transformaci\'{o}n
a los ejes principales}.

\noindent
Igualando las componentes de (\ref{eq17a}) y (\ref{eq23a}), tenemos 
\begin{eqnarray}
L_{1}& =I\omega _{1}=I_{11}\omega _{1}+I_{12}\omega _{2}+I_{13}\omega _{3} \\
L_{2}& =I\omega _{2}=I_{21}\omega _{1}+I_{22}\omega _{2}+I_{23}\omega _{3} \\
L_{3}& =I\omega _{3}=I_{31}\omega _{1}+I_{32}\omega _{2}+I_{33}\omega _{3}
~,
\end{eqnarray}

\noindent
las cuales son un conjunto de ecuaciones que se pueden reescribir 
\begin{eqnarray}
(I_{11}-I)\omega _{1}+I_{12}\omega _{2}+I_{13}\omega _{3}& =0  \label{eq24a}
\\
I_{21}\omega _{1}+(I_{22}-I)\omega _{2}+I_{23}\omega _{3}& =0  \nonumber \\
I_{31}\omega _{1}+I_{32}\omega _{2}+(I_{33}-I)\omega _{3}& =0~.  \nonumber
\end{eqnarray}

\noindent
Para obtener la soluci\'{o}n, el determinante del sistema debe ser cero 
\begin{equation}
\left| 
\begin{array}{ccc}
(I_{11}-I)\omega _{1} & I_{12}\omega _{2} & I_{13}\omega _{3} \\ 
I_{21}\omega _{1} & (I_{22}-I)\omega _{2} & I_{23}\omega _{3} \\ 
I_{31}\omega _{1} & I_{32}\omega _{2} & (I_{33}-I)\omega _{3}
\end{array}
\right| =0~.  \label{eq25a}
\end{equation}

\bigskip

\noindent
El desarrollo de este determinante es un polinomio de grado 3 en $I,$
llamado {\it polinomio caracter\'{\i}stico} y la ecuaci\'{o}n (\ref{eq25a}) se
llama {\it ecuaci\'{o}n secular} o {\it ecuaci\'{o}n caracter\'{\i}stica}. En
la pr\'{a}ctica, los momentos principales de inercia, por ser los valores
propios de ${\bf I}$, se hallan buscando las ra\'{\i}ces de la ecuaci\'{o}n
secular.

\bigskip

\noindent \underline{{\it EJEMPLO}}:

\noindent
Determinar los ejes principales de inercia para el
cubo del ejemplo anterior.

\bigskip

\noindent
Al sustituir los valores obtenidos en el ejemplo anterior en la ecuaci\'{o}n
(\ref{eq25a}) obtenemos: 
\[
\left| \left( 
\begin{array}{ccc}
(\frac{2}{3}\beta -I) & -\frac{1}{4}\beta & -\frac{1}{4}\beta \\ 
-\frac{1}{4}\beta & (\frac{2}{3}\beta -I) & -\frac{1}{4}\beta \\ 
-\frac{1}{4}\beta & -\frac{1}{4}\beta & (\frac{2}{3}\beta -I)
\end{array}
\right) \right| =0 ~,
\]
donde $\beta =Mb^{2}$, de donde obtenemos la caracter\'{\i}stica,
\[
\left( \frac{11}{12}\beta -I\right) \left( \frac{11}{12}\beta -I\right)
\left( \frac{1}{6}\beta -I\right) =0~, 
\]
entonces los eigenvalores o momentos principales de inercia son: 
\[
I_{1}=\frac{1}{6}\beta ,\;\;\;\;I_{2}=I_{3}=\frac{11}{12}\beta~, 
\]

\bigskip
\noindent
cuyos correspondientes eigenvalores son: 
\[
I=\frac{1}{6}\beta \leftrightarrow \frac{1}{\sqrt[2]{3}}\left( 
\begin{array}{c}
1 \\ 
1 \\ 
1
\end{array}
\right) ,\;\;\;\;I_{2},I_{3}=\frac{11}{12}\beta \leftrightarrow \frac{1}{%
\sqrt[2]{2}}\left\{ \left( 
\begin{array}{c}
-1 \\ 
1 \\ 
0
\end{array}
\right) ,\left( 
\begin{array}{c}
-1 \\ 
0 \\ 
1
\end{array}
\right) \right\}~. 
\]
Entonces la matriz que diagonaliza a $\left\{ I_{ij}\right\} $ es: 
\[
\lambda =\sqrt[2]{\frac{1}{3}}\left( 
\begin{array}{ccc}
1 & -\sqrt[2]{\frac{3}{2}} & -\sqrt[2]{\frac{3}{2}} \\ 
1 & \sqrt[2]{\frac{3}{2}} & 0 \\ 
1 & 0 & \sqrt[2]{\frac{3}{2}}
\end{array}
\right)~. 
\]
$\left\{ I_{ij}\right\} $ diagonalizado es: 
\[
\left\{ I_{ij}\right\} _{diag}=\left( \lambda
\right) ^{*}\left\{ I_{ij}\right\} \lambda =\left(
\begin{array}{ccc}
\frac{1}{6}\beta  & 0 & 0 \\ 
0 & \frac{11}{12}\beta  & 0 \\ 
0 & 0 & \frac{11}{12}\beta 
\end{array}
\right) .
\]

\bigskip

\noindent{\bf 3.6 El teorema de los ejes paralelos}.

\noindent
Supongamos que el sistema $x_{1},x_{2},x_{3}$ tiene su origen en el centro
de masas del cuerpo r\'{\i}gido. Un segundo sistema $X_{1},X_{2},X_{3}$,
tiene su origen en otra posici\'{o}n diferente al sistema anterior. la unica
condici\'{o}n es que sean paralelos, definamos los vectores ${\bf r}%
=(x_{1},x_{2},x_{3})$, ${\bf R}=(X_{1},X_{2},X_{3})$ y ${\bf a}%
=(a_{1},a_{2},a_{3})$, de tal manera que ${\bf R}={\bf r}+{\bf a}$ o en
t\'{e}rminos de sus componentes 
\begin{equation}
X_{i}=x_{i}+a_{i}.  \label{eq26a}
\end{equation}

\noindent
Sean $J_{ij}$ las componentes del tensor de inercia respecto al sistema $%
X_{1}X_{2}X_{3}$, 
\begin{equation}
J_{ij}=\sum_{\alpha }m_{\alpha }\left[ \delta _{ij}
\mathop{\textstyle\sum}
_{k}X_{\alpha k}^{2}-X_{\alpha i}X_{\alpha j}\right]~.  \label{eq27a}
\end{equation}
Sustituimos (\ref{eq26a}) en (\ref{eq27a}), 
\[
J_{ij}=\sum_{\alpha }m_{\alpha }\left[ \delta _{ij}
\mathop{\textstyle\sum}
_{k}(x_{\alpha k}+a_{k})^{2}-(x_{\alpha i}+a_{i})(x_{\alpha j}+a_{j})\right] 
\]
\begin{eqnarray}
& =\left[ 
\mathop{\textstyle\sum}
_{\alpha }m_{\alpha }\left( \delta _{ij}
\mathop{\textstyle\sum}
_{k}(x_{\alpha k})^{2}-x_{\alpha i}x_{\alpha j}\right) \right] +
\mathop{\textstyle\sum}
_{\alpha }m_{\alpha }\left( \delta _{ij}
\mathop{\textstyle\sum}
_{k}a_{k}^{2}-a_{i}a_{j}\right)  \label{eq28a} \\
& +\left[ 
\mathop{\textstyle\sum}
_{k}2a_{k}\delta _{ij}\left( 
\mathop{\textstyle\sum}
_{\alpha }m_{\alpha }x_{\alpha k}\right) -a_{j}\left( 
\mathop{\textstyle\sum}
_{\alpha }m_{\alpha }x_{\alpha j}\right) -a_{i}\left( 
\mathop{\textstyle\sum}
_{\alpha }m_{\alpha }x_{\alpha i}\right) \right]~.  \nonumber
\end{eqnarray}
Pero la coordenada del centro de masa se define como 
\[
\bar{x}=\frac{
\mathop{\textstyle\sum}_{\alpha }m_{\alpha }x_{\alpha }}{M}
\]
y como habiamos dicho antes, el origen esta en el centro de masa 
\[
(\bar{x}_{1},\bar{x}_{2},\bar{x}_{3})=(0,0,0)~. 
\]
Ahora si tambien comparamos primer t\'{e}rmino de (\ref{eq28a}) con la
ecuaci\'{o}n (\ref{eq11a}), tendremos: 
\begin{equation}
J_{ij}=I_{ij}+M(a^{2}\delta _{ij}-a_{i}a_{j})  \label{eq29a}
\end{equation}
y entonces los elementos del tensor de inercia $I_{ij}$ para el sistema 
del centro de masa estar\'{a}n dadas por: 
\begin{equation}
I_{ij}=J_{ij}-M(\delta _{ij}a^{2}-a_{i}a_{j})~.  \label{eq30a}
\end{equation}
Este es {\it el Teorema de los Ejes Paralelos}.

\bigskip

\noindent \underline{{\it EJEMPLO}}:

\noindent
Calcular $I_{ij}$ para el cubo anterior respecto a
un sistema paralelo al primer ejemplo y con origen en el centro de masa.

\noindent
Ya sabemos del ejemplo anterior que: 
\[
\left\{ J_{ij}\right\} =\left( 
\begin{array}{ccc}
\frac{2}{3}\beta & -\frac{1}{4}\beta & -\frac{1}{4}\beta \\ 
-\frac{1}{4}\beta & \frac{2}{3}\beta & -\frac{1}{4}\beta \\ 
-\frac{1}{4}\beta & -\frac{1}{4}\beta & \frac{2}{3}\beta
\end{array}
\right)~. 
\]
Ahora el vector ${\bf a}=(\frac{b}{2},\frac{b}{2},\frac{b}{2})$ y ${\bf a}%
^{2}=\frac{3}{4}b^{2}$, entonces usando la ecuaci\'{o}n (\ref{eq30a}) y el
hecho que $\beta =Mb^{2}$ tenemos, 
\begin{eqnarray}
I_{11}& =J_{11}-M(a^{2}-a_{1}^{2})=\frac{1}{6}Mb^{2} \\
I_{22}& =J_{22}-M(a^{2}-a_{2}^{2})=\frac{1}{6}Mb^{2} \\
I_{33}& =J_{33}-M(a^{2}-a_{3}^{2})=\frac{1}{6}Mb^{2} \\
I_{12}& =J_{12}-M(-a_{1}a_{2})=0 \\
I_{12}& =I_{13}=I_{23}=0~,
\end{eqnarray}
por lo tanto 
\[
\left\{ I\right\} =\left( 
\begin{array}{ccc}
\frac{1}{6}Mb^{2} & 0 & 0 \\ 
0 & \frac{1}{6}Mb^{2} & 0 \\ 
0 & 0 & \frac{1}{6}Mb^{2}
\end{array}
\right) . 
\]

\bigskip

{\bf \noindent } \underline{{\it EJEMPLO}}:

\noindent
Consideremos el caso en el que el vector $%
{\bf a}=(0,\frac{b}{2},\frac{b}{2})$ y $a^{2}=\frac{b^{2}}{2}$, entonces
nuestro nuevo tensor de inercia seria: 
\begin{eqnarray}
I_{11}& =J_{11}-M(a^{2}-a_{1}^{2})=\left( \frac{2}{3}Mb^{2}\right) -M\left( 
\frac{b^{2}}{2}-0\right) =\frac{1}{6}Mb^{2} \\
I_{22}& =J_{22}-M(a^{2}-a_{2}^{2})=\left( \frac{2}{3}Mb^{2}\right) -M\left( 
\frac{b^{2}}{2}-\frac{b^{2}}{4}\right) =\frac{5}{12}Mb^{2} \\
I_{33}& =J_{33}-M(a^{2}-a_{3}^{2})=\left( \frac{2}{3}Mb^{2}\right) -M\left( 
\frac{b^{2}}{2}-\frac{b^{2}}{4}\right) =\frac{5}{12}Mb^{2} \\
I_{12}& =J_{12}-M(-a_{1}a_{2})=\left( -\frac{1}{4}Mb^{2}\right) -M(0)
=-\frac{1}{4}Mb^{2} \\
I_{13}& =J_{13}-M(-a_{1}a_{3})=\left( -\frac{1}{4}Mb^{2}\right) -M(0)
=-\frac{1}{4}Mb^{2} \\
I_{23}& =J_{23}-M(-a_{2}a_{3})=\left( -\frac{1}{4}Mb^{2}\right)
-M(\frac{1}{4}Mb^{2})=0~,
\end{eqnarray}
entonces $\{I_{ij}\}$ es igual a: 
\[
\{I_{ij}\}=\left( 
\begin{array}{ccc}
\frac{1}{6}Mb^{2} & -\frac{1}{4}Mb^{2} & -\frac{1}{4}Mb^{2} \\ 
-\frac{1}{4}Mb^{2} & \frac{5}{12}Mb^{2} & 0 \\ 
-\frac{1}{4}Mb^{2} & 0 & \frac{5}{12}Mb^{2}
\end{array}
\right)~. 
\]

\bigskip

\noindent {\bf 3.7 Din\'{a}mica del cuerpo r\'{\i}gido.}

\noindent
La raz\'{o}n de cambio respecto al tiempo del momento angular ${\bf L}$ esta
dado por: 
\begin{equation}
\left( \frac{d{\bf L}}{dt}\right) _{inercial}={\bf N}^{(e)}.  \label{eq31a}
\end{equation}

\noindent
Para la descripci\'{o}n desde el sistema fijo al s\'{o}lido debemos usar la
identidad operadora 
\begin{equation}
{d \overwithdelims() dt}%
_{inercial}=%
{d \overwithdelims() dt}%
_{cuerpo}+{\bf \omega }\times~.  \label{eq32a}
\end{equation}
Aplicando este operador a la ecuaci\'{o}n (\ref{eq31a})
\begin{equation}
{d{\bf L} \overwithdelims() dt}%
_{inercial}=%
{d{\bf L} \overwithdelims() dt}%
_{cuerpo}+{\bf \omega }\times {\bf L.}  \label{eq33a}
\end{equation}
Entonces, en lugar de la ecuaci\'{o}n (\ref{eq31a}) tendremos 
\begin{equation}
{d{\bf L} \overwithdelims() dt}%
_{cuerpo}+{\bf \omega }\times {\bf L}={\bf N.}  \label{eq34a}
\end{equation}

\noindent
Ahora proyectamos la ecuaci\'{o}n (\ref{eq34a}) sobre los ejes principales
de inercia, supongamos que estos son $(x_{1},x_{2},x_{3})$, $T_{rot}$ y $%
{\bf L}$ se simplifican con tal elecci\'{o}n, por ejemplo: 
\begin{equation}
L_{i}=I_{i}\omega _{i}  \label{eq35a}
\end{equation}
la componente $i$-\'esima de (\ref{eq34a}) es
\begin{equation}
\frac{dL_{i}}{dt}+\epsilon _{ijk}\omega _{j}L_{k}=N_{i}~.  \label{eq36a}
\end{equation}
Ahora proyectando sobre los ejes principales de inercia y utilizando la
ecuaci\'{o}n (\ref{eq35a}), la ecuaci\'{o}n (\ref{eq36a}) toma la forma: 
\begin{equation}
I_{i}\frac{d\omega _{i}}{dt}+\epsilon _{ijk}\omega _{j}\omega _{k}I_{k}=N_{i}
\label{eq37a}
\end{equation}
ya que los elementos principales de inercia son independiantes del tiempo.
Entonces, asi obtenemos un sistema de ecuaciones 
\begin{eqnarray}
I_{1}\dot{\omega}_{1}+\omega _{2}\omega _{3}(I_{2}-I_{3})& =N_{1}
\label{eq38a} \\
I_{2}\dot{\omega}_{2}+\omega _{3}\omega _{1}(I_{3}-I_{1})& =N_{2}  \nonumber
\\
I_{3}\dot{\omega}_{3}+\omega _{1}\omega _{2}(I_{1}-I_{2})& =N_{3}~.  
\nonumber
\end{eqnarray}
Estas son las llamadas {\it Ecuaciones de Euler.}

\bigskip

\noindent \underline{{\it EJEMPLO}}:

\noindent
Rodamiento y deslizamiento de una bola de billar.
Demostrar que despues de un golpe horizontal la bola de desplaza resbalando
una distancia 
\[
x_{_{1}}=\frac{12u_{0}^{2}}{49\mu g}~,
\]
para despues empezar a rodar sin resbalar al tiempo 
\[
t_{1}=\frac{2u_{0}}{7\mu g}. 
\]
Al cesar la fuerza impulsiva las condiciones iniciales son: 
\begin{eqnarray*}
x_{0} &=&0,\;\;\;\;\;\;\;\;\;\;\dot{x}_{0}=u_{0} \\
\phi &=&0,\;\;\;\;\;\;\;\;\;\;\dot{\phi}=0~.
\end{eqnarray*}
La fuerza de fricci\'{o}n es
\[
{\bf F}_{f}=-\mu g{\bf \hat{e}}_{1}~,
\]
entonces la ecuaci\'{o}n de movimiento es 
\begin{equation}
\ddot{x}=-\mu gM.  \label{eq39a}
\end{equation}

\noindent
La ecuaci\'{o}n para ${\bf L}$ es 
\begin{equation}
\frac{dL_{3}}{dt}=I_{3}\ddot{\phi}=N_{3}  \label{eq40a}
\end{equation}
donde $I_{3}$ es 
\[
I_{3}=\int \rho ({\bf r})\left[ x_{1}^{2}-x_{2}^{2}\right]
dx_{1}dx_{2}dx_{3}=\frac{2}{5}Ma^{2} 
\]
y 
\[
N_{3}=F_{f}a=\mu Mga~.
\]
Sustituyendo en la ecuaci\'{o}n (\ref{eq40a}), tenemos
\begin{equation}
a\ddot{\phi}=\frac{5}{2}\mu g.  \label{eq41a}
\end{equation}

\noindent
Ahora integrando las ecuaciones (\ref{eq39a}) y (\ref{eq41a}) una sola vez 
\begin{equation}
\dot{x}=-\mu gt+C_{1}  \label{eq42a}
\end{equation}
\begin{equation}
a\dot{\phi}=\frac{5}{2}\mu gt+C_{2}  \label{eq43a}
\end{equation}
y aplicandoles las condiciones iniciales, tales ecuaciones quedan en la
forma 
\begin{equation}
\dot{x}(t)=-\mu gt+u_{0}  \label{eq44a}
\end{equation}
\begin{equation}
a\dot{\phi}(t)=\frac{5}{2}\mu gt.  \label{eq45a}
\end{equation}

\noindent
Para que haya rodamiento puro, sin fricci\'{o}n se necesita que 
\begin{equation}
\dot{x}(t)=a\dot{\phi}(t).  \label{eq46a}
\end{equation}

\noindent
De la ecuaci\'{o}n (\ref{eq45a}) y (\ref{eq46a}) evaluadas en $t_{1}$
\[
\frac{5}{2}\mu gt_{1}=-\mu gt_{1}+u_{0} 
\]
\begin{equation}
\Rightarrow \;t_{1}=\frac{2u_{0}}{7\mu g}.  \label{eq47a}
\end{equation}

\noindent
Ahora integrando de nuevo la ecuaci\'{o}n (\ref{eq44a}) y aplicando las
condiciones iniciales tenemos que 
\begin{equation}
x(t)=-\mu g\frac{t^{2}}{2}+u_{0}t~.  \label{eq48a}
\end{equation}
Evaluando la ecuaci\'{o}n (\ref{eq48a}) y (\ref{eq44a}) en el tiempo $t_{1}$
\[
x=\frac{12u^{2}}{49\mu g} 
\]
\[
\dot{x}=\frac{5}{7}u_{0}~.
\]

\bigskip

{\bf \noindent 3.8 Trompo sim\'{e}trico libre de torcas}.

\noindent
Un trompo sim\'{e}trico es cualquier s\'{o}lido de revoluci\'{o}n. Si los
momentos de inercia son 
$$
I_{1}=I_{2}=I_{3}\qquad \qquad {\rm trompo} \quad {\rm esfe{\acute{}}rico}
$$
$$
I_{1}=I_{2}\neq I_{3}\qquad \qquad {\rm trompo} \quad {\rm sime{\acute{}}trico}
$$
$$
I_{1}\neq I_{2}\neq I_{3}\qquad \qquad {\rm trompo} \quad
{\rm asime{\acute{}}trico.}
$$

\noindent
Tomemos el caso del trompo sim\'{e}trico $I_{1}=I_{2}\neq I_{3}$, en este
caso el eje $X_{3}$ es el eje de sim\'{e}tria. Las ecuaciones de Euler
proyectadas sobre los ejes principales de inercia son: 
\begin{equation}
I_{1}\dot{\omega}_{1}+\omega _{2}\omega _{3}(I_{2}-I_{3})=N_{1}
\label{eq49a}
\end{equation}
\begin{equation}
I_{2}\dot{\omega}_{2}+\omega _{3}\omega _{1}(I_{3}-I_{1})=N_{2}
\label{eq50a}
\end{equation}
\begin{equation}
I_{3}\dot{\omega}_{3}+\omega _{1}\omega _{2}(I_{1}-I_{2})=N_{3}.
\label{eq51a}
\end{equation}

\noindent
Como el sistema que estamos considerando esta libre de torcas 
\begin{equation}
N_{1}=N_{2}=N_{3}=0~,  \label{eq52a}
\end{equation}
utilizando el hecho que $I_{1}=I_{2}$ en la ecuaci\'{o}n (\ref{eq52a}),
obtenemos 
\begin{equation}
I_{1}\dot{\omega}_{1}+\omega _{2}\omega _{3}(I_{2}-I_{3})=0  \label{eq53a}
\end{equation}
\begin{equation}
I_{2}\dot{\omega}_{2}+\omega _{3}\omega _{1}(I_{3}-I_{1})=0  \label{eq54a}
\end{equation}
\begin{equation}
I_{3}\dot{\omega}_{3}=0.  \label{eq55a}
\end{equation}

\noindent
La ecuaci\'{o}n (\ref{eq55a}) implica que 
\[
\omega _{3}=cte. 
\]

\noindent
Las ecuaciones (\ref{eq53a}) y (\ref{eq54a}) las reescribimos como: 
\begin{equation}
\dot{\omega}_{1}=-\Omega \omega _{2}\;\;\;\;\;\;\;{\rm donde}\;\;
\Omega =\omega _{3}\left( \frac{I_{3}-I_{1}}{I_{1}}\right)  \label{eq56a}
\end{equation}
\begin{equation}
\dot{\omega}_{2}=-\Omega \omega _{1}~.  \label{eq57a}
\end{equation}
Multiplicando la ecuaci\'{o}n (\ref{eq57a}) por $i$ y sumandola a la
ecuaci\'{o}n (\ref{eq56a}), tenemos 
\begin{eqnarray*}
(\dot{\omega}_{1}+i\dot{\omega}_{2}) &=&-\Omega (\omega _{2}-i\omega _{1}) \\
(\dot{\omega}_{1}+i\dot{\omega}_{2}) &=&i\Omega (\omega _{1}+i\omega _{2}).
\end{eqnarray*}

\noindent
Sea $\eta (t)=\dot{\omega}_{1}(t)+i\dot{\omega}_{2}(t)$ entonces 
\[
\dot{\eta}(t)-i\Omega \eta (t)=0~, 
\]
cuya soluci\'{o}n es 
\[
\eta (t)=A\exp (i\Omega t)~. 
\]
Esto implica que 
\[
(\omega _{1}+i\omega _{2})=A\cos (\Omega t)+i\sin (\Omega t)~, 
\]
entonces 
\begin{equation}
\omega _{1}=A\cos (\Omega t)  \label{eq58a}
\end{equation}
\begin{equation}
\omega _{2}=A\sin (\Omega t).  \label{eq59a}
\end{equation}

\noindent
La magnitud del vector $\omega$ es
\[
\omega =\left| \left| {\bf \omega }\right| \right| =\sqrt[2]{\omega
_{1}+\omega _{2}+\omega _{3}}=\sqrt[2]{A^{2}+\omega _{3}^{2}}=cte~, 
\]
esto significa que la magnitud de $\omega$ no cambia en el tiempo.
Este vector realiza un movimiento de precesi\'{o}n y la frecuencia de
precesi\'{o}n esta dada por 
\[
\Omega =\omega _{3}\left( \frac{I_{3}-I_{1}}{I_{1}}\right)~,
\]
adem\'{a}s notamos que $\Omega$ es constante.

\noindent
Si le llamamos $\lambda $ al \'{a}ngulo entre ${\bf \omega }$ y $X_{3}$ las
ecuaciones (\ref{eq58a}) y (\ref{eq59a}) toman la forma 
\[
\omega _{1}=\omega \sin \lambda \cos (\Omega t)
\]
\[
\omega _{2}=\omega \sin \lambda \sin (\Omega t)
\]
\[
\omega _{1}=\omega \cos \lambda~,
\]
donde $A=\omega \sin \lambda$.

\noindent
Para un s\'{o}lido de revoluci\'{o}n achatado $I_{1}=I_{2}=I_{12}$
y $I_{3}>I_{1}$. Para el caso de la tierra
\[
\Omega _{\bigoplus}
=\omega _{3}\left( \frac{I_{3}-I_{12}}{I_{12}}\right) \simeq
\frac{\omega _{3}}{305}.
\]

\noindent
Las observaciones indican un valor promedio de 14 meses $\simeq$ 450
d\'{\i}as. (Esto se debe a que no es un solido estrictamente, su estructura
interna es l\'{\i}quida).

\bigskip

\noindent {\bf 3.9 Angulos de Euler}.

\noindent
Como sabemos una rotaci\'{o}n se puede representar por una matriz de
rotaci\'{o}n $\lambda $ a trav\'{e}s de la ecuaci\'{o}n 
\begin{equation}
{\bf x}=\lambda {\bf x{\acute{}}}~.  \label{eq60a}
\end{equation}
${\bf x}$ representa el conjunto de ejes del sistema rotado con respecto al
sistema cuyos ejes representamos por ${\bf x{\acute{}}}$. La
rotaci\'{o}n $\lambda$ puede llevarse a cabo por una sucesi\'on de
rotaciones ``parciales"
$\lambda =\lambda _{1}\lambda _{2}...\lambda _{n}$. Existen muchas
posibilidaddes para elegir estas $\lambda$\'{}s. Una de ellas es el conjunto
de \'{a}ngulos $\phi ,\theta $ y $\varphi$
llamados {\it \'{a}ngulos de Euler}, los cuales se generan a trav\'{e}s de
la siguiente serie de rotaciones:

\begin{itemize}
\item  Una rotaci\'{o}n alrededor del eje $X{\acute{}}
_{3}$ en un \'{a}ngulo $\varphi$ (en sentido positivo). La matriz asociada a
\'esta rotaci\'on es:
\[
\lambda _{\varphi }=\left( 
\begin{array}{ccc}
\cos \varphi & \sin \varphi & 0 \\ 
-\sin \varphi & \cos \varphi & 0 \\ 
0 & 0 & 1
\end{array}
\right)~. 
\]

\item  Una rotaci\'{o}n un angulo $\theta$ alrededor del eje $X
{\acute{}}
{\acute{}}_{1}$ (sentido positivo). La matriz asociada es:
\[
\lambda _{\theta }=\left( 
\begin{array}{ccc}
1 & 0 & 0 \\ 
0 & \cos \theta & \sin \theta \\ 
0 & -\sin \theta & \cos \theta
\end{array}
\right) . 
\]

\item  Una rotaci\'{o}n en un angulo $\phi $ alrededor del
eje $X{\acute{}}{\acute{}}{\acute{}}_{3}$ (sentido positivo), la matriz
asociada a esta rotaci\'on es:
\[
\lambda _{\phi }=\left( 
\begin{array}{ccc}
\cos \phi & \sin \phi & 0 \\ 
-\sin \phi & \cos \phi & 0 \\ 
0 & 0 & 1
\end{array}
\right) . 
\]
\end{itemize}

\noindent
La transformaci\'{o}n completa del sistema de ejes $\left\{ X{\acute{}}
_{1},X{\acute{}}_{2},X{\acute{}}_{3}\right\} $ al
sistema con ejes $\left\{ X_{1},X_{2},X_{3}\right\} $
est\'{a} dada por (\ref{eq60a}), donde 
\[
\lambda =\lambda _{_{\phi }}\lambda _{_{\theta }}\lambda _{_{\varphi }}~. 
\]
Haciendo el producto de matrices 
\[
\lambda _{11}=\cos \varphi \cos \phi -\cos \theta \sin \phi \sin \varphi 
\]
\[
\lambda _{21}=-\sin \varphi \cos \phi -\cos \theta \sin \phi \cos \varphi 
\]
\[
\lambda _{31}=\sin \theta \sin \phi 
\]
\[
\lambda _{12}=\cos \varphi \sin \phi +\cos \theta \cos \phi \sin \varphi 
\]
\[
\lambda _{22}=-\sin \varphi \sin \phi +\cos \theta \cos \phi \sin \varphi 
\]
\[
\lambda _{32}=-\sin \varphi \cos \phi 
\]
\[
\lambda _{13}=\sin \varphi \cos \phi 
\]
\[
\lambda _{21}=\cos \varphi \sin \theta 
\]
\[
\lambda _{33}=\cos \theta 
\]
donde 
\[
\lambda =\left( 
\begin{array}{ccc}
\lambda _{11} & \lambda _{12} & \lambda _{13} \\ 
\lambda _{21} & \lambda _{22} & \lambda _{23} \\ 
\lambda _{31} & \lambda _{32} & \lambda _{33}
\end{array}
\right) . 
\]

\noindent
Ahora, consideremos el hecho de que:

\begin{itemize}
\item  ${\bf \dot{\phi}}$ est\'{a} dirigida a lo largo del eje $X
{\acute{}}
_{3}$ (fijo).

\item  ${\bf \dot{\theta}}$ est\'{a} dirigido a lo largo de la {\it linea de
nodo}.

\item  ${\bf \dot{\varphi}}$ est\'{a} dirigido a lo largo del eje $X_{3}$
(cuerpo).
\end{itemize}

\bigskip

\noindent
Podemos escribir 3 componentes de cada uno de los 3 vectores en el
sistema $\left\{ X_{1},X_{2},X_{3}\right\}$ como:
\[
\begin{array}{ccc}
\dot{\phi}_{1}=\dot{\phi}\sin \theta \sin \varphi , & \dot{\theta}_{1}=\dot{%
\theta}\cos \varphi & \dot{\varphi}_{1}=0 \\ 
\dot{\phi}_{2}=\dot{\phi}\sin \theta \cos \varphi , & \dot{\theta}_{2}=-\dot{%
\theta}\sin \varphi & \dot{\varphi}_{2}=0 \\ 
\dot{\phi}_{1}=\dot{\phi}\cos \theta , & \dot{\theta}_{3}=0 & \dot{\varphi}%
_{3}=\dot{\varphi}~,
\end{array}
\]
entonces 
\begin{eqnarray*}
{\bf \omega } &=&{\bf \dot{\phi}}+{\bf \dot{\theta}}+{\bf \dot{\varphi}} \\
&=&\left[ \left( \dot{\phi}_{1}+\dot{\theta}_{1}+\dot{\varphi}_{1}\right)
,\left( \dot{\phi}_{2}+\dot{\theta}_{2}+\dot{\varphi}_{2}\right) ,\left( 
\dot{\phi}_{3}+\dot{\theta}_{3}+\dot{\varphi}_{3}\right) \right] .
\end{eqnarray*}

\noindent
Entonces las componentes de ${\bf \omega }$ son: 
\begin{eqnarray*}
\omega _{1} &=&\dot{\phi}\sin \theta \sin \varphi +\dot{\theta}\cos \varphi
\\
\omega _{2} &=&\dot{\phi}\sin \theta \cos \varphi -\dot{\theta}\sin \varphi
\\
\omega _{3} &=&\dot{\phi}\cos \theta +\dot{\varphi}.
\end{eqnarray*}

\bigskip

\noindent {\bf 3.10 Trompo sim\'{e}trico con un punto fijo}.

\noindent
Como ejemplo mas complicado de la aplicaci\'{o}n de los m\'{e}todos de la
din\'{a}mica del cuerpo r\'{\i}gido, vamos a considerar el movimiento de un
cuerpo sim\'{e}trico en un campo gravitacional uniforme cuando un punto del
eje de sim\'{e}tria est\'{e} fijo en el espacio.

\noindent
El eje de sim\'{e}tria es, desde luego, uno de los ejes principales y lo
tomaremos como el eje $z$ del sistema de coordenadas solidario al cuerpo.
Como hay un punto fijo, la configuraci\'{o}n del trompo quedar\'{a}
determinada por los tres \'{a}ngulos de Euler: $\theta $ da la
inclinaci\'{o}n del eje $z$ respecto a la vertical, $\phi $ mide el acimut
del trompo respecto a la vertical, mientras que $\varphi $ es el \'{a}ngulo
de rotaci\'{o}n del trompo respecto a su propio eje $z$. La distancia del
centro de gravedad al punto fijo ser\'{a} representada por $l$. para obtener
una soluci\'{o}n del movimiento del trompo vamos a utilizar el m\'{e}todo de
Lagrange en vez de las ecuaciones de Euler.

\noindent
La energ\'{\i}a cin\'{e}tica es: 
\[
T=\frac{1}{2}I_{1}(\omega _{1}^{2}+\omega _{2}^{2})+\frac{1}{2}I_{3}\omega
_{3}^{2}~, 
\]
o bien, en funci\'{o}n de los \'{a}ngulos de Euler 
\[
T=\frac{1}{2}I_{1}(\dot{\phi}^{2}\sin ^{2}\theta +\dot{\theta}^{2})
+\frac{1}{2}I_{3}(\dot{\phi}\cos \theta +\dot{\varphi})^{2}~,
\]
donde han desaparecido los t\'{e}rminos en los que figuraban $\omega
_{1}^{2} $ y $\omega _{2}^{2}.$ Conocemos un teorema elemental seg\'{u}n el
cual en un campo gravitatorio constante la energ\'{\i}a potencial es la
misma que se tendr\'{\i}a si el cuerpo estuviera concentrado en su centro de
masa, pero vamos a dar una demostraci\'{o}n formal del mismo. La
energ\'{\i}a potencial del cuerpo es la suma extendida de todas sus
part\'{\i}culas: 
\begin{equation}
V=-m_{i}{\bf r}_{i}\cdot {\bf g}~,  \label{eq61a}
\end{equation}
donde ${\bf g}$ es el vector constante que representa la aceleraci\'{o}n de
la gravedad, seg\'{u}n como se define el centro de masa, esto es equivalente
a 
\begin{equation}
V=-M{\bf R}_{i}\cdot {\bf g,}  \label{eq62a}
\end{equation}
lo que demuestra el teorema. La energ\'{\i}a potencial en funci\'{o}n de los
\'{a}ngulos de Euler es: 
\begin{equation}
V=Mgl\cos \theta ,  \label{eq63a}
\end{equation}
con lo que la Lagrangiana ser\'{a}
\begin{equation}
L=\frac{1}{2}I_{1}(\dot{\phi}^{2}\sin ^{2}\theta +\dot{\theta}^{2})+\frac{1}{%
2}I_{3}(\dot{\phi}\cos \theta +\dot{\varphi})^{2}-Mgl\cos \theta .
\label{eq64a}
\end{equation}

\noindent
Notamos que $\phi $ y $\varphi $ son coordenadas c\'{\i}clicas, por lo tanto 
$p_{\phi }$ y $p_{\varphi }$ son constantes del movimiento. 
\begin{equation}
p_{\varphi }=\frac{\partial L}{\partial \dot{\varphi}}=I_{3}(\dot{\varphi}+%
\dot{\phi}\cos \theta )=cte  \label{eq65a}
\end{equation}
y 
\begin{equation}
p_{\phi }=\frac{\partial L}{\partial \dot{\phi}}=I_{1}\dot{\phi}\sin
^{2}\theta +I_{3}(\dot{\phi}\cos ^{2}\theta +\dot{\varphi}\cos \theta )=cte.
\label{eq66a}
\end{equation}

\noindent
De la ecuaci\'{o}n (\ref{eq65a}) despejamos $\dot{\varphi}$%
\begin{equation}
\dot{\varphi}=\frac{p_{\varphi }-I_{3}\dot{\phi}\cos \theta }{I_{3}}~,
\label{eq67a}
\end{equation}
sustituimos en la ecuaci\'{o}n (\ref{eq66a}) 
\begin{eqnarray*}
p_{\phi } &=&\frac{\partial L}{\partial \dot{\phi}}=I_{1}\dot{\phi}\sin
^{2}\theta +I_{3}(\dot{\phi}\cos ^{2}\theta +\frac{p_{\varphi }-I_{3}\dot{%
\phi}\cos \theta }{I_{3}}\cos \theta )=cte \\
p_{\phi } &=&I_{1}\dot{\phi}\sin ^{2}\theta +p_{\varphi }\cos \theta~,
\end{eqnarray*}
de donde obtenemos 
\begin{equation}
\dot{\phi}=\frac{p_{\phi }-p_{\varphi }\cos \theta }{I_{1}\sin ^{2}
\theta}~.
\label{eq68a}
\end{equation}
Sustituyendo en (\ref{eq67a}) 
\begin{equation}
\dot{\varphi}=\frac{p_{\varphi }}{I_{3}}-\frac{p_{\phi }-p_{\varphi }\cos
\theta }{I_{1}\sin ^{2}\theta }\cos \theta .  \label{eq69a}
\end{equation}

\noindent
Ahora, como el sistema es conservativo, otra constante de movimiento es la
energ\'{\i}a. 
\[
E=T+V=\frac{1}{2}I_{1}(\dot{\phi}^{2}\sin ^{2}\theta +\dot{\theta}^{2})+%
\frac{1}{2}I_{3}(\dot{\phi}\cos \theta +\dot{\varphi})^{2}+Mgl\cos \theta . 
\]

\noindent
La cantidad $I_{3}\omega _{3}=p_{\varphi }$ es una constante de moviento,
multiplicando esta constante por $p_{\varphi }\omega _{3}$ se obtiene
\begin{eqnarray*}
I_{3}p_{\varphi }\omega _{3}^{2} &=&p_{\varphi }^{2}\omega _{3} \\
I_{3}^{2}\omega _{3}^{3} &=&p_{\varphi }^{2}\omega _{3} \\
\frac{1}{2}I_{3}\omega _{3}^{2} &=&\frac{1}{2}\frac{p_{\varphi }^{2}}{I_{3}}~.
\end{eqnarray*}
La cantidad $\frac{1}{2}I_{3}\omega _{3}^{2}$ es una constante, entonces
podemos definir la cantidad 
\begin{eqnarray*}
E{\acute{}}
&=&E-\frac{1}{2}I_{3}\omega _{3}^{2}=cte \\
&=&\frac{1}{2}I_{1}\dot{\theta}^{2}+\frac{1}{2}I_{1}\dot{\phi}^{2}\sin
^{2}\theta +Mgl\cos \theta ~,
\end{eqnarray*}
de donde identificamos 
\[
V(\theta )=\frac{1}{2}\dot{\phi}^{2}\sin ^{2}\theta +Mgl\cos \theta 
\]
\begin{equation}
V(\theta )=\frac{1}{2}I_{1}\left( \frac{p_{\phi }-p_{\varphi }\cos \theta }{%
I_{1}\sin ^{2}\theta }\right) ^{2}\sin ^{2}\theta +Mgl\cos \theta .
\label{eq70a}
\end{equation}

\noindent
Entonces $E{\acute{}}$ es:
\[
E{\acute{}}
=\frac{1}{2}I_{1}\dot{\theta}^{2}+V(\theta)~.
\]
De esta ecuaci\'{o}n despejamos $\dot{\theta}$ e integramos para obtener
\[
\dot{\theta}=\left[ \frac{2}{I_{1}}\left( E%
{\acute{}}%
-V(\theta )\right) \right] ^{2}=\frac{d\theta }{dt}~,
\]
de donde obtenemos 
\begin{equation}
t(\theta )=%
\displaystyle\int %
\frac{d\theta }{\sqrt[2]{\left( \frac{2}{I_{1}}\right) \left( E%
{\acute{}}%
-V(\theta )\right) }}~.  \label{eq71a}
\end{equation}

\noindent
Al realizar la integral de la ecuaci\'{o}n (\ref{eq71a}) se obtiene
$t=f(\theta)$, de donde en principio podemos despejar y obtener $\theta (t)$.
Entonces $\theta (t)$ se sustituye en las ecuaciones para $\dot{\phi}$
y $\dot{\varphi}$ (ecs. (\ref{eq68a}) y (\ref{eq69a})) y al integrarlas
obtenemos la soluci\'{o}n completa a nuestro problema.


\bigskip
\bigskip

\begin{center} {\bf Bibliogr\'{a}fia}. \end{center}

\bigskip

\begin{itemize}
\item  H. Goldstein, {\it Mec\'{a}nica Cl\'{a}sica}, (Revert\'e, 1992).

\item  L. D. Landau y E. M. Lifshitz, {\it Mec\'{a}nica,} (Revert\'e, 1969).

\item  J. B. Marion, {\it Din\'{a}mica Cl\'{a}sica de las Part\'{\i}culas y
Sistemas}, (Revert\'e, 1995).

\item W. Wrigley \& W.M. Hollister, {\it The Gyroscope: Theory and
application}, Science 149, 713 (Aug. 13, 1965).
\end{itemize}


\newpage
\pagestyle{plain}

\centerline{{\Large 4. OSCILACIONES PEQUE\~{N}AS.}}

\bigskip
\bigskip

\noindent
{\bf Pr\'ologo}:
Una forma muy com\'{u}n de movimienteo en los sistemas mec\'{a}nicos, son las
peque\~{n}as oscilaciones. Estas las encontramos en sistemas tales como
vibraciones at\'{o}micas, moleculares, circuitos el\'{e}ctricos, acustica.
Todo movimiento alrededor de las posici\'{o}nes de equilibrio estable, es el
llamado vibratorio.

\bigskip
\bigskip

{\bf CONTENIDO}:

\bigskip

4.1 OSCILADOR ARMONICO SIMPLE.

\bigskip

4.2 OSCILADOR ARMONICO FORZADO.

\bigskip

4.3 OSCILADORES ARMONICOS AMORTIGUADOS.

\bigskip

4.4 MODOS NORMALES.

\bigskip

4.5 RESONANCIA PARAMETRICA.

\newpage

\section*{\protect\smallskip 4.1 OSCILADOR ARMONICO SIMPLE.}

Un sistema se encuentra en equilibrio estable cuando su energ\'{\i}a
potencial $U(q)$ es m\'{\i}nima; al separarlo de esta posici\'{o}n se origina
una fuerza $-dU/dq$ que tiende a devolver al sistema al equilibrio. Sea $q_0$
el valor de la coordenada generalizada correspondiente a la posici\'{o}n de
equilibrio. Al desarrollar $U(q) - U(q_{0})$ en serie de
potencias de Taylor de $q - q_0$ para peque\~{n}as desviaciones de la
posici\'{o}n de equilibrio
\setcounter{equation} {0}\\
\[
U(q)-U(q_0)\cong \frac 12k(q-q_0)^2~,
\]

\noindent
donde:
\begin{eqnarray*}
\frac{\partial U}{\partial q} &=&0 \\
U(q) &=&0~,
\end{eqnarray*}

\noindent
es decir: no hay fuerzas externas que act\'{u}an sobre el sistema y se ha
escogido el nivel de referencia de tal modo que coincide con la posici\'{o}n
de equilibrio; adem\'{a}s de despreciar terminos de orden superior. El
coeficiente $k$ representa el valor de la segunda derivada U(q) para q=q$_0$.
Por simplificaci\'{o}n haremos la siguiente designaci\'{o}n

\[
x=q-q_0
\]

\noindent
con lo que la ecuaci\'{o}n de energ\'{\i}a potencial toma la forma:

\begin{equation}
U(x)=\frac 12kx^2~.  \label{4.1.1}
\end{equation}

\noindent
La energ\'{\i}a cin\'{e}tica de un sistema es en general de la forma

\begin{equation}
T=\frac 12m\stackrel{\cdot }{x}^2  \label{4.1.2}
\end{equation}

\noindent
con (\ref{4.1.1}) y (\ref{4.1.2}) obtemos la expresi\'{o}n para la
Lagrangiana de un sistema que realiza oscilaciones lineales (a tal sistema
se le llama frecuentemente oscilador lineal):

\begin{equation}
L=\frac 12m\stackrel{\cdot }{x}^2-\frac 12kx^2~.  \label{4.1.3}
\end{equation}

\noindent
La ecuaci\'{o}n de movimiento correspondiente a esta $L$ es:

\[
m\stackrel{\cdot \cdot }{x}+kx=0~,
\]

\noindent
o bien

\begin{equation}
\stackrel{\cdot \cdot }{x}+w^2x=0~,  \label{4.1.4}
\end{equation}

\noindent
donde $w^2=\sqrt{k/m}$. La ecuaci\'{o}n diferencial tiene dos soluciones
independientes: $\cos wt$ y ${\rm sen} wt$, as\'{\i}, formamos la
soluci\'{o}n general:

\begin{equation}
x=c_1\cos wt+c_2{\rm sen} wt~,  \label{4.1.5}
\end{equation}

\noindent
o bien, podemos expresar la soluci\'{o}n de la forma:

\begin{equation}
x=a\cos (wt+\alpha )~.  \label{4.1.6}
\end{equation}

\noindent
Puesto que $\cos(wt+\alpha )=\cos wt\cos \alpha -{\rm sen} wt
{\rm sen} \alpha$, la
comparaci\'{o}n con (\ref{4.1.5}) muestra que las constantes arbitrarias $a$
y $\alpha $ est\'{a}n relacionadas con los coeficientes $c_1$ y $c_2$ de la
forma:
$$a=\sqrt{(c_1^2+c_2^2)},\;\;\;\; {\rm y}\;\;\;\; {\rm tan}\alpha =-c_1/c_2~.$$

\noindent
As\'{\i}: un sistema en las proximidades de su posici\'{o}n de equilibrio
estable, ejecuta un movimiento oscilatorio arm\'{o}nico. El coeficiente $a$
en (\ref{4.1.6}) es la amplitud de las oscilaciones, y el argumento del
coseno su fase; $\alpha$ es el valor inicial de la fase, y depende
evidentemente de la elecci\'{o}n del origen de tiempos. La magnitud $w$ es
la frecuencia angular de las oscilaciones, esta no depende de las
condiciones inicales del sistema por lo cual es la caracter\'{\i}stica
fundamental de las oscilaciones.

\noindent
A menudo la soluci\'{o}n es expresada de la forma

\[
x=re\left[ A\exp(iwt)\right]
\]

\noindent
donde $A$ es la amplitud compleja, su modulo es la amplitud ordinaria:

\[
A=a\exp(i\alpha )~.
\]

\noindent
La energ\'{\i}a de un sistema que realiza peque\~{n}as oscilaciones es:

\[
E=\frac 12m\stackrel{\cdot }{x}^2+\frac 12kx^2~,
\]

\noindent
o sustituyendo (\ref{4.1.6})

\[
E=\frac 12mw^2a^2~.
\]

\noindent
Ahora, consideremos el caso para un n\'{u}mero $n$ de grados de libertad. En
este caso seguiremos considerando que la suma de las fuerzas que actuan es
cero, con lo que

\begin{equation}
Q_i=-\frac{\partial U}{\partial q_i}=0~.  \label{4.1.7}
\end{equation}

\noindent
Procediendo de igual forma que en el caso de un s\'{o}lo grado de libertad,
realizamos una expanci\'{o}n en series de Taylor para la energ\'{\i}a
potencial donde ahora consideramos un m\'{\i}nimo para $q_i=q_{i0}$.
Introduciendo peque\~{n}os desplazamientos

\[
x_i=q_i-q_{i0}~,
\]

\noindent
al realizar la expanci\'{o}n en series

\begin{equation}
U(q_1,q_2,...,q_n)=U(q_{10},q_{20},...,q_{n0})+\sum \left( \frac{\partial U}
{\partial q_i}\right) _0x_i+\frac 1{2!}\sum \left( \frac{\partial ^2U}{
\partial q_i\partial q_j}\right) _0x_ix_j+....  \label{4.1.8}
\end{equation}

\noindent
Bajo las mismas consideraciones que se hicieron en (\ref{4.1.1}), llegamos a
la siguiente relaci\'{o}n:

\begin{equation}
U(q_1,q_2,...,q_n)=U=\frac 12\sum_{i,j}k_{ij}x_ix_j~.  \label{4.1.9}
\end{equation}

\noindent
De la ecuaci\'{o}n (\ref{4.1.8}) se puede notar que $k_{ij}=k_{ji}$, es
decir son sim\'{e}tricos con respecto a sus \'{\i}ndices. Ahora consideremos
la situaci\'{o}n para la energ\'{\i}a cin\'{e}tica. Esta en general es de la
forma

\[
\frac 12a_{ij}(q)\stackrel{\cdot }{x}_i\stackrel{\cdot }{x}_j~,
\]

\noindent
donde las $a_{ij}$ s\'{o}lo son funciones de las coordenadas. Al designar
estas $a_{ij}=m_{ij}$ la energ\'{\i}a cin\'{e}tica es de la forma

\begin{equation}
T=\frac 12\sum_{i,j}m_{ij}\stackrel{\cdot }{x}_i\stackrel{\cdot }{x}_j~.
\label{4.1.10}
\end{equation}

\noindent
Una vez conociendo las energias tenemos que la Lagrangiana para
sistemas con n grados de libertad es de la forma

\begin{equation}
L=T-U=\frac 12\sum_{i,j}(m_{ij}\stackrel{\cdot }{x}_i\stackrel{\cdot }{x}
_j-k_{ij}x_ix_j)~.  \label{4.1.11}
\end{equation}

\noindent
Esta Lagrangiana lleva a las ecuaciones diferenciales de movimientos
simultaneas

\begin{equation}
\frac d{dt}\frac{\partial L}{\partial \stackrel{\cdot }{x}_i}-\frac{\partial
L}{\partial x_i}=0  \label{4.1.12}
\end{equation}

\noindent
o bien

\begin{equation}
\sum (m_{ij}\stackrel{\cdot \cdot }{x}_j+k_{ij}x_j)=0~.  \label{4.1.13}
\end{equation}

\noindent
Tenemos as\'{\i} un sistema de ecuaciones diferenciales lineales y
homog\'{e}neas, las cuales pueden pueden ser considerasdas como las n
componenetes de la ecuaci\'{o}n matricial

\begin{equation}
(M)(\stackrel{\cdot \cdot }{X})+(K)(X)=0~,  \label{4.1.14}
\end{equation}

\noindent
donde las matrices est\'{a}n definidas por:

\begin{equation}
(M)=\left(
\begin{array}{llll}
m_{11} & m_{12} & ... & m_{1n} \\
m_{21} & m_{22} & ... & m_{2n} \\
\vdots &  &  & \vdots \\
m_{n1} & m_{n2} & ... & m_{nn}
\end{array}
\right)  \label{4.1.15}
\end{equation}
\begin{equation}
(K)=\left(
\begin{array}{llll}
k_{11} & k_{12} & ... & k_{1n} \\
k_{21} & k_{22} & ... & k_{2n} \\
\vdots &  &  & \vdots \\
k_{n1} & k_{n2} & ... & k_{nn}
\end{array}
\right)  \label{4.1.16}
\end{equation}

\begin{equation}
(\stackrel{\cdot \cdot }{X})=\frac{d^2}{dt^2}\left(
\begin{array}{l}
x_1 \\
x_2 \\
\vdots \\
x_n
\end{array}
\right)  \label{4.1.17}
\end{equation}

\begin{equation}
(X)=\left(
\begin{array}{l}
x_1 \\
x_2 \\
\vdots \\
x_n
\end{array}
\right)~.  \label{4.1.18}
\end{equation}

\noindent
De manera anal\'{o}ga al sistema con un grado de libertad, buscamos n
funciones inc\'{o}gnitas $x_j(t)$ de la forma

\begin{equation}
x_j=A_j\exp(iwt)~,  \label{4.1.19}
\end{equation}

\noindent
siendo A$_j$\ constantes a determinar. Sustituyendo (\ref{4.1.19}) en (\ref
{4.1.13}) y dividiendo todo entre $\exp(iwt)$, se obtiene un sistema de
ecuaciones algebraicas lineales y homog\'{e}neas, a las que deben satisfacer
A$_j$.

\begin{equation}
\sum_j(-w^2m_{ik}+k_{ik})A_k=0~.  \label{4.1.20}
\end{equation}

\noindent
Para que este sistema tenga soluciones distintas de cero, el determinante de
sus coeficientes debe anularse.

\begin{equation}
\left| k_{ij}-w^2m_{ij}\right| ^2=0~.  \label{4.1.21}
\end{equation}

\noindent
Esta es la ecuaci\'{o}n caracter\'{\i}stica y es de grado n con respecto a
$w^2$. En general, tiene n raices distintas reales y positivas $w_{\alpha}$
($\alpha =1,2,...,n$). Las magnitudes $w_{\alpha}$ se llaman frecuencias
propias del sistema. Multiplicando por A$_i^{*}$ y sumando sobre $i$ se tiene

\[
\sum_j(-w^2m_{ij}+k_{ij})A_i^{*}A_j=0~,
\]

\noindent
de donde

\[
w^2=\sum k_{ij}A_i^{*}A_i/\sum m_{ij}A_i^{*}A_i~.
\]

\noindent
Como los coeficientes $k_{ij}$ y $m_{ij}$ son reales y sim\'{e}tricos, las
formas cuadr\'{a}ticas del numerador y denominador de esta expresi\'{o}n son
reales, y al ser esencialmente positivas, $w^2$ es igualmente positivo.

\underline{EJEMPLO}

\noindent
Como ejemplo se modelar\'{a}n las ecuaciones del movimiento del p\'{e}ndulo
doble. La energ\'{\i}a potencial para este es (el problema posee dos grados
de libertad)

\[
U=m_1gl_1(1-\cos\theta _1)+m_2gl_1(1-\cos\theta _1)+m_2gl_2(1-\cos\theta _2)~.
\]

\noindent
Al aplicar la expanci\'{o}n (\ref{4.1.8}), se tiene

\[
U=\frac 12(m_1+m_2)gl_1\theta _1^2+\frac 12m_2gl_2\theta _2^2~.
\]

\noindent
Al comparar con (\ref{4.1.9}), identificamos
\begin{eqnarray*}
k_{11} &=&(m_1+m_2)l_1^2 \\
k_{12} &=&k_{21}=0 \\
k_{22} &=&m_2gl_2~.
\end{eqnarray*}

\noindent
Para la energ\'{\i}a cin\'{e}tica se encontr\'{o}

\[
T=\frac 12(m_1+m_2)l_1^2\stackrel{.}{\theta }_1^2+\frac 12m_2l_2^2\stackrel{.%
}{\theta }_2^2+m_2l_1l_2\stackrel{.}{\theta }_1\stackrel{.}{\theta }_2~.
\]

\noindent
Identificando t\'{e}rminos al comparar con (\ref{4.1.10})
\begin{eqnarray*}
m_{11} & = & (m_1+m_2)l_1^2 \\
m_{12} & = & m_{21}=m_2l_1l_2 \\
m_{22} & = & m_2l_2^2~.
\end{eqnarray*}

\noindent
Al sustituir las ecuaciones de las energ\'{\i}as en (\ref{4.1.11}) se obtiene
la Lagrangiana para el oscilador de p\'{e}ndulo doble y como resultado final:

\[
\left(
\begin{array}{ll}
m_{11} & m_{12} \\
m_{21} & m_{22}
\end{array}
\right) \left(
\begin{array}{l}
\stackrel{..}{\theta }_1 \\
\stackrel{..}{\theta }_2
\end{array}
\right) +\left(
\begin{array}{ll}
k_{11} & 0 \\
0 & k_{22}
\end{array}
\right) \left(
\begin{array}{l}
\theta _1 \\
\theta _2
\end{array}
\right) =0~.
\]

\noindent
Estas son las ecuaciones de movimiento en este caso.

\section*{4.2 OSCILADOR ARMONICO FORZADO.}

\noindent
Si un sistema oscilatorio se somete a la acci\'{o}n de un campo externo
variable, son las llamadas oscilaciones forzadas. Como consideramos
peque\~{n}as oscilaciones, entonces esperamos que la acci\'{o}n del campo
exterior sea d\'{e}bil.

\noindent
Adem\'{a}s de su energ\'{\i}a potencial propia, el sistema posee en este caso
una energ\'{\i}a potencial $U_{e}(x,t)$ debida al campo exterior. Desarrollando
esta \'{u}ltima en serie de potencias de la peque\~{n}a magnitud $x$:

\[
U_e(x,t)\cong U_e(0,t)+x\left[ \frac{\partial U_e}{\partial x}\right] _{x=0}~.
\]

\noindent
El segundo t\'{e}rmino, es la fueza exterior que act\'{u}a sobre el sistema
en su posici\'{o}n de equilibrio, desisgnado este t\'{e}rmino como $F(t)$.
Con esto, la Lagrangiana para este sistema es de la forma

\begin{equation}
L=\frac 12m\stackrel{\cdot }{x}^2-\frac 12kx^2+xF(t)~.  \label{4.2.1}
\end{equation}

\noindent
La ecuaci\'{o}n de movimiento correspondiente es

\[
m\stackrel{\cdot \cdot }{x}+kx=F(t)~,
\]

\noindent
o bien

\begin{equation}
\stackrel{\cdot \cdot }{x}+w^2x=F(t)/m~,  \label{4.2.2}
\end{equation}

\noindent
donde $w$ es la frecuencia para las oscilaciones libres. La soluci\'{o}n
general a esta ecuaci\'{o}n es de la forma
\[
x=x_h+x_p~,
\]

\noindent
es decir, de una parte homog\'{e}nea y una soluci\'{o}n correspondiente a un
caso particular. Analizando el caso para el cual la fuerza exterior es
funci\'{o}n peri\'{o}dica simple del tiempo, de frecuencia $\gamma $ de la
forma

\[
F(t)=f\cos(\gamma t+\beta )~.
\]

\noindent
Al hallar la integral particular para la ecuaci\'{o}n  \ref{4.2.2} en la
forma $x_1=b\cos(\gamma t+\beta )$ y al sustituir, se tiene que
$b=f/m(w^2-\gamma ^2)$, que al juntar ambas soluciones, tenemos que la
soluci\'{o}n total es

\begin{equation}
x=a\cos(wt+\alpha )+\left[ f/m(w^2-\gamma ^2)\right] \cos(\gamma t+\beta )~.
\label{4.2.3}
\end{equation}

\noindent
El resultado muestra una suma de dos oscilaciones: una debida a la
frecuencia propia y otra con la frecuencia de la fuerza exterior.

\noindent
La ecuaci\'{o}n (\ref{4.2.2}) puede ser integrada en forma general para una
fuerza exterior arbitraria. Escribiendo la ecuaci\'{o}n de la forma

\[
\frac d{dt}(\stackrel{\cdot }{x}+iwx)-iw(\stackrel{\cdot }{x}+iwx)=\frac
1mF(t)~,
\]

\noindent
haciendo $\xi =\stackrel{\cdot }{x}+iwx$, se tiene

\[
\frac d{dt}\xi -iw\xi =F(t)/m~.
\]

\noindent
La soluci\'{o}n a esta \'{u}ltima es del tipo $\xi =A(t)\exp(iwt)$; para la
funci\'{o}n $A(t)$ se obtiene

\[
\stackrel{\cdot }{A}=F(t)\exp(-iwt)/m~.
\]

\noindent
Al integrarla se obtiene la soluci\'{o}n

\begin{equation}
\xi =\exp(iwt)\int_{0}^{t}\frac{1}{m}F(t)\exp(-iwt)dt+\xi _o~.  \label{4.2.4}
\end{equation}

\noindent
Esta es la soluci\'{o}n general buscada; la funci\'{o}n x(t) est\'{a} dada
por la parte imaginaria de esta \'{u}ltima dividiendo por w.

\underline{EJEMPLO}

\noindent
Como empleo de la ecuaci\'{o}n anterior se muestra el siguiente ejemplo.

\noindent
Determinar la amplitud final de las oscilaciones de un sistema bajo la
acci\'{o}n de una fuerza exterior tal que $F_0=cte$ durante un tiempo
l\'{\i}mitado T. Para este intervalo de tiempo se tiene
$$\xi =  \frac{F_0}{m}\exp(iwt)\int_{0}^{T}\exp(-iwt)dt$$
$$\xi =  \frac{F_0}{iwm}[1-\exp(-iwt)]\exp(iwt) \nonumber$$
y el m\'{o}dulo cuadrado da la amplitud. De la relaci\'{o}n $\left| \xi
\right| ^2=a^2w^2$ con lo cual

\[
a=\frac{2F_0}{mw^2}{\rm sen}(\frac 12wT)~.
\]

\section*{4.3 OSCILADOR ARMONICO AMORTIGUADO.}

\noindent
En las secciones anteriores consideramos s\'{o}lo la carencia o la presencia
de fuerzas externas, para los casos del oscilador arm\'{o}nico simple y
forzado; respectivamente. Es decir, el movimiento ten\'{\i}a lugar en el
vacio o bien, que la influencia del medio en el movimiento era despreciable.
En la realidad, cuando un sistema se mueve a traves de un medio, \'{e}ste
ofrece resistencia que tiende a retardar el movimiento. La energ\'{\i}a del
sistema se disipa (ya sea en forma de calor \'{o} de algun otra forma de
energ\'{\i}a). Primero analizamos como afecta este fen\'omeno a las
oscilaciones simples.

\noindent
El modo en que este medio afecta al movimiento es por fuerzas de rozamiento.
Si esta fuerza disipativa es lo suficientemente peque\~{n}a, podemos
desarrollarla en potencias de la velocidad. El t\'{e}rmino de orden cero del
desarrollo es nulo, ya que ninguna fuerza de rozamiento act\'{u}a sobre
un cuerpo enreposo, por lo que el primer t\'{e}rmino que no se anula es
proporcional a la velocidad, adem\'{a}s despreciando t\'{e}rminos de orden
superior.

\[
f_r=-\alpha \stackrel{\cdot }{x}~,
\]

\noindent
donde $x$ es la coordenada generalizada y $\alpha$ un coeficiente positivo;
el signo menos indica que es en sentido opuesto al movimiento. A\~{n}adiendo
esta fuerza a la ecuaci\'{o}n de movimiento

\[
m\stackrel{..}{x}=-kx-\alpha \stackrel{\cdot }{x}~,
\]

\noindent
o bien

\begin{equation}
\stackrel{..}{x}=-kx/m-\alpha \stackrel{\cdot }{x}/m~.  \label{4.3.1}
\end{equation}

\noindent
Haciendo $k/m=w_o^2$ y $\alpha /m=2\lambda $; donde $w_o$ es la frecuencia
de las oscilaciones libres del sistema y $\lambda $ es el coeficiente de
amortiguamiento. Con lo anterior

\[
\stackrel{..}{x}+2\lambda \stackrel{\cdot }{x}+w_o^2x=0~.
\]

\noindent
La soluci\'{o}n para la ecuaci\'{o}n anterior es de la forma $x=\exp(rt)$; al
sustituir esta en la ecuaci\'{o}n anterior obtenemos la ecuaci\'{o}n
caracter\'{\i}stica para $r$. De tal modo

\[
r^2+2\lambda +w_o^2=0~,
\]

\noindent
de donde

\[
r_{1,2}=-\lambda \pm \sqrt{\left( \lambda ^2-w_o^2\right) }~,
\]

\noindent
con lo que la soluci\'{o}n general a la ecuaci\'{o}n de movimiento es

\[
x=c_1\exp(r_1t)+c_2\exp(r_2t)~. \]

\noindent
De las raices de $r$ podemos considerar los siguientes casos especiales:

(i) $\lambda <w_o$. Se tienen raices imaginarias conjugadas. Con lo que
la soluci\'{o}n es

\[
x=re\left\{ Aexp\left[ -\lambda t+i\sqrt{(w_o^2-\lambda ^2)}\right] \right\}~,
\]

\noindent
siendo $A$ una constante compleja arbitraria. La soluci\'{o}n puede ser
escrita en la forma

\begin{equation}
x=a\exp(-\lambda t)\cos(wt+\alpha )
\;\;\;\;{\rm siendo}\;\;\;\;w=\sqrt{\left( w_o^2-\lambda
^2\right)}~,  \label{4.3.2}
\end{equation}

\noindent
donde $a$ y $\alpha$ son constantes reales. De este modo, puede decirse que
una oscilaci\'{o}n amortiguada es como una oscilaci\'{o}n arm\'{o}nica cuya
amplitud decrece exponenecialmente. La rapidez de disminuci\'{o}n de la
amplitud est\'{a} determinada por el exponente $\lambda$ y la frecuencia w
es menor que las oscilaciones libres en ausencia de rozamiento.

(ii) $\lambda > w_o$. Entonces los dos valores de $r$ son reales y
negativos. La forma general de la soluci\'{o}n es:

\[
x=c_1\exp\left\{ -\left[ \lambda -\sqrt{\left( \lambda ^2-w_o^2\right) }%
\right] t\right\} +c_2\exp\left\{ -\left[ \lambda +\sqrt{\left( \lambda
^2-w_o^2\right) }\right] t\right\}~.
\]

\noindent
Si el rozamiento es muy grande, el movimiento consiste en una
disminuci\'{o}n mon\'{o}tona, que tiende asint\'{o}ticamente (cuando $t
\rightarrow \infty$) a la posici\'{o}n de equilibrio (sin oscilaci\'{o}n).
Este tipo de movimiento se llama aperi\'{o}dico.

(iii)
$\lambda =w_o$. Se tiene que $r=-\lambda$, cuya soluci\'{o}n general es
de la forma

\[
x=(c_1+c_2t)\exp(-\lambda t)~.
\]

\noindent
Si generalizamos para sistemas con n grados de libertad, las fuerzas de
rozamiento generalizadas correspondientes a las coordenadas $x_i$ son
funciones lineales de las velocidades

\begin{equation}
f_{r,i}=\sum_{j}\alpha _{ij}\stackrel{.}{x}_i~,  \label{4.3.3}
\end{equation}

\noindent
con $\alpha _{ik}=\alpha _{ki}$ se puede escribir como

\[
f_{r,i}=-\frac{\partial F}{\partial \stackrel{.}{x}_i}
\]~,

\noindent
donde $F=\frac{1}{2}\sum_{i,j}\alpha _{ij}\stackrel{.}{x}_i
\stackrel{.}{x}_j$ y se le llama funci\'{o}n disipativa. La ecuaci\'{o}n
diferencial se obtiene al sumar estas fuerzas a la ecuaci\'{o}n (\ref{4.1.13}
)

\begin{equation}
\sum (m_{ij}\stackrel{\cdot
\cdot }{x}_j+k_{ij}x_j)
=-\sum_{j}\alpha _{ij}\stackrel{.}{x}_i~.  \label{4.3.4}
\end{equation}

\noindent
Haciendo en estas ecuaciones

\[
x_k=A_k\exp(rt)
\]

\noindent
y al sustituir esta \'{u}ltima en (\ref{4.3.4}) y dividiendo por $\exp(rt)$,
se tiene el siguiente sistema de ecuaciones algebraicas lineales para las
constantes $A_j$

\[
\sum_{j}(m_{ij}r^2+\alpha _{ij}r+k_{ij})A_j=0~.
\]

\noindent
Igualando a cero el determinante de este sistema, se encuentra la
ecuaci\'{o}n caracter\'{\i}stica para este sistema.

\begin{equation}
\left| m_{ij}r^2+\alpha _{ij}r+k_{ij}\right| =0~.  \label{4.3.5}
\end{equation}

\noindent
Esta es una ecuaci\'{o}n en $r$ de grado 2n.
\section*{4.4 MODOS NORMALES.}

\noindent
Antes de definir los modos normales, reescribiremos la ecuaci\'{o}n (\ref
{4.1.14}) de la siguiente manera

\[
M\left| \stackrel{..}{X}\right\rangle +K\left| X\right\rangle =0~,
\]

\noindent
donde $\left| X\right\rangle $ es el vector n-dimensional cuya matrix de
representaci\'{o}n es (\ref{4.1.18}); $M$ y $K$ son dos operadores que
tienen la representaci\'{o}n matricial definidas por (\ref{4.1.15}) y (\ref
{4.1.16}) respectivamente. La ecuaci\'{o}n antes descrita es una
ecuaci\'{o}n con operadores. Dado que M es un operador no sigular y
sim\'{e}trico para el cual el operador inverso M$^{-1}$ y el operador M$
^{1/2}$ y M$^{-1/2}$ ex\'{\i}sten. Con lo anterior podemos expresar la
ecuaci\'{o}n de operadores en la forma

\[
\frac{d^2}{dt^2}M^{1/2}\left| X\right\rangle =-M^{-1/2}KM^{-1/2}M^{1/2}\left|
X\right\rangle~,
\]

\noindent
o de la forma m\'{a}s compacta

\begin{equation}
\frac{d^2}{dt^2}\left| \stackrel{\_}{X}\right\rangle =-\lambda \left|
\stackrel{\_}{X}\right\rangle~,  \label{4.4.1}
\end{equation}

\noindent
donde

\[
\left| \stackrel{\_}{X}\right\rangle =M^{1/2}\left| X\right\rangle
\]

\noindent
y
\[
\lambda =M^{-1/2}KM^{-1/2}~.
\]

\noindent
Como M$^{-1/2}$ y K son operadores sim\'{e}tricos, entonces $\lambda$ es
igualmente sim\'{e}trico. Si empleamos eigenvectores ortogonales como base
vectorial (por ejemplo el espacio tridimencional), la representaci\'{o}n
matricial del operador puede ser diagonal de la forma

\[
\lambda _{ij}=\lambda _i\delta _{ij}~.
\]

\noindent
Trataremos el siguiente problemas de eigenvalores

\begin{equation}
\lambda \left| \rho _i\right\rangle =\lambda _i\left| \rho _i\right\rangle~,
\label{4.4.2}
\end{equation}

\noindent
donde $\left| \rho _i\right\rangle $ representa un conjunto de eigenvectores
mutuamente ortogonales; o bien

\[
M^{-1/2}KM^{-1/2}\left| \rho _i\right\rangle =\lambda _i\left| \rho
_i\right\rangle~.
\]

\noindent
Los eigenvalores son obtenidos al multiplicar ambos lados por $\left\langle
\rho _i\right| $, con lo cual

\[
\lambda _i=\frac{\left\langle \rho _i\right| M^{-1/2}KM^{-1/2}\left| \rho
_i\right\rangle }{\langle \rho _i\left| \rho _i\right\rangle }~.
\]

\noindent
Dado que las energ\'{\i}as potencial y cin\'{e}tica son cantidades positivas,
se tine que

\[
\left\langle \rho _i\right| M^{-1/2}KM^{-1/2}\left| \rho _i\right\rangle
\rangle 0
\]

\noindent
y por lo tanto

\[
\lambda _i> 0~.
\]

\noindent
Esto nos p\'{e}rmite el conjunto

\[
\lambda _i=w_i^2~.
\]

\noindent
Si expresamos el vector $\left| \stackrel{\_}{X}\right\rangle $\ en
t\'{e}rminos de estos eigenvectores de $\lambda$,

\[
\left| \stackrel{\_}{X}\right\rangle =\sum_{i}y_i\left|
\stackrel{\_}{X}\right\rangle~,
\]

\noindent
donde

\begin{equation}
y_i=\langle \rho _i\left| \stackrel{\_}{X}\right\rangle~.
\label{4.4.3}
\end{equation}

\noindent
Al insertar este resultado en la ecuaci\'{o}n de movimiento (\ref{4.4.1}),
se tiene

\[
\frac{d^2}{dt^2}\sum_{i}y_i\left| \rho _i\right\rangle
=-\lambda \left| \stackrel{\_}{X}\right\rangle =-\sum_{i}
\lambda _iy_i\left| \rho _i\right\rangle~.
\]

\noindent
El producto escalar de esta ecuaci\'{o}n con el eigenvector constante
$\left\langle \rho _j\right|$, produce la ecuaci\'{o}n de movimiento para
las coordenadas generalizadas $y_j$

\[
\frac{d^2}{dt^2}y_j=-w_j^2y_j~.
\]

\noindent
La soluci\'{o}n para esta ecuaci\'{o}n es de la forma

\begin{equation}
y_j=A_j\cos(w_jt+\phi _j)~.  \label{4.4.4}
\end{equation}

\noindent
En base a estas nuevas coordenadas para el movimiento arm\'{o}nico de un
sistema de particulas, pueden obtenerse un conjunto de ecuaciones de
movimiento generalizadas e independientes. La relaci\'{o}n entre
estas $y_j\acute{}s$ y $\stackrel{\_}{x}_i\acute{}s$ est\'{a} dada
por (\ref{4.4.3})

\[
y_j=\rho _{j1}\stackrel{\_}{x}_1+\rho _{j2}\stackrel{\_}{x}_2+...+\rho _{jn}%
\stackrel{\_}{x}_n~.
\]

\noindent
Los componentes $\rho _{jl}$ ($l=1,2,..,n$) son determinados al resolver el
problema de eigenvalores de la ecuaci\'{o}n (\ref{4.4.2}). Las nuevas
coordenadas son referidas como las coordenadas normales, y las $w_j\acute{}s$
como las frecuencias normales. La forma equivalente (\ref{4.4.4}) en
matrices es

\begin{equation}
\left(
\begin{array}{l}
\stackrel{\_}{x}_1^{(j)} \\
\stackrel{\_}{x}_2^{(j)} \\
\vdots  \\
\stackrel{\_}{x}_n^{(j)}
\end{array}
\right) =A_j\cos(w_jt+\phi _j)\left(
\begin{array}{l}
\rho _{j1} \\
\rho _{j2} \\
\vdots  \\
\rho _{jn}
\end{array}
\right)~.   \label{4.4.5}
\end{equation}

\noindent
Esta \'{u}ltima refiere los modos normales de vibraci\'{o}n del sistema. Una
de las razones de haber introducido coordenadas $y_j\acute{}s$ se aprecia al
observar que la expresi\'{o}n para la energ\'{\i}a cin\'{e}tica no cambia si
se rotaran los ejes del nuevo sistema no cambian

\[
T=\frac{1}{2}\sum_{j=1}^{n}M_{j}\stackrel{.}{y}_j^2~.
\]

\underline{EJEMPLO}

\noindent
Suponiendo un arreglo matricial de la forma mostrada, la cual
representa las ecuaciones de movimiento obtenidas para un sistema

\[
\frac{d^2}{dt^{2}}\left(
\begin{array}{c}
\stackrel{\_}{x}_{1} \\
\stackrel{\_}{x}_{2} \\
\stackrel{\_}{x}_{3}
\end{array}
\right) =-\left(
\begin{array}{ccc}
5 & 0 & 1 \\
0 & 2 & 0 \\
1 & 0 & 5
\end{array}
\right) \left(
\begin{array}{c}
\stackrel{\_}{x}_{1} \\
\stackrel{\_}{x}_{2} \\
\stackrel{\_}{x}_{3}
\end{array}
\right)~.
\]

\noindent
Comparando con (\ref{4.4.1}), identificamos el valor del
operador $\lambda$. Al encontrar los egienvectores
empleamos (\ref{4.4.2}) con lo cual

\[
\left(
\begin{array}{ccc}
5 & 0 & 1 \\
0 & 2 & 0 \\
1 & 0 & 5
\end{array}
\right) \left(
\begin{array}{c}
\rho _{1} \\
\rho _{2} \\
\rho _{3}
\end{array}
\right) =\lambda _{i}\left(
\begin{array}{c}
\rho _{1} \\
\rho _{2} \\
\rho _{3}
\end{array}
\right)~.
\]

\noindent
Determinamos la ecuaci\'{o}n caracter\'{\i}stica para las $\lambda _{i}$. De 
manera que

\[
\det (\lambda -\lambda _{i}I)=0~,
\]

\noindent
al sustituir valores

\[
\left|
\begin{array}{ccc}
5-\lambda  & 0 & 1 \\
0 & 2-\lambda  & 0 \\
1 & 0 & 5-\lambda
\end{array}
\right| =0~.
\]
\newline

\noindent
Al resolver la ecuaci\'{o}n se tiene que $\lambda _{i}=2,4,6$.
Para $\lambda=4$

\[
\left(
\begin{array}{ccc}
5 & 0 & 1 \\
0 & 2 & 0 \\
1 & 0 & 5
\end{array}
\right) \left(
\begin{array}{c}
\rho _{1} \\
\rho _{2} \\
\rho _{3}
\end{array}
\right) =4\left(
\begin{array}{c}
\rho _{1} \\
\rho _{2} \\
\rho _{3}
\end{array}
\right)
\]

\noindent
se tiene el siguiente conjunto de ecuaciones

\begin{eqnarray*}
(5-4)\rho _{1}+\rho _{3} &=&0 \\
2\rho _{2}-4\rho _{2} &=&0 \\
\rho _{1}+(5-4)\rho _{3} &=&0~.
\end{eqnarray*}

\noindent
Teniendo en cuenta la condici\'{o}n de normalizaci\'{o}n, se obtienen los
valores

\begin{eqnarray*}
\rho _{1} &=&-\rho _{3}=\frac{1}{\sqrt{2}} \\
\rho _{2} &=&0~.
\end{eqnarray*}

\noindent
Con lo anterior

\[
\left| \rho _{\lambda =4}\right\rangle =\frac{1}{\sqrt{2}}\left(
\begin{array}{c}
1 \\
0\\
-1
\end{array}
\right)~.
\]

\noindent
De igual modo se obtienen

\begin{eqnarray*}
\left| \rho _{\lambda =6}\right\rangle  &=&\frac{1}{\sqrt{2}}\left(
\begin{array}{c}
1 \\
0 \\
1
\end{array}
\right)  \\
\left| \rho _{\lambda =2}\right\rangle  &=&\left(
\begin{array}{c}
0 \\
1 \\
0
\end{array}
\right)~.
\end{eqnarray*}

\noindent
As\'{\i}, el nuevo espacio vectorial est\'{a} determinado por

\[
\left| \rho _{i}\right\rangle =\left(
\begin{array}{ccc}
\frac{1}{\sqrt{2}} & \frac{1}{\sqrt{2}} & 0 \\
0 & 0 & 1 \\
-\frac{1}{\sqrt{2}} & \frac{1}{\sqrt{2}} & 0
\end{array}
\right)~,
\]

\noindent
de donde

\[
\left\langle \rho _{i}\right| =\left(
\begin{array}{ccc}
\frac{1}{\sqrt{2}} & 0 & -\frac{1}{\sqrt{2}} \\
\frac{1}{\sqrt{2}} & 0 & \frac{1}{\sqrt{2}} \\
0 & 1 & 0
\end{array}
\right)~.
\]

\noindent
As\'{\i}, las coordenadas normales son dadas por (\ref{4.4.3})

\[
\left(
\begin{array}{c}
y_{1} \\
y_{2} \\
y_{3}
\end{array}
\right) =\left(
\begin{array}{ccc}
\frac{1}{\sqrt{2}} & 0 & -\frac{1}{\sqrt{2}} \\
\frac{1}{\sqrt{2}} & 0 & \frac{1}{\sqrt{2}} \\
0 & 1 & 0
\end{array}
\right) \left(
\begin{array}{c}
\stackrel{\_}{x}_{1} \\
\stackrel{\_}{x}_{2} \\
\stackrel{\_}{x}_{3}
\end{array}
\right)~.
\]
\section*{4.5 RESONANCIA PARAMETRICA.}

\noindent
El importante fen\'{o}meno de resonancia param\'{e}trica se presenta al tener 
un sistema que se se encuentra en un estado de reposo (en la posici\'{o}n de
equilibrio $x=0$) y es inestable; es decir, bastar\'{a} una separaci\'{o}n de
esta posici\'{o}n por peque\~{n}a que sea para provocar un desplazamiento
$x$ r\'{a}pidamente creciente con el tiempo. Se diferencia de las resonancias
ordinarias, en las cuales el desplazamiento crece con el tiempo
(proporcional a $t$).

\noindent
Los par\'{a}metros de un sistema lineal son los coeficientes $m$ y $k$ de la
Lagrangiana (\ref{4.1.3}); si estos son funci\'{o}n del tiempo, la
ecuaci\'{o}n del movimiento es:

\begin{equation}
\frac d{dt}(m\stackrel{\cdot }{x})+kx=0~.  \label{5.7.1}
\end{equation}

\noindent
Si consideramos la masa constante, la ecuaci\'{o}n anterior toma la forma

\begin{equation}
\frac{d^2x}{dt^2}+w^2(t)x=0~.  \label{4.6.2}
\end{equation}

\noindent
La forma de la funci\'{o}n $w(t)$ est\'{a} dada por las condiciones del
problema; suponiendo que la funci\'{o}n es peri\'{o}dica de frecuencia
$\gamma$ (y de periodo $T=2\pi/\gamma $). Lo que significa

\[
w(t+T)=w(t)~,
\]

\noindent
por lo cual, toda ecuaci\'{o}n del tipo (\ref{4.6.2}) es invariante con
respecto a la transformaci\'{o}n $t\rightarrow t + T$. As\'{\i}, si $x(t)$ es
una soluci\'{o}n de esta, la funci\'{o}n $x(t+T)$ es tambien 
soluci\'{o}n. Con
lo anterior, sean $x_1(t)$ y $x_2(t)$ dos integrales independientes de (\ref
{4.6.2}), estas deben transformarse en ellas mismas en combinaci\'{o}n
lineal cuando se sustituye $t\rightarrow t + T$. La forma de ello es

\begin{eqnarray}
x_1(t+T) &=&\mu _1x(t)  \label{4.6.3} \\
x_2(t+T) &=&\mu _2x(t)~,  \nonumber
\end{eqnarray}

\noindent
o en forma general

\begin{eqnarray*}
x_1(t) &=&\mu _1^{t/T}F(t) \\
x_2(t) &=&\mu _2^{t/T}G(t)~,
\end{eqnarray*}

\noindent
donde $F(t)$ y $G(t)$ son funciones puramente periodicas del tiempo
(de per\'{\i}odo $T$). La relaci\'{o}n entre estas constantes se obtiene al
manipular las ecuaciones siguientes

\begin{eqnarray*}
\stackrel{..}{x}_1+w^2(t)x_1 &=&0 \\
\stackrel{..}{x}_2+w^2(t)x_2 &=&0~.
\end{eqnarray*}

\noindent
Multiplicando por $x_2$ y $x_1$ respectivamente, y rest\'{a}ndolas miembro a
miembro, se obtiene

\[
\stackrel{..}{x_1}x_2-\stackrel{..}{x}_2x_1=\frac d{dt}(\stackrel{.}{x_1}x_2-%
\stackrel{.}{x}_2x_1)=0~,
\]

\noindent
o tambien

\[
\stackrel{.}{x_1}x_2-\stackrel{.}{x}_2x_1=cte~.
\]

\noindent
Al sustituir $t$ por $t+T$ en la ecuaci\'{o}n anterior, el miembro
derecho est\'{a} multiplicado por $\mu _1\mu _2$ (debido
a \ref{4.6.3}); por lo que es evidente que

\begin{equation}
\mu _1\mu _2=1~,  \label{4.6.4}
\end{equation}

\noindent
teniendo en cuenta (\ref{4.6.2}) y sabiendo que los coeficientes son reales.
Si $x(t)$ es una integral de esta ecuaci\'{o}n, la funci\'{o}n $x*(t)$
tambi\'{e}n lo es. Lo anterior conduce a que $\mu _1$, $\mu _2$ deben
coincidir con $\mu _1^{*}$, $\mu _2^{*}$, es decir,
o $\mu _1$=$\mu _2^{*}$ o tambi\'{e}n $
\mu _1$ y  $\mu _2$ son reales. En el primer caso y teniendo en
cuenta (\ref{4.6.4}) resulta que $\mu _1=1/$ $\mu _1^{*}$ lo que es
igual $\left| \mu _1\right| ^2=\left| \mu _2\right| ^2=1$. En el segundo
caso, las dos integrales son de la forma
\begin{eqnarray*}
x_1(t) &=&\mu ^{t/T}F(t) \\
x_2(t) &=&\mu ^{-t/T}G(t)~.
\end{eqnarray*}

\noindent
Una de estas funciones crece exponencialmente con el tiempo.

\bigskip
\bigskip

\begin{center} BIBLIOGRAFIA. \end{center}

\bigskip

* H. Goldstein, {\it Classical mechanics}, Second ed. (Addison-Wesley, 1981).

\bigskip

* L. D. Landau \& E. M. Lifshitz, {\it Mec\'{a}nica}, (Revert\'e, 1969).

\bigskip

* W. Hauser, {\it Introduction to the principles of mechanics}, (Wesley, 1965).

\bigskip

* E.I. Butikov, {\it Parametric Resonance}, Computing
in Science \&

Engineering, May/June 1999, pp. 76-83 (http://computer.org).


\newpage




\centerline{\Large 5. TRANSFORMACIONES CANONICAS}

\bigskip
\bigskip

\noindent
{\bf Pr\'ologo}:
La idea principal de las transformaciones can\'onicas  es encontrar todos 
aquellos sistemas de coordenas (en el espacio de fases) los cuales 
{\it preserven la forma de las 
ecuaciones de Hamilton, independientemente de qu\'e Hamiltoniano se trate}.
Posteriormente se 
escoge de entre todos esos sistemas de coordenadas aquel en el cual se
facilite la resoluci\'on del problema en particular.\\

\bigskip

{\bf CONTENIDO:}

\bigskip

5.1 Definiciones, Hamiltoniano y Kamiltoniano

\bigskip

5.2 Condiciones necesarias y suficientes para que una transf. sea can\'onica

\bigskip

5.3 Ejemplo de aplicaci\'on de transf. can\'onica

\newpage

\noindent
\section*{5.1 Definiciones, Hamiltoniano y Kamiltoniano}
Se define una transformaci\'on can\'onica,
considerando los casos en que la 
transformaci\'on depende explicitamente (o no) del tiempo, de la siguiente
manera: 

\noindent
{\bf Definici\'on 1:} Una transformaci\'on independiente del 
tiempo $Q=Q(q,p)$, y $P=P(q,p)$ se dice 
que es can\'onica si y solo si existe una funci\'on $F(q,p)$ tal que
\setcounter{equation} {0}\\
$$dF(q,p)=\sum_{i}p_{i}dq_{i}-\sum_{i}P_{i}(q,p)dQ_{i}(q,p)~.$$
{\bf Definici\'on 2:} Una transformaci\'on dependiente del tiempo $Q=Q(q,p,t)$, y $P=P(q,p,t)$ se dice 
que es can\'onica si y solo si existe una funci\'on $F(q,p,t)$ tal que para
un tiempo arbitrario fijo 
$t=t_{0}$ 
$$dF(p,q,t_{0})= \sum_{i}p_{i}dq_{i}-\sum_{i}P_{i}(q,p,t_{0})
dQ_{i}(p,q,t_{0})~,$$
donde
$$dF(p,q,t_{0})= \sum_{i}\frac{\partial F(p,q,t_{0})}{\partial q_{i}}dq_{i}
+ \sum_{i}\frac{\partial 
F(p,q,t_{0})}{\partial p_{i}}dp_{i}$$
y
$$dQ(p,q,t_{0})= \sum_{i}\frac{\partial Q(p,q,t_{0})}{\partial q_{i}}dq_{i}
+ \sum_{i}\frac{\partial
Q(p,q,t_{0})}{\partial p_{i}}dp_{i}$$

\noindent
{\bf Ejemplo:}
Demostrar que la siguiente transformaci\'on es can\'onica:
\begin{eqnarray*}
P &=& \frac{1}{2}(p^2+q^2)\\
Q &=& Tan^{-1}\left( \frac{q}{p}\right)~.
\end{eqnarray*}
{\bf Soluci\'on:}
De acuerdo a la definici\'on 1, debemos verificar que $pdq-PdQ$ es una
diferencial exacta. Sustituyendo $P$ y $Q$ de la transformaci\'on dada en el
problema obtenemos:
$$
pdq-PdQ = pdq - \frac{1}{2}(p^2+q^2)\frac{pdq-qdq}{p^2+q^2}=
d\left(\frac{pq}{2}\right)~.
$$
De lo anterior concluimos que efectivamente, la transformaci\'on dada en el
problema es can\'onica.
Sabemos que un sistema din\'amico se encuentra caracterizado por su 
Hamiltoniano $H=H(q,p,t)$, en donde $q=q(q_{1},q_{2},...,q_{n})$, y 
$p=p(p_{1},p_{2},...,p_{n})$, y que por tanto el sistema tiene asociado un 
conjunto de $2n$ ecuaciones diferenciales de primer orden, dadas por las 
ecuaciones de Hamilton:
\begin{eqnarray}
\dot{q_{i}}=\frac{\partial H}{\partial p_{i}}\\
-\dot{p_{i}}=\frac{\partial H}{\partial q_{i}}~.
\end{eqnarray}
Sean las transformaciones de coordenadas en el espacio de fase, denotadas
como 
\begin{eqnarray}
Q_{j}=Q_{j}(q,p,t)\\
P_{j}=P_{j}(q,p,t)~,
\end{eqnarray}
entonces, de acuerdo a lo dicho al principio, denotaremos como  
transformaciones can\'onicas al conjunto de transformaciones de la forma de 
$(3)$ y $(4)$ para las cuales, 
an\'alogamente a $(1)$ y $(2)$, exista una funci\'on $K=K(Q,P,t)$ tales 
que podamos escribir
\begin{eqnarray}
\dot{Q_{i}}=\frac{\partial K}{\partial P_{i}}\\
-\dot{P_{i}}=\frac{\partial K}{\partial Q_{i}}~.
\end{eqnarray}
La relaci\'on existente entre el Hamiltoniano $H$ y el nuevo Kamiltoniano 
$K$\footnote{Aqui seguiremos la misma terminologia empleada por 
Goldstein al referirse a la nueva funci\'on $K=K(Q,P,t)$, la cual difiere 
del Hamiltoniano $H=H(p,q,t)$ por una derivada temporal aditiva, como la 
\it{Kamiltoniana}.} se puede obtener a partir de las siguientes
consideraciones\footnote{Una derivaci\'on alternativa a la presente ha 
sido dada por G. S. S. Ludford and D. W. Yannitell, Am. J. Phys. 36,
231 (1968).}.\\
$\;$\\
De acuerdo al principio de Hamilton, la trayectoria real que un sistema
cl\'asico
describir\'a se puede obtener a partir de variaci\'on de la integral de
acci\'on dada por 
\begin{equation}
\delta \int (\sum_{i}p_{i}dq_{i} - Hdt) = 0~.
\end{equation}
Si la transformaci\'on es can\'onica, entonces el nuevo
Kamiltoniano $K$ debe tambi\'en
cumplir una relaci\'on similar a (7), es decir, con el nuevo conjunto de
coordenadas $Q$ y $P$ tambi\'en es v\'alido que
\begin{equation}
\delta \int (\sum_{i}P_{i}dQ_{i} - Kdt) = 0~.
\end{equation}
Sabemos adem\'as que, de acuerdo a la transformaci\'on de 
Legendre, $\sum_{i}p_{i}dq_{i} - Hdt= L(q,\dot{q},t)dt$, por lo 
que (7) -lo mismo que (8)- equivale a 
\begin{equation}
\delta \int_{t_{1}}^{t_{2}} L(q,\dot{q},t)dt = 0
\end{equation}
y que (9) no se altera si $L$ es sustituido por ${\cal L}= L +
\frac{dF(q,t)}{dt}$ porque en este caso,
\begin{eqnarray}
\delta \int_{t_{1}}^{t_{2}} {\cal L}dt = \delta \int_{t_{1}}^{t_{2}} (L +
\frac{dF(q,t)}{dt})dt~,
\end{eqnarray}
o en forma equivalente
\begin{eqnarray}
\delta \int_{t_{1}}^{t_{2}} {\cal L}dt = \delta \int_{t_{1}}^{t_{2}}
L(q,\dot{q},t)dt + 
\delta F(q_{(2)},t_{2}) - \delta F(q_{(1)},t_{1})
\end{eqnarray}~,
por lo que (10) y (11) difieren solamente por terminos constantes los
cuales dan como resultado 
una variaci\'on nula al momento de aplicar el principio de Hamilton.\\
$\;$
De acuerdo a lo anterior podemos exigir que el Hamiltoniano $H$ y el
Kamiltoniano $K$ se 
encuentren relacionados por la ecuaci\'on \footnote{Algunos autores
agregan al lado derecho de esta ecuaci\'on un factor multiplicativo
constante $A$ ya que
en este caso tampoco se ve alterado (1.9). Nosotros escogimos
arbitrariamente $A=1$, es decir,
decidimos trabajar con {\it transformaciones can\'onicas restringidas}, ya
que este caso es 
suficiente para mostrar la estructura de las transformaciones can\'onicas.}
\begin{eqnarray}
p_{i}\dot{q_{i}} - H = P_{i}\dot{Q_{i}} - K + \frac{dF}{dt}~.
\end{eqnarray}
La funci\'on $F$ es llamada {\bf funci\'on generadora}. Esta puede ser
expresada como una
funci\'on de cualquier conjunto arbitrario de variables independientes.
Sin embargo, algunos
resultados muy convenientes son obtenidos si $F$ es expresada como
funci\'on de las $n$ 
viejas variables y las $n$ nuevas variables, m\'as el tiempo.  Los
resultados son 
especialmente convenientes si las $n$ viejas variables son en su totalidad
las $n$ $q_{i}$ 
- o las $n$ $p_{i}$-, y si las nuevas variables son en su totalidad
las $n$ $Q_{i}$ -
o las $n$ $P_{i}$.\\
$\;$\\
Atendiendo a lo dicho en el p\'arrafo anterior, las posibles combinaciones
de $n$ variables viejas y $n$ 
variables nuevas -incluyendo a $t$- en la funci\'on generadora
son: \footnote{Usaremos la misma 
convenci\'on de Goldstein para denotar cada una de las diferentes
combinaciones de las variables nuevas y 
viejas en la funci\'on generadora.}
\begin{eqnarray}
F_{1} & = & F_{1}(Q,q,t) \\ 
F_{2} & = & F_{2}(P,q,t) \nonumber \\
F_{3} & = & F_{3}(Q,p,t) \nonumber \\
F_{4} & = & F_{4}(P,p,t)~. \nonumber
\end{eqnarray}
Por otro lado, si multiplicamos $(1.12)$ por $dt$ obtenemos:
\begin{equation}
p_{i}dq_{i} - Hdt = PdQ_{i} - Kdt + dF~.
\end{equation}
Y haciendo el cambio $F \rightarrow F_{1}$ en la relaci\'on anterior, y
recordando que $dQ_{i}$, 
$dq_{i}$, y $dt$ son variables independientes, obtenemos:
\begin{eqnarray}
P_{i} &=& -\frac{\partial F_{1}}{\partial Q_{i}} \nonumber \\
p_{i} &=& \frac{\partial F_{1}}{\partial q_{i}} \nonumber \\
K     &=& H+\frac{\partial F_{1}}{\partial t}~. \nonumber
\end{eqnarray}
Mediante manipuleos algebraicos es posible obtener expresiones  an\'alogas a
la anterior que involucren a 
las restantes funciones generadores.  Los resultados que se obtienen se
muestran a continuaci\'on:
\begin{center}
\begin{tabular}{clll}
$F_{2}:$ & $Q_{i} = \;\; \frac{\partial F_{2}}{\partial P_{i}}$ & $p_{i}
= \;\; \frac{\partial 
F_{2}}{\partial q_{i}}$ 
        & $K     = H+\frac{\partial F_{2}}{\partial t}$ \\
&&&\\
$F_{3}:$ & $P_{i} = -\frac{\partial F_{3}}{\partial Q_{i}}$ & $q_{i}
= -\frac{\partial F_{3}}{\partial 
p_{i}}$  
        & $K     = H+\frac{\partial F_{3}}{\partial t}$ \\
&&&\\
$F_{4}:$ & $Q_{i} = \;\; \frac{\partial F_{4}}{\partial P_{i}}$  & $q_{i}
= -\frac{\partial F_{4}}{\partial 
p_{i}}$ 
        & $K     = H+\frac{\partial F_{4}}{\partial t}~.$ \\
\end{tabular}
\end{center}
En la practica resulta \'util el siguiente teorema, el cual permite, junto con
las definiciones dadas 
en la introducci\'on para que una transformaci\'on sea can\'onica (ya sea
que la 
transforamaci\'on dependa o no expl\'{\i}citamente del tiempo), resolver
cualquier problema mec\'anico 
de inter\'es \footnote{Para un ejemplo, ver la secci\'on final de
este cap\'{\i}tulo.}.

\noindent
{\bf Teorema 1.1.}:
Consideremos un sistema sobre el cual se ejerce una fuerza
neta dada. Supongamos 
adem\'as que el estado din\'amico del sistema est\'a definido por un conjunto
de variables $q,p= 
q_{1},q_{2},...,q_{n},p_{1},p_{2},...,p_{n}$ y que el Hamiltoniano del sistema
es $H=H(q,p,t)$ tal que 
el comportamiento de las variables $q$ y $p$ est\'e determinado por las
ecuaciones de Hamilton
\begin{eqnarray}
\dot{q_{i}} &=& \;\; \frac{\partial H(q,p,t)}{\partial p_{i}} \nonumber \\
\dot{p_{i}} &=& -\frac{\partial H(q,p,t)}{\partial q_{i}}~. \nonumber 
\end{eqnarray}
Si nosotros hacemos una transformaci\'on a las nuevas variables 
$$Q=Q(q,p,t) \qquad ;\qquad P=P(q,p,t)$$
y si la transformaci\'on es can\'onica, es decir, si existe una
funci\'on $F(q,p,t)$ tal que para un 
tiempo arbitrario fijo $t=t_{0}$,
$$dF(q,p,t_{0})= \sum_{i}y_{i}dx_{i}-\sum_{i}Y_{i}dX_{i}~,$$
donde $x_{i},y_{i}=q_{i},p_{i}$ o $p_{i}, -q_{i}$ y $X_{i},Y_{i}=Q_{i},P_{i}$,
o $P_{i}, -Q_{i}$, 
entonces las ecuaciones de movimiento en t\'erminos de las
variables $Q$ y $P$ son
\begin{eqnarray}
\dot{Q_{i}} &=& \;\; \frac{\partial K(Q,P,t)}{\partial P_{i}} \nonumber \\
\dot{P_{i}} &=& -\frac{\partial K(Q,P,t)}{\partial Q_{i}}~, \nonumber
\end{eqnarray}
donde
$$K \equiv H + \frac{\partial F(q,p,t)}{\partial t} + \sum_{i}Y_{i}
\frac{\partial X_{i}(q,p,t)}{\partial t}~.$$
Adem\'as si el determinante de la matriz $[\frac{\partial X_{i}}
{\partial y_j}]$ es distinto
de cero, entonces la ecuaci\'on anterior se reduce a
$$K \equiv H + \frac{\partial F(x,X,t)}{\partial t}~.$$
\section*{5.2 Condiciones necesarias y suficientes para que una
transformaci\'on sea can\'onica}
Hemos mencionado al principio que por transformaci\'on can\'onica entenderemos
aquella transformaci\'on 
que, independientemente de cual sea la forma del Hamiltoniano, preserva la
forma de las ecuaciones de 
Hamilton.  Se debe ser muy cuidadoso respecto a este punto ya que es posible
que existan
transformaciones que preservan la forma de las ecuaciones de 
Hamilton, pero para {\it un Hamiltoniano particular} \footnote{Ver por
ejemplo, J. Hurley, Am. J. Phys. {\bf 40}, 533 (1972).}. Algunos autores le denominan a este tipo de
transformaci\'on {\bf 
transformaci\'on can\'onica respecto a H}
\footnote{Ver por ejemplo, R. A. Matzner y
L. C. Shepley, {\it Classical Mechanics}
(Prentice Hall, 1991).}.\\ $\;$\\
Para ilustrar este punto utilizaremos el siguiente ejemplo, sugerido en
el art\'{\i}culo de J. Hurley:
Consideremos un {\it sistema f\'{\i}sico particular} cuyo Hamiltoniano sea
\begin{eqnarray}
H = \frac{p^2}{2m}
\end{eqnarray}
y consideremos las transformaciones 
\begin{eqnarray} 
\begin{array}{lll}
P & = & p^{2}  \\
Q & = & q~. 
\end{array}
\end{eqnarray}
Es f\'acil demostrar que el Kamiltoniano $K$ dado por 
\begin{eqnarray}
K=\frac{2P^{3/2}}{3m} \nonumber
\end{eqnarray}
conduce a
\begin{eqnarray}
\dot{P}=2p\dot{p}=0=-\frac{\partial K}{\partial Q} \nonumber 
\end{eqnarray}
y 
\begin{equation}
\dot{Q}=\dot{q}=\frac{p}{m}=\frac{P^{1/2}}{m}=\frac{\partial K}{\partial P}~.
\nonumber
\end{equation}
Por otro lado, si escogemos el Hamiltoniano
\begin{eqnarray}
H = \frac{p^2}{2m} + q^2~, \nonumber
\end{eqnarray}
entonces es imposible encontrar un Kamiltoniano $K$ tal que al usar las
ecuaciones de 
transformaci\'on $(16)$ se preserve la forma de las
ecuaciones de Hamilton. Como vemos, en el 
anterior ejemplo las ecuaciones $(16)$ preservan la
forma de las ecuaciones de Hamilton, pero 
para un {\it Hamiltoniano en particular}.

\noindent
Se puede demostrar 
que las condiciones necesarias y suficientes 
para que transformaciones de la forma $(3)$ y $(4)$ sean can\'onicas,
es decir, 
que preserven la forma de las ecuaciones de Hamilton {\it independientemente
del Hamiltoniano que se considere}, son:
\begin{equation}
[Q_{i},P_{j}]= \alpha 
\end{equation}
\begin{equation}
[P_{i},P_{j}]  =  0\\
\end{equation}
\begin{equation}
[Q_{i},Q_{j}]  =  0~,
\end{equation}
en donde $\alpha$ es una constante cualquiera, relacionada con cambios de
escala.
Algunos comentarios merecen ser mencionados antes de finalizar esta secci\'on.  En primer lugar, 
debemos mantener en mente que $Q$ y $P$ {\it no constituyen variables que
definan la configuraci\'on 
del sistema}, es decir, no constituyen en general un conjunto de coordenadas
generalizadas 
\footnote{Salvo el caso trivial en que la transformaci\'on can\'onica
sea $Q=q$ y $P=p$.}.  Para 
distinguir a $Q$ y $P$, de las coordenadas generalizadas $q$ y $p$, se les
denomina {\it variables 
can\'onicas}. Y a las ecuaciones de movimiento -similares en forma a las
ecuaciones de Hamilton para 
las coordenadas generalizas $q$ y $p$- que se obtienen para $Q$ y $P$ se les
denomina {\it ecuaciones can\'onicas de Hamilton}.
En segundo lugar, aunque no lo probamos aqu\'{\i}, si la
transformaci\'on $Q=Q(q,p,t)$ y 
$P=P(q,p,t)$ es can\'onica, entonces la transformaci\'on inversa $q=q(Q,P,t)$
y $p=p(Q,P,t)$ es 
tambi\'en can\'onica \footnote{Para una demostraci\'on sobre este punto, ver
por ejemplo, E. A. Desloge,
{\it Classical Mechanics, Volume 2} (John Wiley \& Sons, 1982).}.
\section*{5.3 Ejemplo de aplicaci\'on de TC}
Como se mencion\'o en la introducci\'on a este cap\'{\i}tulo, la idea principal
de realizar una
transformaci\'on can\'onica es encontrar sistemas de coordenadas (en el
espacio de fases) los cuales 
preserven la forma de las ecuaciones de Hamilton, independientemente de la
forma del Hamiltoniano, y 
escoger de entre todas ellos, aquel que {\it facilite la resoluci\'on del
problema en particular}. 
Vamos a ilustrar este punto con un ejemplo.

\noindent
\underline{{\it EJEMPLO}}:

\noindent
El Hamiltoniano de cierto sistema f\'{\i}sico esta dado
por $H=\omega^{2}p(q+t)^{2}$, donde
$\omega$ es una constante. Determine $q$ como una funci\'on del tiempo.

\noindent
{\bf Soluci\'on:}

\noindent
1. {\it Resolviendo las ecuaciones de Hamilton para las
variables $q$ y $p$.} Al aplicar las
ecuaciones de Hamilton $(1)$ y $(2)$ al Hamiltoniano dado en este
problema, se obtiene
\begin{eqnarray}
\omega^{2}(q+t)^{2}=\dot{q}, & \;\;\;\;\;\;\;\; 2\omega^{2}p(q+t)=-\dot{p}~.
\nonumber
\end{eqnarray} 
Este sistema no es de soluci\'on f\'acil. Sin embargo, este problema se
puede resolver f\'acilmente
con una adecuada transformaci\'on can\'onica, tal como se muestra a
continuaci\'on.

\noindent
2. {\it Haciendo uso de la transformaci\'on can\'onica $Q=q+t$, $P=p$.}
De acuerdo al teorema dado en la secci\'on $(1.1)$, puesto que
\begin{eqnarray}
\frac{\partial Q}{\partial  p} &=& 0  \nonumber \\
\frac{\partial P}{\partial (-q)} &=& 0~, \nonumber
\end{eqnarray}
entonces el Kamiltoniano $K$ del sistema est\'a dado por
\begin{equation}
K = H + \frac{\partial F(q,p,t)}{\partial t} + P\frac{\partial Q}{\partial t}
 - Q\frac{\partial P}{\partial t}~. 
\end{equation}
La forma de la funci\'on $F(q,p,t)$ la encontramos a partir de la definici\'on
de transformaci\'on
can\'onica dada en la secci\'on 1 -este caso corresponde a una transformaci\'on
can\'onica
dependiente expl\'{\i}citamente del tiempo-. Es decir, a partir de
$$dF(q,p,t)= pdq - PdQ~.$$
Y sustituyendo $Q=q+t$, $P=p$ en la relaci\'on anterior se obtiene sin ninguna
dificultad que
$$F(q,p,t)= c, \;\;\;\; \mbox{c= constante}~.$$
Por otro lado
\begin{eqnarray}
\frac{\partial P}{\partial t} &=& 0  \nonumber \\
\frac{\partial Q}{\partial t} &=& 1~.  \nonumber
\end{eqnarray}
Finalmente, sustituyendo estos resultados
(y sustituyendo $Q=q+t$, $P=p$ en el Hamiltoniano $H$)
en $(21)$ obtenemos
$$K= P(\omega^{2}Q^{2}+1)$$
Y de $(5)$ obtenemos
$$\dot{Q}= \omega^{2}Q^{2}+1~.$$
Esta ecuaci\'on diferencial se resuelve f\'acilmente, y se obtiene como
resultado final
$$q=\frac{1}{\omega}{\rm tan}(\omega t + \phi)-t~,$$
en donde $\phi$ es una fase arbitraria.
Como se puede observar, la correcta elecci\'on de la transformaci\'on
can\'onica puede
facilitar la soluci\'on de cualquier problema mec\'anico.


\newpage



\centerline{\Large 6. PARENTESIS DE POISSON}

\bigskip

\noindent
{\bf Pr\'ologo}:
Los par\'entesis de Poisson son herramientas anal\'{\i}ticas muy \'utiles para
estudiar el
comportamiento de cualquier sistema din\'amico.  Nosotros definiremos en este
cap\'{\i}tulo
lo que se entiende por par\'entesis de Poisson; daremos algunas de sus
propiedades, y finalmente
presentaremos algunas aplicaciones de los mismos en el estudio de sistemas
din\'amicos.

\bigskip
\bigskip

{\bf CONTENIDO:}

\bigskip



1. Definici\'on y propriedades

\bigskip

2. Formulaci\'on de Poisson para las ecs. de movimiento

\bigskip

3. Las constantes de movimiento en la formulaci\'on de Poisson

\newpage


\noindent
{\bf 1. Definici\'on y propiedades de los par\'entesis de Poisson}
\bigskip

\noindent
Si $u$ y $v$ son cualesquiera dos cantidades que dependen del estado
din\'amico del sistema (es
decir, de $p$ y de $q$) y posiblemente del tiempo, el par\'entesis de Poisson
de $u$ y $v$
con respecto a un conjunto de variables can\'onicas $q$ y $p$ \footnote{Como
en el cap\'{\i}tulo
anterior, entendemos a $q$ y $p$ como $q=q_{1}, q_{2},...,q_{n}$ y $p=p_{1},
p_{2},..., p_{n}.$} 
es definido como
\setcounter{equation}{0}\\
\begin{equation}
[u,v] \equiv \sum_{i} \left( \frac{\partial u(q,p,t)}{\partial q_{i}}
\frac{\partial v(q,p,t)}{\partial p_{i}}
- \frac{\partial u(q,p,t)}{\partial p_{i}}\frac{\partial v(q,p,t)}{\partial 
q_{i}} \right)~.
\end{equation}
Los par\'entesis de Poisson tienen las siguientes propiedades (donde $u$, $v$,
y $w$ son 
funciones arbitrarias de $q$, $p$, y de $t$;
$a$ es una constante arbitraria, y $r$
es cualquiera de las variables $q_{i}$, $p_{i}$ o $t$) \footnote{La prueba
de estas propiedades
se logra utilizando la definici\'on de los par\'entesis de Poisson para
expresar cada uno de los t\'erminos
de estas identidades en t\'erminos de las derivadas parciales
de $u$, $v$, y $w$, y notando 
por inspecci\'on la valides de las ecuaciones resultantes.}:
\begin{description}
\item[{\it Propiedad 1.}] \hspace{2in} $[u,v] \equiv - [v,u]$
\item[{\it Propiedad 2.}] \hspace{2in} $[u,u] \equiv 0$
\item[{\it Propiedad 3.}] \hspace{2in} $[u,v+w] \equiv [u,v] + [u,w]$
\item[{\it Propiedad 4.}] \hspace{2in} $[u,vw] \equiv v[u,w] + [u,v]w$
\item[{\it Propiedad 5.}] \hspace{2in} $a[u,v] \equiv [au,v] \equiv [u,av]$
\item[{\it Propiedad 6.}] \hspace{2in} $\frac{\partial [u,v]}{\partial r}
\equiv [\frac{\partial u}{\partial r},
v]+[u,\frac{\partial v}{\partial r}]$
\item[{\it Propiedad 7.}] {\it Identidad de Jacobi},
\hspace{0.3in}  $[u,[v,w]]+[v,[w,u]]+[w,[u,v]] \equiv 0~.$
\end{description}
Otra propiedad muy importante de los par\'entesis de Poisson es enunciada
en el siguiente teorema:

\noindent
{\bf Teorema 6.1} Si la transformaci\'on $Q=Q(q,p,t)$, $P=P(q,p,t)$
es una transformaci\'on can\'onica, el
par\'entesis de Poisson de dos cantidades $u$ y $v$ con respecto al conjunto de variables $q$, $p$, es igual 
al par\'entesis de Poisson de $u$ y $v$ con respecto al conjunto de variables $Q$, $P$, es decir
$$
\sum_{i} \left( \frac{\partial u(q,p,t)}{\partial q_{i}}
\frac{\partial v(q,p,t)}{\partial p_{i}}
- \frac{\partial u(q,p,t)}{\partial p_{i}}\frac{\partial v(q,p,t)}{\partial q_{i}}
\right)=
$$
$$
\sum_{i} \left( \frac{\partial u(q,p,t)}{\partial Q_{i}}
\frac{\partial v(q,p,t)}{\partial P_{i}}
- \frac{\partial u(q,p,t)}{\partial P_{i}}
\frac{\partial v(q,p,t)}{\partial Q_{i}} \right)~.
$$

\bigskip
\noindent
{\bf 2. La formulaci\'on de Poisson para las ecuaciones de movimiento}
\bigskip

\noindent
Resumimos a continuaci\'on,  en forma de teoremas, los resultados m\'as
importantes sobre los 
par\'entesis de Poisson en el an\'alisis de el movimiento de cualquier sistema
din\'amico \footnote{
Las demostraciones han sido omitidas por ser muy conocidas.  Ver por ejemplo,
E. A. Desloge,
{\it Classical Mechanics, Volume 2} (John Wiley \& Sons, 1982).}:

\noindent
{\bf Teorema 6.2}: Consideremos un sistema cuyo estado din\'amico est\'a
definido por las variables
can\'onicas $q$, $p$ y cuyo comportamiento din\'amico est\'a definido
por la Hamiltoniana 
$H=H(q,p,t)$.  Sea $F$ una cantidad arbitraria que depende del estado
din\'amico del sistema (es
decir, de $q$, $p$, y posiblemente $t$). La raz\'on de cambio en el
tiempo de $F$ est\'a dada por
$$\dot{F}=[F,H]+\frac{\partial F(q,p,t)}{\partial t}~,$$
donde $[F,H]$ es el par\'entesis de Poisson de $F$ con $H$.

\noindent
{\bf Teorema 6.3} {\bf (Formulaci\'on de Poisson de las ecs.
de movimiento)}. Consideremos
un sistema cuyo estado din\'amico est\'a definido
por las variables can\'onicas $q$, $p$, y 
cuyo comportamiento din\'amico est\'a dado por la
Hamiltoniana $H=H(q,p,t)$.  El movimiento
del sistema est\'a gobernado por las ecuaciones
\begin{eqnarray}
\dot{q_{i}} &=& [q_{i},H] \nonumber \\
\dot{p_{i}} &=& [p_{I},H]~. \nonumber
\end{eqnarray}

\noindent
{\bf 3. Las constantes de movimiento en la formulaci\'on de Poisson}

\bigskip

\noindent
Nuevamente enunciaremos los resultados m\'as importantes sobre la
formulaci\'on de Poisson 
de las constantes de movimiento, en forma de teoremas. Estos son los
siguientes:

\noindent
{\bf Teorema 6.4} Si una  cantidad din\'amica $F$ no es una funci\'on
expl\'{\i}cita de el tiempo, y si el par\'entesis de Poisson
de $F$ con $H$ es nulo,
es decir, $[F,H]=0$, entonces $F$ es una constante del movimiento.

\noindent
{\bf Corolario 6.4.a} Si el Hamiltoniano no es una funci\'on expl\'{\i}cita
del tiempo, entonces
es una constante de movimiento.



\newpage


\centerline{\Large 7. LAS ECUACIONES DE HAMILTON-JACOBI}

\bigskip
\bigskip

\noindent
{\bf Pr\'ologo}:
Sabemos de los cap\'{\i}tulos anteriores que podemos, en principio, disminuir
la complejidad de cualquier problema de din\'amica escogiendo una adecuada
transformaci\'on can\'onica.
En particular, podemos tratar de buscar aquellas transformaciones can\'onicas
que hagan que el Kamiltoniano $K$ sea nulo, lo que da las ecs. de
Hamilton-Jacobi.

\bigskip
\bigskip

{\bf CONTENIDO:}

\bigskip

7.1 Introducci\'on

\bigskip

7.2 Ec. de Hamilton-Jacobi dependiente del tiempo

\bigskip

7.3 Ec. de Hamilton-Jacobi independiente del tiempo

\bigskip

7.4 Generalizacion de la ec. de Hamilton-Jacobi

\bigskip

7.5 Ejemplo de aplicacion de la ec. de Hamilton-Jacobi

\newpage

\section*{7.1 Introducci\'on.}
Para lograr nuestros prop\'ositos en este cap\'{\i}tulo, necesitamos
hacer uso del
siguiente resultado, el cual nos permite encontrar el conjunto de variables
can\'onicas que hacen que
el Kamiltoniano asuma {\it una forma particular}.

\noindent
{\bf Teorema 7.1.} Consideremos un sistema cuyo estado din\'amico est\'a
definido por las 
variables $p$, $q$ y cuyo comportamiento bajo la acci\'on de una fuerza dada
est\'a dado por 
la Hamiltoniana $H=H(q,p,t)$. Sea $K=K(Q,P,t)$ una {\it funci\'on conocida} de
las variables
can\'onicas $Q$, $P$, y tambi\'en del tiempo.  Entonces cualquier
funci\'on $F(q,Q,t)$ que 
satisfaga la ecuaci\'on diferencial parcial
\setcounter{equation}{0}\\
\begin{eqnarray}
K\left[Q,-\frac{\partial F(q,Q,t)}{\partial Q},t\right]= H\left[q,\frac{\partial F(q,Q,t)}{\partial q},t\right]+
\frac{\partial F(q,Q,t)}{\partial t} \nonumber
\end{eqnarray}
y tambi\'en satisface la condici\'on 
\begin{eqnarray}
\left| \frac{\partial ^{2} F(q,Q,t)}{\partial q_{j}\partial Q_{j}}\right|
\neq 0 \nonumber
\end{eqnarray}
es una funci\'on generadora para una transformaci\'on can\'onica de
las variables $q$, $p$ a las variables $Q$, $P$, y la correspondiente
Kamiltoniana es justamente $K=K(Q,P,t)$.

\noindent
En las siguientes secciones haremos uso del anterior problema para encontrar
aquellas transformaciones
can\'onicas cuyo kamiltoniano sea nulo \footnote{Es
decir, $K\left[Q,-\frac{\partial F(q,Q,t)}{\partial Q},t\right]=0$.};
esto nos conducir\'a a las ecuaciones de Hamilton-Jacobi.
\section*{7.2 Ecuaci\'on de HJ dependiente del tiempo.}
Como una consecuencia del Teorema 7.1 y de exigir que el Kamiltoniano $K$ sea
nulo, obtenemos el siguiente teorema:

\noindent
{\bf Teorema 7.2.} Consideremos un sistema de $f$ grados de
libertad cuyo estado 
din\'amico est\'e definido por el conjunto de variables $q$, $p$ y cuyo
Hamiltoniano sea 
$H=H(q,p,t)$.  Si construimos la ecuaci\'on diferencial parcial
\begin{eqnarray}
H\left[ q,\frac{\partial S(q,t)}{\partial q}, t\right]
+ \frac{\partial S(q,t)}{\partial t}=0
\end{eqnarray}
y si encontramos una soluci\'on a \'esta ecuaci\'on de la forma 
$$S=S(q,\alpha,t)~,$$
donde $\alpha=\alpha_{1}, \alpha_{2},...,\alpha_{f}$ es un conjunto de
constantes, y si la 
soluci\'on satisface la condici\'on
$$\left| \frac{\partial^{2}S(q,\alpha,t)}{\partial q_{i}\partial \alpha_{i}}
\right| \neq 0~,$$
entonces $q(t)$ puede ser obtenido de las ecuaciones
\begin{equation}
\frac{\partial S(q,\alpha,t)}{\partial \alpha_{i}}= \beta_{i}~,
\end{equation}
donde $\beta= \beta_{1},\beta_{2},...,\beta_{f}$ es un conjunto de constantes.
El conjunto de ecuaciones $(2)$ nos proporciona $f$ ecuaciones algebraicas
en las $f$ variables desconocidas $q_{1}, q_{2},...,q_{f}$.  Los valores
de las constantes $\alpha$ y
$\beta$ son determinados por las condiciones de frontera.
Adem\'as es posible, dada $q(t)$, encontrar $p(t)$ a partir de la relaci\'on
\begin{eqnarray}
p_{i} &=& \frac{\partial S(q,\alpha,t)}{\partial q_{i}}~.
\end{eqnarray}
A la ecuaci\'on diferencial parcial $(1)$ se le denomina {\it ecuaci\'on
de Hamilton-Jacobi 
dependiente del tiempo}.  Y a la funci\'on $S(q,\alpha,t)$ se
le denomina {\it funci\'on 
principal de Hamilton}.

\noindent
Para lograr una mayor comprensi\'on de este teorema,
as\'{\i} como del significado de las 
constantes $\alpha$ y $\beta$, procedemos a continuaci\'on a desarrollar
su demostraci\'on.

\noindent
{\bf Demostraci\'on del Teorema 7.2.}
De acuerdo al Teorema 7.1, cualquier funci\'on $F(q,Q,t)$
que satisface la ecuaci\'on diferencial parcial
$$H\left[q,\frac{\partial F(q,Q,t)}{\partial q},t\right]
+\frac{\partial F(q,Q,t)}{\partial t}=0$$
y tambi\'en satisface la condici\'on 
$$\left| \frac{\partial ^{2} F(q,Q,t)}{\partial q_{j}
\partial Q_{j}}\right| \neq 0 $$
deber\'a ser una funci\'on generadora para una transformaci\'on
can\'onica a un conjunto de 
variables can\'ononicas $Q$, $P$ para las cuales el
Kamiltoniano $K$ sea nulo, es decir
$K(Q,P,t)=0$.  La funci\'on
$$F(q,Q,t)=[S(q,\alpha,t)]_{\alpha=Q}\equiv S(q,Q,t)$$
es tal funci\'on.  Entonces $S(q,Q,t)$ es la funci\'on generadora para
una transformaci\'on 
can\'onica que lleva al nuevo conjunto de variables $Q$, $P$, para las
cuales el Kamiltoniano 
$K$ es identicamente igual a cero. Las ecuaciones de
transformaci\'on asociadas con la 
funci\'on generadora $S(q,Q,t)$ est\'an definidas por las ecuaciones
\begin{eqnarray}
p_{i} &=& \;\; \frac{\partial S(q,Q,t)}{\partial q_{i}} \\
P_{i} &=& -\frac{\partial S(q,Q,t)}{\partial Q_{i}}
\end{eqnarray}
y como $K(Q,P,t)\equiv 0$, las ecuaciones de movimiento son
$$\begin{array}{lllll}
\dot{Q}_{i} &=& \;\; \frac{\partial K(Q,P,t)}{\partial P_{i}} &=& 0 \\
&&&& \\
\dot{P}_{i} &=& -\frac{\partial K(Q,P,t)}{\partial Q_{i}} &=& 0
\end{array}~.$$
De las anteriores ecuaciones inferimos que 
\begin{eqnarray}
Q_{i} &=& \;\; \alpha_{i} \\
P_{i} &=& -\beta~,
\end{eqnarray}
donde $\alpha_{i}$ y $\beta_{i}$ son constantes.  La elecci\'on
el signo negativo para $\beta$ 
en la ecuaci\'on $(7)$ se hizo por mera
conveniencia. Si ahora sustituimos las ecuaciones
$(6)$ y $(7)$ en la ecuaci\'on $(5)$ obtenemos
$$ -\beta_{i} = -\left[ \frac{\partial S(q,Q,t)}{\partial Q_{i}}
\right]_{Q=\alpha}=  
   -\frac{\partial S(q,\alpha,t)}{\partial \alpha_{i}}~,$$
la cual se reduce a $(2)$.  Si adem\'as sustituimos $(6)$ en $(4)$
obtenemos $(2)$. Esto completa la prueba.
\section*{7.3 Ecuaci\'on HJ independiente del tiempo.}
Si la funci\'on Hamiltoniana no depende expl\'{\i}citamente del tiempo,
nosotros podemos 
parcialmente resolver la ecuaci\'on de Hamilton-Jacobi dependiente del
tiempo. Este resultado se enuncia en el siguiente teorema:

\noindent
{\bf Teorema 7.3.} Consideremos un sistema de $f$ grados de
libertad cuyo estado 
din\'amico est\'a definido por el conjunto de variables $q$, $p$, y cuyo
comportamiento bajo la 
acci\'on de una fuerza definido por la funci\'on Hamiltoniana
independiente del tiempo $H(q,p)$.

\noindent
Si construimos la ecuaci\'on diferencial parcial 
\begin{equation}
H\left[q,\frac{\partial W(q)}{\partial q}\right]= E~,
\end{equation}
donde $E$ es una constante cuyo valor para un conjunto particular de
condiciones es igual al 
valor de la constante de movimiento $H(q,p)$ para las condiciones
de frontera dadas, y si 
nosotros podemos encontrar una soluci\'on a esta ecuaci\'on, de la forma
$$W=W(q,\alpha)~,$$
donde $\alpha \equiv \alpha_{1},\alpha_{2},...,\alpha_{f}$ es un conjunto
de constantes que 
expl\'{\i}cita o impl\'{\i}citamente incluye a la constante $E$,
es decir, $E=E(\alpha)$, y
si la soluci\'on satisface la condici\'on
$$\left| \frac{\partial^{2}W(q,\alpha)}{\partial q_{i} \partial
\alpha_{j}}\right| \neq 0~,$$
entonces las ecuaciones de movimiento est\'an dadas por 
\begin{equation}
\frac{\partial S(q,\alpha,t)}{\partial \alpha_{i}}= \beta_{i}
\end{equation}
donde
$$S(q,\alpha,t) \equiv W(q,\alpha)-E(\alpha)t$$
y $\beta = \beta_{1}, \beta_{2},...,\beta_{f}$ es un conjunto de
constantes. El conjunto de
ecuaciones $(9)$ proporciona $f$ ecuaciones algebraicas en las $f$
variables desconocidas $q_{1}, q_{2},...,q_{f}$.
Los valores de las constantes $\alpha$ y
$\beta$ son determinados por las condiciones de
frontera. La ecuaci\'on diferencial
parcial $(8)$ recibe el nombre de {\it ecuaci\'on de Hamilton-Jacobi
independiente del tiempo}, y a la funci\'on $W(q,\alpha)$ se le conoce
como {\it funci\'on caracter\'{\i}stica de Hamilton}.
\section*{7.4 Generalizaci\'on de la ecuaci\'on de HJ.}
La ecuaci\'on de Hamilton-Jacobi puede ser generalizada como se establece
en el siguiente teorema, el cual permite, algunas veces, el simplificar
algunos problemas de Hamilton-Jacobi.

\noindent
{\bf Teorema 7.4.} Consideremos un sistema de $f$ grados de
libertad cuyo estado 
din\'amico est\'a definido por un conjunto de varibles $x$, $y$, donde $x_{i},
y_{i}= q_{i}, p_{i}$
o $p_{i}, -q_{i}$, y cuyo comportamiento bajo la acci\'on de una fuerza dada
est\'a dado por 
la Hamiltoniana $H(x,y,t)$.  Si construimos la ecuaci\'on diferencial parcial
$$ H\left[ x,\frac{\partial S(x,t)}{\partial x},t\right]+\frac{\partial
S(x,t)}{\partial t}=0$$
y si podemos encontrar una soluci\'on a esta ecuaci\'on de la forma
$$S=S(x,\alpha,t)$$
donde $\alpha \equiv \alpha_{1},\alpha_{2},...,\alpha_{f}$ es un conjunto de
constantes, y si la soluci\'on satisface la condici\'on
$$\left|\frac{\partial^{2}S(x,\alpha,t)}{\partial x_{j}\partial\alpha_{j}}
\right|\neq  0$$
entonces el movimiento del sistema puede ser obtenido de las ecuaciones
\begin{eqnarray}
\frac{\partial S(x,\alpha,t)}{\partial x_{i}} &=&y_{i} \\
\frac{\partial S(x,\alpha,t)}{\partial \alpha_{i}} &=&\beta_{i}
\end{eqnarray}
donde $\beta \equiv \beta_{1},\beta_{2},...,\beta_{f}$ es un conjunto
de constantes.

\noindent
En la siguiente secci\'on mostraremos algunos ejemplos donde hacemos
uso de todos los resultados obtenidos hasta aqu\'{\i} en este cap\'{\i}tulo.
\section*{7.5 Ejemplo de aplicaci\'on de la ecuaci\'on de HJ}
Vamos a resolver el problema del oscilador arm\'onico unidimensional de
masa $m$, utilizando el m\'etodo de Hamilton-Jacobi.

\noindent
Sabemos que la Hamiltoniana del sistema es 
\begin{equation}
H=\frac{p^2}{2m}+\frac{kx^2}{2}
\end{equation}
De acuerdo al Teorema 7.2 la ecuaci\'on de Hamilton-Jacobi para el sistema es
\begin{equation}
\frac{1}{2m}\left( \frac{\partial F}{\partial q}\right)+\frac{kq^2}{2}
+\frac{\partial F}{\partial t}=0
\end{equation}
Supongamos una soluci\'on a $(13)$ de la forma $F=F_{1}(q)+F_{2}(t)$.
Por lo tanto, $(13)$ se convierte en
\begin{equation}
\frac{1}{2m}\left( \frac{dF_1}{dq} \right) ^{2} + \frac{kq^2}{2}
= -\frac{dF_2}{dt}
\end{equation}
Haciendo cada lado de la ecuaci\'on anterior igual a $\alpha$, encontramos
\begin{eqnarray}
\frac{1}{2m}\left( \frac{dF_1}{dq} \right) ^{2} + \frac{kq^2}{2}&=&\alpha \\
\frac{dF_2}{dt}&=& -\alpha
\end{eqnarray}
Omitiendo las constantes de integraci\'on, las soluciones son
\begin{eqnarray}
F_{1}&=& \int \sqrt{2m(\alpha - \frac{kq^2}{2})}dq \\
F_{2}&=& -\alpha t
\end{eqnarray}
Por tanto, la funci\'on generadora $F$ es
\begin{equation}
F= \int \sqrt{2m(\alpha - \frac{kq^2}{2})}dq  -\alpha t
\end{equation}
Y de acuerdo a $(2)$, $q(t)$ se obtiene a partir de
\begin{eqnarray}
\beta &=& \frac{\partial}{\partial \alpha}\left\{ \int \sqrt{2m(\alpha
- \frac{kq^2}{2})}dq  -\alpha t \right\}\\
&=& \frac{\sqrt{2m}}{2}\int \frac{dq}{\sqrt{\alpha - \frac{kq^2}{2}}} - t
\end{eqnarray}
Y realizando la integral obtenemos que
\begin{equation}
\sqrt{\frac{m}{k}}sen^{-1}(q\sqrt{k/2\alpha})= t+\beta
\end{equation}
Y despejando $q$ obtenemos finalmente que
\begin{equation}
q= \sqrt{\frac{2\alpha}{k}}sen\sqrt{k/m}(t+\beta)~.
\end{equation}
Podemos adem\'as dar una interpretaci\'on f\'{\i}sica a la
constante $\alpha$ siguiendo el siguiente razonamiento:

\noindent
El factor $\sqrt{\frac{2\alpha}{k}}$ debe corresponder justamente a la
m\'axima amplitud $A$ que el oscilador puede
tener.  Por otro lado, la energ\'{\i}a total $E$ de un oscilador arm\'onico
undimensional de amplitud $A$ est\'a dada por 
$$
E = \frac{1}{2}kA^2
  = \frac{1}{2}k\left( \sqrt{\frac{2\alpha}{k}}\right) ^2
  = \alpha~.
$$

\noindent
Es decir, $\alpha$ es f\'{\i}sicamente igual a la energ\'{\i}a total $E$
del oscilador arm\'onico unidimensional.

\bigskip
\bigskip

\begin{center} BIBLIOGRAFIA COMPLEMENTARIA \end{center}

\bigskip

\noindent
C.C. Yan, {\it Simplified derivation of the HJ eq.}, Am. J. Phys. 52, 555
(1984)

\bigskip
\noindent
N. Anderson \& A.M. Arthurs, {\it Note on a HJ approach to the rocket pb.},
Eur. J. Phys. 18, 404 (1997)

\bigskip
\noindent
M.A. Peterson, {\it Analogy between thermodynamics and mechanics},
Am. J. Phys. 47, 488 (1979)

\bigskip
\noindent
Y. Hosotani \& R. Nakayama, {\it The HJ eqs for strings and p-branes},
hep-th/9903193 (1999)


\newpage


\centerline{\Large 8. VARIABLES ACCION-ANGULO}

\bigskip
\bigskip

\noindent {\bf Pr\'ologo}:
La ecuaci\'on de Hamilton-Jacobi nos proporciona un m\'etodo de
transformar un conjunto de variables $q$, $p$ a un segundo conjunto de
variables can\'onicas
$Q$, $P$, siendo cada una de las variables can\'onicas una constante
de movimiento.

\noindent
Ahora, en este cap\'{\i}tulo nosotros consideraremos un m\'etodo, v\'alido
para ciertos 
tipos de movimiento, para transformar un conjunto de variables $q$, $p$,
a un segundo conjunto 
de variables $Q$, $P$, no siendo al mismo tiempo las dos variables can\'onicas
constantes de movimiento.

\bigskip
\bigskip

{\bf CONTENIDO:}

\bigskip

8.1 Sistemas separables

\bigskip

8.2 Sistemas c\'{\i}clicos

\bigskip

8.3 Variables acci\'on-\'angulo

\bigskip

8.4 Movimiento en variables acci\'on-\'angulo

\bigskip

8.5 Importancia de las variables acci\'on-\'angulo

\bigskip

8.6 Ejemplo: el oscilador arm\'onico

\newpage

\section*{8.1 Sistemas separables.}
Por sistemas separables nosotros entenderemos sistemas cuya
Hamiltoniana no sea una funci\'on expl\'{\i}cita del tiempo, es decir
\setcounter{equation}{0}\\
$$H=H(q,p)$$
y para los cuales es posible encontrar una soluci\'on a la ecuaci\'on de
Hamilton-Jacobi independiente del tiempo, de la forma
$$W(q,\alpha)=\sum_{i}W_{i}(q_{i},\alpha)~.$$
\section*{8.2 Sistemas c\'{\i}clicos.} Sabemos que el estado de un sistema
se encuentra caracterizado
por un conjunto de coordenadas generalizadas $q\equiv q_{1}, q_{2},..., q_{f}$
y por el conjunto
de momenta generalizados $p\equiv p_{1}, p_{2},..., p_{f}$.  Conforme el
sistema se mueve, \'este traza una \'orbita en el espacio $q$, $p$ (el
espacio cuyas coordenadas son las variables
$q_{1}, q_{2},..., q_{f}, p_{1}, p_{2},..., p_{f}$). De igual forma, el sistema traza una 
\'orbita en cada uno de los subespacios $q_{i}, p_{i}$.  La \'orbita en cada
uno de los planos
$q_{i}, p_{i}$ puede ser representada por una ecuaci\'on de la
forma $p_{i}= p_{i}(q_{i})$ o por 
un par de ecuaciones $p_{i}= p_{i}(t)$, $q_{i}= q_{i}(t)$.  Si
para cada valor de $i$,
la \'orbita $p_{i}= p_{i}(q_{i})$  en el plano $q_{i}-p_{i}$ es una curva
cerrada, entonces 
nos referimos a tal sistema como un {\it sistema c\'{\i}clico}.  En la
siguiente figura
mostramos las dos posibilidades por las cuales un sistema puede
ser c\'{\i}clico.  En la Figura 
$8.1a$, el sistema es c\'{\i}clico porque $q_{i}$ oscila entre los
l\'{\i}mites definidos $q_{i}=a$ y $q_{i}=b$; en la Figura $8.1b$, el sistema
es c\'{\i}clico porque $q_{i}$
se mueve desde $q_{i}=a$ hasta $q_{i}=b$, y luego inicia nuevamente
en $q_{i}=a$.
\vskip 1ex
\centerline{
\epsfxsize=190pt
\epsfbox{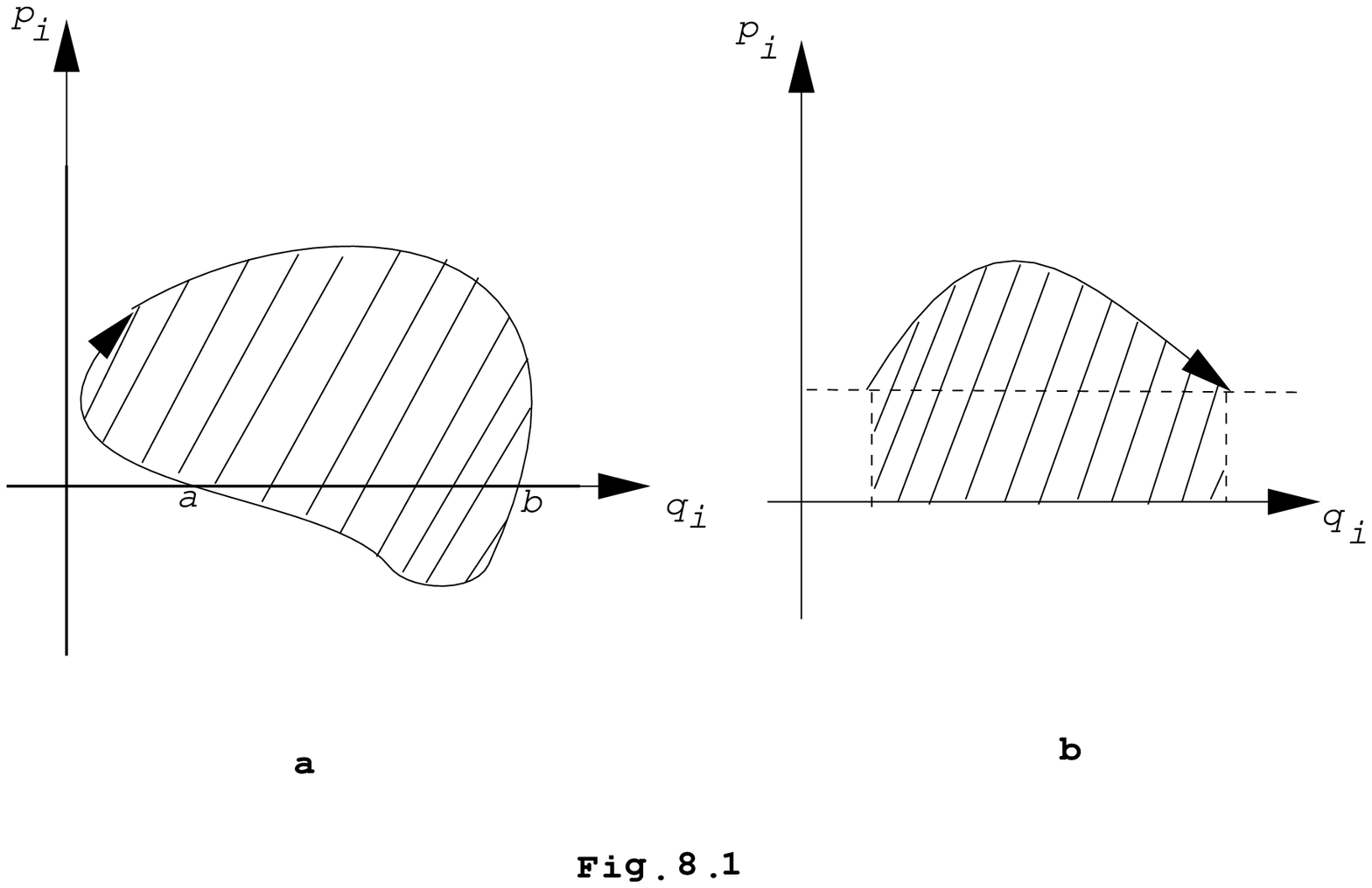}}
\vskip 2ex


\noindent
Es necesario en este punto hacer dos importantes aclaraciones:

{\bf Aclaraci\'on 1}: El t\'ermino c\'{\i}clico ha sido introducido
solamente para simplificar
la notaci\'on en las siguientes secciones.  No debe interpretarse
este t\'ermino pensando que
si el sistema es c\'{\i}clico en cada uno de los
subespacios $q_{i}, p{i}$, entonces el sistema
deba regresar a su estado original en el espacio $q, p$.

{\bf Aclaraci\'on 2}: Si un sistema c\'{\i}clico solamente tiene un grado de
libertad, el tiempo
requerido por el sistema para ejecutar un ciclo en el plano $q-p$ es constante; por tanto el 
movimiento en el plano $q-p$ ser\'a per\'{\i}odico en el tiempo.  Si el
sistema tiene mas de
un grado de libertad, entonces, en general, el tiempo requerido para ejecutar
un ciclo particular 
en uno de los espacios $q_{i}, p_{i}$ no ser\'a una constante, sino que
depender\'a del movimiento
de las otras coordenadas; por tanto, el movimiento en el espacio
dado $q_{i}, p_{i}$ no ser\'a
per\'{\i}odico en el tiempo. Debe de tenerse cuidado en este punto, ya
que {\it no es cierto 
que cada uno de los movimientos en cada plano $q_{i}, p_{i}$ es
per\'{\i}odico en el tiempo.}
\section*{8.3 Variables acci\'on-\'angulo.}
Consideremos un sistema c\'{\i}clico de $f$ grados de libertad, cuyo estado
din\'amico est\'a 
caracterizado por las variables can\'onicas $q$, $p$.  Sea $H(q,p)$ la
Hamiltoniana del sistema, y sea
$$W(q,\alpha)\equiv \sum_{i} W_{i}(q_{i},\alpha)~,$$
(donde $\alpha = \alpha_{1}, \alpha_{2},...,\alpha_{f}$ son constantes)
una soluci\'on a la ecuaci\'on de Hamilton-Jacobi independiente del tiempo
$$H(q,\frac{\partial W}{\partial q}) = E~.$$
Sea $J\equiv J_{1}, J_{2},..., J_{f}$ el conjunto de constantes definidas
por las ecuaciones
\begin{equation}
J_{i}(\alpha) = \oint \frac{\partial W_{i}(q_{i},\alpha)}{\partial
q_{i}}dq_{i}~,
\end{equation}
donde la integral es sobre un cilo completo para la variable $q_{i}$.
Si nosotros usamos la funci\'on
\begin{eqnarray*}
W(q,\alpha) &\equiv & W[q,\alpha (J)]\\
            &\equiv & \sum_{i}W_{i}[q_{i},\alpha (J)] \\
            &\equiv & \sum_{i}W_{i}(q_{i},J)
\end{eqnarray*}
como una funci\'on generadora para una transformaci\'on can\'onica de las
variables $q$, $p$ a un nuevo 
conjunto de coordenadas $w$, y momenta $J$, es decir, si
definimos las variables $w$ y $J$ por las 
ecuaciones de transformaci\'on
\begin{equation}
p_{i}  =  \frac{\partial W(q,\alpha)}{\partial q_{i}}  =  \frac{\partial 
W_{i}(q_{i},J)}{\partial q_{i}} 
\end{equation}
\begin{equation}
w_{i}   =  \frac{\partial W(q,J)}{\partial J_{i}}  
\end{equation}
entonces las nuevas coordenadas $w_{1}, w_{2},..., w_{f}$ son
llamadas {\it variables \'angulo}, y los 
nuevos momenta $J_{1}, J_{2},..., J_{f}$ son
llamadas {\it variables acci\'on}.

\noindent
De la ecuaci\'on $(2)$ obtenemos que
\begin{equation}
p_{i}(q_{i},\alpha)= \frac{\partial W_{i}[q_{i},J(\alpha)]}{\partial q_{i}}
=\frac{\partial 
W_{i}(q_{i},\alpha)}{\partial q_{i}}~.
\end{equation}
Sustituyendo la ecuaci\'on $(4)$ en $(1)$ obtenemos
\begin{equation}
J_{i}(\alpha)=\oint p_{i}(q_{i},\alpha)dq_{i}~.
\end{equation}
La ecuaci\'on $p_{i}=p_{i}(q_{i},\alpha)$ es la ecuaci\'on de la
proyecci\'on de la \'orbita
$p=p(q)$ sobre el subespacio $p_{i}, q_{i}$.  La integral en el lado derecho
de la ecuaci\'on 
$(5)$ es por tanto el area encerrada dentro de la \'orbita, o bajo
la \'orbita, como se ilustra 
por las regiones sombreadas en la Figura 8.1. Por tanto,
la funci\'on $J_{i}(\alpha)$ puede ser 
interpretada geometricamente como el area barrida en el
subespacio $q_{i}, p_{i}$ durante un ciclo
completo en este subespacio. Esta \'area depende de las constantes $\alpha$ o
equivalentemente de las condiciones iniciales y puede en general asumir
cualquier valor \footnote{Historicamente,
el primer intento de pasar de la mec\'anica cl\'asica a la
mec\'anica cu\'antica consisti\'o en asumir que el valor que $J_{i}$ podia
tomar no era completamente arbitrario, sino que debe ser un
m\'ultiplo de $h/2\pi$, donde $h$ es la constante de Planck.}.
\section*{8.4 El movimiento de un sistema en t\'erminos de las variables
de acci\'on-\'angulo.}
Resumimos el movimiento de un sistema en t\'erminos de las variables
de acci\'on-\'angulo en la siguiente teorema:

\noindent
{\bf Teorema 8.4.}

\noindent
Consideremos un sistema ciclico separable de $f$ grados
de libertad cuyo estado 
din\'amico est\'a dado por las variables $q, p \equiv q_{1}, q_{2},...,q_{f},
p_{1}, p_{2},..., p_{f}$, y cuyo comportamiento din\'amico est\'a dado por
el Hamiltoniano $H(q,p)$. Si nosotros
transformamos a las variables de acci\'on-\'angulo $J, w$, entonces
el Hamiltoniano $H$ es una funci\'on de $J$ solamente, es decir,
$$H=H(J)$$
y el movimiento del sistema est\'a dado por 
\begin{eqnarray*}
J_{i} &=& \gamma_{i} \\
w_{i} &=& \nu_{i}t+\phi_{i}~,
\end{eqnarray*}
donde $\gamma_{i}$ y $\phi_{i}$ son constantes que est\'an determinadas por
las condiciones 
iniciales, y las $\nu_{i}$ son constantes, llamadas
{\it las frecuencias del sistema} y se encuentran definidas como
\begin{eqnarray*}
\nu_{i} = \left[ \frac{\partial H(J)}{\partial J_{i}}\right]_{J=\gamma_{i}}~.
\end{eqnarray*}
\begin{center}
\section*{8.5 Importancia de las variables acci\'on-\'angulo}
\end{center}
La importancia de las variables acci\'on-\'angulo radica en que proporcionan
una t\'ecnica potente para la obtenci\'on de la frecuencia de un movimiento
peri\'odico de un movimiento sin hallar una soluci\'on completa del movimiento
del sistema.

\noindent
Lo anterior puede ser visto del siguiente argumento:
Consideremos el cambio de $w$ cuando $q$ describe un ciclo completo, dado por
\begin{eqnarray*}
\Delta w= \oint \frac{\partial w}{\partial q}dq
\end{eqnarray*}
Por otro lado, sabemos que
\begin{eqnarray*}
w=\frac{\partial W}{\partial J}~,
\end{eqnarray*}
por lo que
\begin{eqnarray*}
\Delta w &=& \oint \frac{\partial^{2}W}{\partial q\partial J}dq\\
         &=& \frac{d}{dJ}\oint \frac{\partial W}{\partial q}dq \\
         &=& \frac{d}{dJ}\oint pdq\\
         &=& 1~.
\end{eqnarray*}
El anterior resultado nos indica que $w$ cambia en una unidad cuando $q$ varia
a lo largo de un per\'{\i}odo completo.

\noindent
De la relaci\'on
\begin{eqnarray*}
w = \nu t + \phi~,
\end{eqnarray*}
concluimos que en un per\'{\i}odo $\tau$
\begin{eqnarray*}
\Delta w &=& 1 \\
         &=& \nu \tau~,
\end{eqnarray*}
es decir, podemos identificar a la constante $\nu$ con el rec\'{\i}proco del
per\'{\i}odo,
\begin{eqnarray*}
\nu = \frac{1}{\tau}~.
\end{eqnarray*}
\section*{8.6 Ejemplo: El oscilador arm\'onico simple}
A partir del formalisto de las variables acci\'on \'angulo vamos a demostrar
que la frecuencia $\nu$ del oscilador arm\'onico simple unidimensional est\'a
dada por $\nu = \sqrt{k/m}/2\pi$.

\noindent
Dado que $H$ es una constante de movimiento, la \'orbita en el
espacio $q-p$ est\'a dada por
\begin{eqnarray*}
\frac{p^2}{2m} + \frac{kq^2}{2} = E~,
\end{eqnarray*}
donde $E$ es la energ\'{\i}a. Esta es la ecuaci\'on de una elipse con
semiejes $\sqrt{2mE}$ y $\sqrt{2E/k}$.  El \'area encerrada por la elipse es
igual al valor de la variable de acci\'on $J$. Por tanto,
\begin{displaymath}
J=\pi \sqrt{2mE}\sqrt{\frac{2E}{k}}= 2\pi \sqrt{\frac{m}{k}}E~.
\end{displaymath}
Se sigue entonces que
\begin{displaymath}
H(J)=E=\frac{\sqrt{k/m}}{2\pi}J
\end{displaymath}
y la frecuencia est\'a dada por
\begin{displaymath}
\nu= \frac{\partial H(J)}{\partial J}= \frac{\sqrt{k/m}}{2\pi}~.
\end{displaymath}



\newpage
\oddsidemargin  0in
\evensidemargin 0in
\marginparwidth 1.2in
\marginparsep   0.1in
\marginparpush  5pt
\topmargin  0in
\headheight   0pt
\headsep    0in
\topskip=10pt
\footheight  12pt
\footskip   0.6in
\textheight   8.5in
\textwidth  6.5in
\columnsep 10pt
\columnseprule 0pt
\renewcommand{\baselinestretch}{1.2}
\def\pa{\\$\;\;\;$\\}

\def\t{\'}
\def\ti{\'{\i}}
\def\n{\~n}



\newpage
$\;$\\
\vspace{0.3in}

\begin{center}
{\Large  9. TEOR\'IA CAN\'ONICA DE PERTURBACIONES}
\end{center}

\vspace{0.4in}


\noindent
{\bf Pr\'ologo}:
Existen muchos problemas en la naturaleza que no pueden ser resueltos
de manera exacta. Por esta raz\'on,
y tomando en cuenta el gran desarrollo experimentado por la inform\'atica,
se ha puesto mucho inter\'es 
en el desarrollo de m\'etodos para hallar soluciones aproximadas. El
m\'etodo de perturbaciones 
se aplica cuando se tiene un problema f\'{\i}sico que no se puede resolver
exactamente, pero cuya 
Hamiltoniana difiere s\'olo ligeramente de la Hamiltoniana correspondiente a
un problema que puede 
resolverse de manera exacta. A la diferencia entre ambas Hamiltonianas se
le conoce como la {\it 
Hamiltoniana de la perturbaci\'on} y la teor\'{\i}a de perturbaciones est\'a
basada en la pequ\~nez de la misma.

$\;\;$\\

\bigskip
\bigskip

{\bf CONTENIDO:}

\bigskip

9.1 Teor\'{\i}a de perturbaciones dependiente del tiempo (con dos ejemplos)

\bigskip

9.2 Teor\'{\i}a de perturbaciones independiente del tiempo (con un ejemplo)

\newpage
$\;\;$\\
{\Large \bf 9.1 Teor\'{\i}a de pertubaciones dependiente del tiempo}

$\;\;$\\ 
La formulaci\'on de la mec\'anica cl\'asica que simplifica m\'as el
desarrollo de la teor\'{\i}a de 
perturbaciones es la de Hamilton-Jacobi.  Consideremos entonces que $H_0 (p,
q,t)$ es la Hamiltoniana 
correspondiente al problema soluble o no perturbado, y que se ha solucionado
mediante la funci\'on 
principal de Hamilton $S(q, \alpha _0 , t)$, la cual genera una
transformaci\'on can\'onica, de las 
coordenadas $(p,q)$ a 
$(\alpha _0 , \beta _0)$, en la que la nueva
Hamiltoniana (o Kamiltoniana) $K_0 $ del sistema no perturbado, es nula. En 
s\'{\i}mbolos:
\setcounter{equation} {0}\\
\begin{eqnarray}
\frac{\partial S}{\partial t} + H_0(\frac{\partial S}{\partial q},q,t)
= K_0 = 0 \; .
\end{eqnarray}

$\;$\\
\'Esta es  la ecuaci\'on de Hamilton-Jacobi y hemos usado que $p
= \partial S / \partial q$. Las coordenadas can\'onicas transformadas
$(\alpha _0 , \beta _0 )$ son entonces todas constantes en el caso no
perturbado, ya que $K_0 = 0$ y:

\begin{eqnarray}
\dot{\alpha} _0 &=& - \frac{\partial K_0}{\partial \beta _0 }, \nonumber \\ 
\dot{\beta} _0 &=& \;\; \frac{\partial K_0}{\partial \alpha _0 }. 
\end{eqnarray}  

$\;$\\
Consideremos ahora la hamiltoniana del sistema perturbado como:

\begin{eqnarray}
H(q,p,t) = H_0 (q,p,t) + \Delta H(q,p,t) ;  \;\;\;\;\;\;\;\;  (\Delta H \ll
H_0).
\end{eqnarray}

$\;$\\
Aunque $(\alpha _0 , \beta _0 )$ siguen siendo coordenadas can\'onicas (pues
la transformaci\'on generada por $S$ 
es independiente de la forma particular de la hamiltoniana),
ya no ser\'an constantes y la  
Kamiltoniana del sistema perturbado ($K$) no ser\'a nula.  Para no olvidar
que en el sistema perturbado las coordenadas transformadas ya no son constantes, las
denotaremos como $\alpha$ y $\beta$, en vez de $\alpha _0$ y $\beta _0$,
que ser\'{\i}an sus correspondientes valores constantes
en el sistema no perturbado. En el sistema perturbado tenemos pues que:

\begin{eqnarray}
K(\alpha, \beta, t) = H + \frac{\partial S}{\partial t} = (H_0 + \frac{\partial S}{\partial t}) + 
\Delta H = \Delta H(\alpha, \beta, t).
\end{eqnarray}

$\;$\\
Las ecuaciones de movimiento que satisfacen las variables transformadas del
sistema pertubado son entonces:

\begin{eqnarray}
\dot{\alpha_i} &=& - \frac{\partial \Delta H(\alpha,\beta,t)}{\partial \beta_i}
\nonumber \\ 
\dot{\beta_i} &=& \;\; \frac{\partial \Delta H(\alpha, \beta, t)}{\partial
\alpha_i},
\end{eqnarray}
 
$\;$\\
donde $i = 1, 2, ... , n$ y $n$ es el n\'umero de grados de libertad del
sistema.  Las ecuaciones anteriores 
son rigurosas.  Si del sistema de $2n$ ecuaciones 
se pudieran obtener $\alpha_i$ y $\beta_i$ como funci\'on del tiempo, las
ecuaciones de transformaci\'on 
$(p,q) \rightarrow (\alpha,\beta)$ dar\'{\i}an $p _i$ y $q _i$ en funci\'on
del tiempo y el problema 
estar\'{\i}a resulto.  Sin embargo, la soluci\'on exacta de las ecuaciones
en $(5)$ no suele ser menos 
dif\'{\i}cil que la soluci\'on de las ecuaciones originales.
De $(5)$ vemos que a\'un cuando $\alpha$ y $\beta$ ya no son 
constantes, su variaci\'on con el tiempo es lenta, si suponemos
que $\Delta H$ cambia
poco respecto a $\alpha$ y $\beta$. Una primera
aproximaci\'on a la variaci\'on temporal de 
$(\alpha, \beta)$ se obtiene sustituyendo en los segundos
miembros de $(5)$ a $\alpha$ y $\beta$ por sus 
valores constantes o no perturbados; es decir:

\begin{eqnarray}
\dot{\alpha} _{i1} &=& - \left. \frac{\partial \Delta H(\alpha,
\beta, t)}{\partial \beta _i} \right 
\vert _0 \nonumber \\
\dot{\beta} _{i1} &=& \;\;\; \left. \frac{\partial
\Delta H(\alpha, \beta, t)}{\partial \alpha _i} \right 
\vert _0,
\end{eqnarray}

$\;$\\
donde $\alpha_{i1}$ y $\beta_{i1}$ representan las soluciones a primer
orden de perturbaci\'on para 
$\alpha_i$ y $\beta_i$, y las barras verticales con sub\'{\i}ndece
cero indican que despu\'es de la 
derivaci\'on deben sustituirse $\alpha$ y $\beta$ por sus valores  
constantes no perturbados. Hecho esto, las ecuaciones en $(6)$ se pueden
integrar para dar las $\alpha_i$ 
y las $\beta_i$ en funci\'on del tiempo (a primer orden).
Luego, mendiante las ecuaciones de transformaci\'on se obtienen 
$p$ y $q$ como funci\'on del tiempo en una primera aproximaci\'on.
La aproximaci\'on de segundo 
orden se obtiene sustituyendo en los segundos miembros de $(6)$ la
primera aproximaci\'on de la 
dependencia de $\alpha$ y $\beta$ con respecto al tiempo. En general,
la soluci\'on de perturbaci\'on de 
orden $N$ se obtiene integrando las ecuaciones:

\begin{eqnarray}
\dot{\alpha} _{iN} &=& - \left. \frac{\partial \Delta
H(\alpha, \beta, t)}{\partial \beta _i} \right 
\vert _{N-1} \nonumber \\
\dot{\beta} _{iN} &=& \;\;\; \left. \frac{\partial \Delta
H(\alpha, \beta, t)}{\partial \alpha _i} \right 
\vert _{N-1}~.
\end{eqnarray}

$\;$\\
\underline{{\bf Ejemplo 1}}

$\;$\\
Consideraremos aqu\'{\i} el caso simple de una part\'{\i}cula libre,
que m\'as adelante sujetaremos  
a una perturbaci\'on arm\'onica simple. Este ejemplo, aunque trivial,
servir\'a para ilustrar el 
procedimiento delineado antes. La Hamiltoniana no pertubada es:

\begin{eqnarray}
H_0 = \frac{p^2}{2m}.
\end{eqnarray}

$\;$\\
Puesto que $H_0 \neq H_0 (x)$, es decir, como $x$ es c\'{\i}clica, $p=
\alpha _0$ es una constante en el sistema no perturbado. 
Recordando que 
$p = \partial S / \partial x$ y sustituyendo en $(1)$:

\begin{eqnarray}
\frac{1}{2m} \left( \frac{\partial S}{\partial x} \right)^2
+ \frac{\partial S}{\partial t} = 0.
\end{eqnarray}

$\;$\\
Ahora bien, dado que el sistema es conservativo, resulta conveniente
considerar una funci\'on principal de la forma: 

\begin{eqnarray}
S = {\cal \bf S} (x) + F(t). 
\end{eqnarray}

$\;$\\
Esta clase de separaci\'on de variables, es especialmente \'util cuando la
Hamiltoniana no depende expl\'{\i}citamente del tiempo, en donde se propone
que $F(t) = -Et$, con $E$ como
la energ\'{\i}a total del sistema \footnote{V\'ease: Spiegel, Murrary R.
{\it Mec\'anica Te\'orica}, pp. 315, 316.}. Con $(10)$ en $(9)$ tenemos que:

\begin{equation} \label{1}
\frac{1}{2m} \left( \frac{d {\cal \bf S}}{dx} \right)^2 = E,\qquad \qquad
{\cal \bf S} = \sqrt{2mE} x = \alpha _0 x.
\end{equation}

$\;$\\
Y sustituyendo $(11)$ en $(10)$, junto con el hecho de que en este caso
la Hamiltoniana es igual a la
energ\'{\i}a, tenemos que la funci\'on principal de Hamilton es:

\begin{eqnarray}
S = \alpha _0 x - \frac{\alpha _0 ^2 t}{2m}.
\end{eqnarray}

$\;$\\
Si el momentum transformado es $\alpha_0$, la coordenada
transformada (que tambi\'en es constante en el 
sistema no perturbado) es:

\begin{eqnarray}
\beta _0 = \frac{\partial S}{\partial \alpha _0} = x - \frac{\alpha _0 t}{m}~,
\nonumber
\end{eqnarray}

$\;$\\
de modo que la transformaci\'on generada por $S$ est\'a dada por las
ecuaciones:

\begin{eqnarray}
p &=& \alpha _0, \nonumber \\
x &=& \frac{\alpha _0 t}{m} + \beta _0,
\end{eqnarray}

$\;$\\
que es la soluci\'on esperada para el movimiento de una part\'{\i}cula libre.
Lo realizado hasta 
aqu\'{\i} 
s\'olo muestra el procedimiento para hallar las ecuaciones del
movimiento mediante la formulaci\'on de 
Hamilton-Jacobi.  Es hasta ahora que introduciremos una perturbaci\'on de la
forma: 

\begin{eqnarray}
\Delta H = \frac{kx^2}{2} = \frac{m \omega ^2 x^2}{2},
\end{eqnarray}

$\;$\\
o bien, en t\'erminos de las coordenadas transformadas, utilizando $(13)$:

\begin{eqnarray}
\Delta H = \frac{m \omega^2}{2} \left( \frac{\alpha t}{m} + \beta \right) ^2.
\end{eqnarray}

$\;$\\
N\'otese que en la expresi\'on anterior hemos suprimido ya
los sub\'{\i}ndices $0$ de las coordenadas
transformadas, pues estamos ya ocup\'andonos del sistema perturbado.

$\;$\\
Sustituyendo $(15)$ en $(5)$:

\begin{eqnarray}
\dot{\alpha} &=& - m \omega^2 \left( \frac{\alpha t}{m} + \beta \right),
\nonumber \\
\dot{\beta} &=& \;\; \omega^2 t \left( \frac{\alpha t}{m} + \beta \right).
\end{eqnarray}

$\;$\\
Las ecuaciones anteriores tienen una soluci\'on exacta y es de
forma arm\'onica, como cabe esperar. Para
asegurarnos s\'olo derivamos respecto al tiempo la primera de las ecuaciones
y llegaremos a que $\alpha$ 
tiene una variaci\'on arm\'onica siemple, lo que tambi\'en puede
asegurarse de $x$ en virtud de las ecuaciones de transformaci\'on
dadas en $(13)$, que siguen manteniendo su forma en el sistema perturbado
(omitiendo, claro, los sub\'{\i}ndices de las coordenadas transformadas).  
Sin embargo, nos interesa ilustrar el m\'etodo de perturbaciones, as\'{\i}
que consideremos que $k$ (la 
constante el\'astica) es un par\'ametro peque\~no y busquemos soluciones
aproximadas de distintos \'ordenes 
de perturbaci\'on, sin perder de vista que las variables
transformadas $(\alpha, \beta)$ en el sistema 
perturbado dejar\'an de ser constantes.  Dicho de otra manera, a\'un
cuando $(\alpha, \beta)$ contienen 
informaci\'on referente a los par\'ametros del sistema no perturbado, el efecto de la perturbaci\'on es 
hacer variar estos par\'ametros con el tiempo.

$\;$\\
La perturbaci\'on de primer orden se obtiene seg\'un est\'a indicado
de manera general en $(6)$. 
As\'{\i} que debemos sustituir en los segundos miembros de $(16)$, $\alpha$ y 
$\beta$ por sus valores no perturbados.  Para simplificar 
consideremos que $x(t=0)=0$ y por tanto que $\beta_0 = 0$, entonces:

\begin{eqnarray}
\dot{\alpha}_1 &=& - \omega^2 \alpha_0 t, \nonumber \\ 
\dot{\beta}_1 &=& \alpha_0 \frac{\omega^2 t^2}{m}, 
\end{eqnarray}

$\;$\\
que luego de integrar nos conduce a:

\begin{eqnarray}
\alpha_1 &=& \alpha_0 - \frac{\omega^2 \alpha_0 t^2}{2}, \nonumber \\ 
\beta_1 &=& \frac{\alpha_0 \omega^2 t^3}{3m}.
\end{eqnarray}

$\;$\\
Las soluciones para $x$ y $p$ a primer orden las obtenemos
sustituyendo $\alpha_1$ y $\beta_1$ en las 
ecuaciones de transformaci\'on $(13)$, de donde:

\begin{eqnarray}
x &=& \frac{\alpha_0}{m \omega} \left ( \omega t - \frac{\omega^3 t^3}{6}
\right), \nonumber \\
p &=& \alpha_0 \left( 1 - \frac{\omega^2 t^2}{2} \right).
\end{eqnarray}

$\;$\\
Para generar la soluci\'on aproximada a un segundo orden de perturbaci\'on
debemos hallar 
$\dot{\alpha_2}$ y $\dot{\beta_2}$, seg\'un se indic\'o en $(7)$,
sustituyendo en los 
segundos miembros de $(16)$, $\alpha$ y $\beta$ por $\alpha_1$ y $\beta_1$ como fueron dadas en $(18)$. 
Integrando $\dot{\alpha_2}$ y $\dot{\beta_2}$ y luego utilizando de nuevo las
ecuaciones de 
transformaci\'on $(13)$, llegamos a las soluciones de segundo orden
para $x$ y $p$:

\begin{eqnarray}
x &=& \frac{\alpha_0}{m \omega} \left( \omega t - \frac{\omega^3 t^3}{3!}
+ \frac{\omega^5 t^5}{5!} 
\right), \nonumber \\
p &=& \alpha_0 \left( 1 - \frac{\omega^2 t^2}{2!}
+ \frac{\omega^4 t^4}{4!} \right).
\end{eqnarray}

$\;$\\
En el l\'{\i}mite en que el orden de perturbaci\'on $N$ tiende a infinito,
obtenemos las soluciones 
esperadas y compatibles con las condiciones iniciales:

\begin{eqnarray}
x  \rightarrow  \frac{\alpha_0}{m \omega} \sin{\omega t}, & \;\;\;\; p
\rightarrow \alpha_0 \cos{\omega t}.
\end{eqnarray}

$\;$\\
Las variables transformadas $(\alpha, \beta)$ contienen informaci\'on
referente a los par\'ametros de la 
\'orbita sin perturbar.  Por ejemplo, si consideramos como sistema no
perturbado aqu\'el correspondiente 
al problema de Kepler, un sistema adecuado de coordnadas $(\alpha, \beta)$
podr\'{\i}an ser las 
variables $(J, \delta)$ que son respectivamente la variable acci\'on y
el \'angulo de 
fase que aparece en en la variable \'angulo $w$ (recordar que $w = \nu
t + \delta$, donde $\nu$ es la 
frecuencia).  Estas variables est\'an relacionadas con los p\'arametros
orbitales tales como el 
semieje mayor, la excentricidad, la inclinaci\'on, etc. 

\noindent
\underline{El
efecto de perturbaci\'on es hacer 
variar estos par\'ametros con el tiempo}.  Si la perturbaci\'on es
peque\~na, la variaci\'on de 
los par\'ametros durante un per\'{\i}odo del movimiento no perturbado
tambi\'en ser\'a peque\~na. 
De modo que por cortos intervalos de tiempo el sistema se mover\'a a lo largo
de una \'orbita, llamada 
{\it \'orbita osculatriz}, que es de 
la misma forma funcional que la del sistema no perturbado; sin embargo,
los par\'ametros de esta \'orbita 
var\'{\i}an con el tiempo.

$\;$\\
Los par\'ametros de la \'orbita osculatriz pueden variar con el tiempo
de dos maneras:

\begin{itemize}

\item Variaci\'on peri\'odica:  el par\'ametro vuelve a su valor inicial
despu\'es de un intervalo de 
tiempo que en primera aproximaci\'on suele ser el per\'{\i}odo del
movimiento no perturbado.  Estos 
efectos de la perturbaci\'on no alteran los valores medios de
los par\'ametros y por tanto la trayectoria 
sigue siendo muy parecida a la \'orbita no perturbada.
Estos efectos se pueden eliminar promediando la 
perturbaci\'on sobre un per\'{\i}odo del movimiento no perturbado.\\

\item Variaci\'on secular:  Al final de cada uno de los per\'{\i}odos
orbitales sucesivos hay un incremento 
neto del valor del par\'ametro.  Al cabo de muchos per\'{\i}odos los
par\'ametros orbitales pueden ser 
muy diferentes de sus valores no perturbados.  Rara vez interesa el valor
instant\'aneo de la variaci\'on de alg\'un par\'ametro, digamos la
frecuencia, porque su variaci\'on es muy peque\~na en
la mayor\'{\i}a de casos en que funciona el formalismo de la teor\'{\i}a de las
perturbaciones. (Esta variaci\'on es tan peque\~na que resulta dif\'{\i}cil,
sino imposible, percibirla en uno solo per\'{\i}odo orbital, y por eso
se mide s\'olo la variaci\'on secular despu\'es de varios per\'{\i}odos.)\\

\end{itemize}

$\;$\\
\underline{{\bf Ejemplo 2}}

$\;$\\
Del problema de dos cuerpos tenemos que si al potencial
de Kepler se le suma un potencial de la
forma $1/r^2$, la \'orbita del problema acotado es una elipse que gira y cuyo
peri\'apsis est\'a animado de precesi\'on. Encontraremos aqu\'{\i} la velocidad
de precesi\'on, considerando un potencial perturbador algo m\'as general:

\begin{eqnarray}
V = - \frac{k}{r} - \frac{h}{r^n} \; ,
\end{eqnarray}

$\;$\\
donde $n (\geq 2)$ es un entero, y $h$ es tal que el segundo t\'ermino
del potencial sea una peque\~na perturbaci\'on
del primero. La Hamiltoniana de la perturbaci\'on ser\'a pues: 

\begin{eqnarray}
\Delta H = - \frac{h}{r^n} \; .
\end{eqnarray}

$\;$\\
En el problema sin perturbar, la posici\'on angular del
peri\'apsis en el plano de la \'orbita
viene dada por la constante $\omega = 2 \pi w_2$. En el caso perturbado:

\begin{eqnarray}
\dot{\omega} = 2 \pi \frac{\partial \Delta H}{\partial J_2}
= \frac{\partial \Delta H}{\partial l} \; , 
\end{eqnarray}

$\;$\\
donde hemos usado $J_2 = 2 \pi l$. Adem\'as $J_2$ y $w_2$ son
dos de las cinco constantes del movimiento a las que lleva el tratamiento
del problema de Kepler mediante las variables de acci\'on \'angulo.

$\;$\\
Necesitamos conocer el promedio de $\dot{\omega}$ en un per\'{\i}odo de
la \'orbita no perturbada $\tau$:

\begin{eqnarray}
\langle \dot{\omega} \rangle \equiv \frac{1}{\tau} \int _0 ^{\tau}
\frac{\partial \Delta H}{\partial l} dt =
\frac{\partial }{\partial l} \left( \frac{1}{\tau} \int _0 ^{\tau}
\Delta H \; dt \right)
= \frac{\partial \langle \Delta H \rangle}{\partial l} \; .
\end{eqnarray}

$\;$\\
Pero el promedio temporal de la Hamiltoniana no perturbada es:

\begin{eqnarray}
\langle \Delta H \rangle = - h \langle \frac{1}{r^n} \rangle =
- \frac{h}{\tau} \int _0 ^{\tau} \frac{dt}{r^n} \;.
\end{eqnarray}

$\;$\\
Por otro lado, sabemos que $l = mr^2 (d\theta / dt)$, de donde
podemos despejar $dt$ y sustituirlo en $(27)$, con lo que 

\begin{eqnarray}
\langle \Delta H \rangle &=& - \frac{mh}{l\tau} \int _0 ^{2 \pi}
\frac{d\theta}{r ^{n-2}} \nonumber \\
&=& - \frac{mh}{l\tau} \left( \frac{mk}{l^2} \right) ^{n-2} \int _0 ^{2 \pi}
[1 + e \cos{(\theta - \eta)} ] ^{n-2} d \theta \; .
\end{eqnarray}

$\;$\\
donde $\eta$ es una fase constante, $e$ es la excentricidad,  y donde se ha
expresado $r$ en funci\'on de $\theta$ haciendo uso de la ecuaci\'on general
de la \'orbita (con el origen en un foco de la c\'onica correspondiente):

\begin{eqnarray}
\frac{1}{r} = \frac{mk}{l^2} [1 + e \cos{(\theta - \eta)} ]
\end{eqnarray}~.

$\;$\\
En el caso en que $n=2$:

\begin{eqnarray}
\langle \Delta H \rangle &=& - \frac{2 \pi mh}{l \tau} \; , \nonumber \\
\langle \dot{\omega} \rangle &=& \frac{2 \pi mh}{l^2 \tau} \; .
\end{eqnarray}

$\;$\\
En el caso en que $n=3$:

\begin{eqnarray}
\langle \Delta H \rangle &=& - \frac{2 \pi m^2 hk}{l^3 \tau} \; , \nonumber \\
\langle \dot{\omega} \rangle &=& \frac{6 \pi m^2 hk}{l^4 \tau} \; .
\end{eqnarray}
 
$\;$\\
Este \'ultimo caso, $n=3$, reviste una importancia especial, ya
que la teor\'{\i}a de la Relatividad General
predice una correcci\'on del movimiento newtoniano del
orden  de $r^{-3}$ precisamente. Tal
predicci\'on se someti\'o a prueba con el c\'elebre problema de
la precesi\'on de la \'orbita de Mercurio. Sustituyendo los
apropiados valores de per\'{\i}odo,
masa, semieje mayor  de la \'orbita (que va incluido en $h$), etc.,
la ecuaci\'on $(30)$ predice una velocidad media de precesi\'on:

$$ \langle \dot{\omega} \rangle = 42.98 \;\; {\rm arcsegundos/siglo} \;. $$

$\;$\\
El valor medido es mucho mayor que el mencionado arriba
(por un factor mayor que 100). Pero antes
de hacer cualquier comparaci\'on deben eliminarse del valor medido,
las contribuciones
debidas a: {\it a)} el efecto conocido como la precesi\'on
de los equinoccios (movimiento del punto de
referencia de longitudes respecto a la galaxia), {\it b)} las perturbaciones
de la \'orbita de Mercurio debido a la interacci\'on con los
otros planetas.  Una vez eliminados estos efectos (de los cuales el
primero es el de 
mayor peso), se debe obtener lo que ser\'{\i}a la contribuci\'on  al valor
medido de $\dot{\omega}$, debido al efecto relativista.  En 1973 se calcul\'o
esta \'ultima contribuci\'on en $(41.4 \pm 0.9)$ ${\rm arcsegundos/siglo}$,
que es consistente con la predicci\'on que se obtiene de $(30)$.

$\;$\\
{\Large \bf 9.2 Teor\'{\i}a de perturbaciones independiente del tiempo}

$\;$\\
Mientras que la teor\'{\i}a de perturbaciones dependiente del tiempo
busca la dependecia temporal de los, 
en un principio constantes, par\'ametros del sistema no perturbado,
la teor\'{\i}a independiente del 
tiempo pretende hallar las cantidades que son constantes en el sistema
pertubado.  Esta teor\'{\i}a se 
aplica s\'olo a sistemas conservativos y peri\'odicos (tanto en el estado 
perturbado, como en el no perturbado). Por ejemplo, se aplica en
el caso de movimientos planetarios 
cuando se introduce cualquier perturbaci\'on conservativa al
problema de Kepler (m\'etodo de von Zeipel o de Poincar\'e).

$\;$\\
Consideraremos aqu\'{\i} el caso de sistemas de un solo grado de libertad.
Consideremos un sistema peri\'odico con una Hamiltoniana indepentiente 
del tiempo de la forma:

\begin{eqnarray}
H = H(p,q, \lambda), 
\end{eqnarray}

$\;$\\
donde $\lambda$ es una constante que especifica la magnitud de la
perturbaci\'on y se supone suficientemente peque\~na.  Asumimos que

\begin{eqnarray}
H_0 (p,q) = H(p,q,0)
\end{eqnarray}

$\;$\\
corresponde a un sistema que puede resolverse exactamente (sistema
no perturbado), por medio del uso de las
convenientes variables acci\'on-\'angulo  $(J_0 , w_0 )$:  

\begin{eqnarray}
H_0 (p,q) &=& K_0 (J_0) \nonumber \\
\nu_0 &=& \dot{w}_0 = \frac{\partial K_0}{\partial J_0} \; ;  \;\;\;\;\;
(w_0 = \nu_0 t + \delta_0 ). 
\end{eqnarray}

$\;$\\
La transformaci\'on can\'onica que nos lleva de $(p,q)$ a $(J_0 , w_0 )$ es
independiente de la forma 
particular de la Hamiltoniana.  Entonces, la Hamiltoniana perturbada
$H(p,q,\lambda)$ puede escribirse 
como $H(J_0, w_0, \lambda)$.  Debido a que la Hamiltoniana perturbada
s\'{\i} depende de $w_0$, $J_0$ ya 
no es constante. Por otro lado, en principio, uno puede obtener nuevas
variables acci\'on-\'angulo $(J,w)$ apropiadas para el sistema perturbado,
tales que:

\begin{eqnarray}
H(p,q,\lambda) &=& E(J,\lambda) \nonumber \\
\nu &=& \dot{w} = \frac{\partial E}{\partial J} \\
\dot{J} &=& - \frac{\partial E}{\partial w} = 0 \; ;  \;\;\; (J= constante).
\nonumber 
\end{eqnarray}

$\;$\\
Puesto que la transformaci\'on que conecta $(p,q)$ a $(J_0, w_0)$ es
conocida, ahora debemos encontrar la 
transformaci\'on can\'onica $S$, que conecta $(J_0, w_0)$ a $(J,w)$.
Si asumimos que 
$\lambda$ es peque\~na, 
la transformaci\'on que buscamos no debe diferir mucho de la transformaci\'on
identidad. La expansi\'on de la funci\'on generadora que buscamos ser\'a
entonces:

\begin{eqnarray}
S = S(w_0, J, \lambda) = S_0 (w_0, J) + \lambda S_1 (w_0, J)
+ \lambda ^2 S_2 (w_0, J) + ...~.
\end{eqnarray}

$\;$\\
Para $\lambda = 0$ requerimos que $S$ sea la identidad; entonces hacemos:

\begin{eqnarray}
S_0 = w_0 J~.
\end{eqnarray}

$\;$\\
Las transformaciones can\'onicas generadas por $S$ son:

\begin{eqnarray}
w &=& \frac{\partial S}{\partial J} = w_0 + \lambda \frac{\partial
S_1}{\partial J} (w_0,J) + \lambda ^2 \frac{\partial S_2}
{\partial J} (w_0,J) + ...~. \nonumber \\
J_0 &=& \frac{\partial S}{\partial w_0} = J + \lambda \frac{\partial
S_1}{\partial w_0} (w_0,J) + \lambda ^2 \frac{\partial S_2}
{\partial w_0} (w_0,J) + ...~. 
\end{eqnarray}

$\;$\\
Debido a que $w_0$ es una variable \'angulo (del sistema
no perturbado), sabemos que 
$\Delta w_0 = 1$ sobre un ciclo.  Por otro lado, sabemos que las
transformaciones can\'onicas tienen la propiedad de
conservar el volumen en el espacio de fase. Por tanto, podemos escribir: 

\begin{eqnarray}
J = \oint pdq = \oint J_0 dw_0~.
\end{eqnarray}

$\;$\\
Integrando la segunda ecuaci\'on 
en $(37)$ sobre una \'orbita del sistema perturbado, tenemos:

\begin{eqnarray}
\oint J_0 dw_0 = \oint J dw_0 + \sum_{n=1} \lambda^n \oint \frac{\partial S_n}
{\partial w_0} dw_0 ,
\end{eqnarray}

$\;$\\
y sustituyendo $(39)$ en $(38)$:

\begin{eqnarray}
J = J \Delta w_0 + \sum_{n=1} \lambda^n \oint \frac{\partial S_n}
{\partial w_0} dw_0.
\end{eqnarray}

$\;$\\
En vista de que $\Delta w_0 = 1$, debe cumplirse que:

\begin{eqnarray}
\sum_{n=1} \lambda ^n \oint \frac{\partial S_n}{\partial w_0} dw_0 = 0, 
\end{eqnarray}   

$\;$\\
o bien que:

\begin{eqnarray}
\oint \frac{\partial S_n}{\partial w_0} dw_0 = 0.
\end{eqnarray}

$\;$\\
Adem\'as, la Hamiltoniana puede expanderse en $\lambda$ como funci\'on
de $w_0$ y $J_0$:

\begin{eqnarray}
H(w_0, J_0, \lambda) = K_0(J_0) + \lambda K_1(w_0,J_0)
+ \lambda^2 K_2 (w_0,J_0) + ... \; ,
\end{eqnarray}

$\;$\\
donde las $K_i$ son conocidas, ya que $H$ es una funci\'on
conocida de $w_0$ y $J_0$ para una 
$\lambda$ dada.  Por otro lado tenemos que:

\begin{eqnarray}
H(p,q, \lambda) &=& H(w_0, J_0, \lambda) \nonumber \\
&=& E(J,\lambda) \; ,
\end{eqnarray}

$\;$\\
que es la expresi\'on para la energ\'{\i}a in las nuevas coordenadas de
acci\'on \'angulo (donde $J$ ser\'a constante y $w$ 
ser\'a funci\'on lineal del tiempo).  

$\;$\\
Podemos tambi\'en expandir $E$ en potencias de $\lambda$:

\begin{eqnarray}
E(J, \lambda) = E_0 (J) + \lambda E_1 (J) + \lambda ^2 E_2 (J) + ... \;.
\end{eqnarray}

$\;$\\
En virtud de $(44)$ podemos igualar los coeficientes de las distintas
potencias de $\lambda$ en 
$(43)$ y $(45)$.  Sin embargo, estas dos expresiones para la energ\'{\i}a
dependen de dos diferentes conjuntos de variables.  Para
resolver esto expresaremos $H_0$ en t\'erminos de $J$, y esto se logra haciendo 
una expansi\'on de Taylor de $H (w_0,J_0,\lambda)$ respecto de $J_0$
alrededor de $J$:

\begin{eqnarray}
H(w_0,J_0,\lambda) = H(w_0, J, \lambda) + (J_0 - J) \frac{\partial H}
{\partial J}
+ \frac{(J_0  
- J)^2}{2} \frac{\partial ^2 H}{\partial J^2} + ... \; ,
\end{eqnarray}

$\;$\\
Las derivadas del desarrollo de Taylor son, hablando con propiedad,
derivadas respecto a $J_0$ calculadas
en $J_0 = J$, si bien podemos escribirlas sin p\'erdidas de rigor,
en la forma de
derivadas con respecto a $J$, una vez sustituida $J_0$ por $J$ en $H_0 (J_0)$.  
En la ecuaci\'on anterior, todo t\'ermino que contiene a $J_0$ debe
reescribirse en t\'erminos 
de $J$ haciendo uso de la transformaci\'on definida por $(37)$ que conecta
las coordenadas 
$(J_0, w_0)$ con $(J,w)$.  As\'{\i} que, de la segunda ecuaci\'on en $(37)$
obtenemos 
$(J_0 - J)$, que sustitu\'{\i}do en $(46)$ da:

\begin{eqnarray}
H(w_0,J_0,\lambda) = H(w_0, J, \lambda) + \frac{\partial H}
{\partial J} \left( \lambda 
\frac{\partial S_1}{\partial w_0} + \lambda ^2
\frac{\partial S_2}{\partial w_0} + ... \right)
+ \frac{1}{2} \frac{\partial ^2 H}{\partial J^2} \lambda ^2
\left( \frac{\partial S_1}{\partial w_0} \right) ^2 
+ O (\lambda ^3)  \; .
\end{eqnarray}

$\;$\\
Luego, podemos hacer uso de $(43)$ para expresar $H(w_0,J,\lambda) =
H(w_0, J_0, \lambda) 
\mid _{J_0 = J}$, y sustituir luego en la ecuaci\'on $(47)$ para obtener:

\begin{eqnarray}
H(w_0,J_0,\lambda) &=& K_0(J) + \lambda K_1(w_0,J) + \lambda ^2 K_2 (w_0,J)
+ ... \nonumber \\
&+&  \lambda \frac{\partial S_1}{\partial w_0}
\left( \frac{\partial K_0(J)}{\partial J}
+ \lambda \frac{\partial K_1(w_0,J)}{\partial J} + ... \right) \nonumber \\
&+&  \lambda ^2 \left[ \frac{\partial K_0(J)}{\partial J} \frac{\partial S_2}
{\partial w_0} + \frac{1}{2} \frac{\partial ^2 K_0}{\partial J^2} \left(
\frac{\partial S_1}{\partial w_0} \right) ^2 + ... \right] \nonumber \\
&\equiv& E(J,\lambda) \\
&=& E_0 (J) + \lambda E_1 (J) + \lambda ^2 E_2(J) + ... \; . \nonumber
\end{eqnarray}

$\;$\\
Ahora, podemos resolver para los coeficientes $E_i(J)$; esto nos dar\'a la
posibilidad de calcular la 
frecuencia del movimiento perturbado a distintos \'ordenes de perturbaci\'on.   
Puesto que la expansi\'on de los t\'erminos de $E_i$ no involucra 
una dependencia en $w_0$, entonces la aparici\'on de $w_0$ en $(48)$ debe
ser espuria.  Las $K_i 
(w_0,J)$ de $(48)$ son funciones conocidas, mientras que las $S_i (w_0,J)$ y
las $E_i(J)$ son desconocidas.

$\;$\\
Igualando potencias de $\lambda$ tenemos:

\begin{eqnarray}
E_0 (J) &=& K_0 (J) \nonumber \\
E_1 (J) &=& K_1 (w_0,J) + \frac{\partial S_1}{\partial w_0} \frac{\partial
K_0 (J)}{\partial J} \nonumber \\
E_2 (J) &=& K_2 (w_0,J) + \frac{\partial S_1}{\partial w_0} \frac{\partial
K_1 (w_0,J)}{\partial J} \nonumber \\
&+& \frac{1}{2} \left( \frac{\partial S_1}{\partial w_0} \right) ^2
\frac{\partial ^2 K_0(J)}
{\partial J^2} + \frac{\partial S_2}{\partial w_0} \frac{\partial K_0 (J)}
{\partial J} \;~. 
\end{eqnarray}

\vspace{0.5in}

$\;$\\
Vemos que, para determinar $E_1$ necesitamos conocer $S_1$ adem\'as de $K_1$.
No debemos perder 
de vista que la $E_i$ son constantes, pues son s\'olo funciones de $J$ (que
es una constante del movimiento).  Tambi\'en debemos notar
que $\partial K_0 / \partial J$ es un t\'ermino
independiente de $w_0$ (pues $K_0 = K_0 (J)= K_0 (J_0) \mid _{J_0=J}$).  
Si promediamos sobre $w_0$ en ambos lados de la segunda ecuaci\'on en $(49)$,
obtenemos:

\begin{eqnarray}
E_1 &=& \langle E_1 \rangle \nonumber \\
&=& \langle K_1 \rangle + \frac{\partial K_0}{\partial J} \langle
\frac{\partial S_1}{\partial 
w_0} \rangle \; . 
\end{eqnarray}

$\;$\\
Pero hemos visto ya que $\langle \partial S_i / \partial w_0 \rangle
= \oint (\partial S_i / 
\partial w_0 ) dw_0 = 0$.  Luego,

\begin{eqnarray}
E_1 = \langle E_1 \rangle = \langle K_1 \rangle~. 
\end{eqnarray}

$\;$\\
Sustituyendo $(51)$ en el miembro izquierdo de la segunda ecuaci\'on de $(49)$
y despejando luego $(\partial S_1 / \partial w_0)$, tenemos que:

\begin{eqnarray}
\frac{\partial S_1}{\partial w_0} = \frac{\langle K_1 \rangle - K_1}{\nu _0
(J)} \; ,
\end{eqnarray}

$\;$\\
en donde hemos usado que $\nu _0 = \partial K_0 / \partial J$.

$\;$\\
La soluci\'on para $S_1$ se encuentra por integraci\'on directa.  En general,
vemos que el procedimiento para hallar $E_n$ es el que sigue (una vez asumido
que se ha resuelto ya para $E_{n-1}$):  

\begin{itemize}

\item 
Promediar en ambos lados de la $n$-\'esima ecuaci\'on de $(49)$.

\item 
Insertar el valor promediado de $E_n$ que se hall\'o, en la ecuaci\'on completa
que se daba en $(49)$ para $E_n$ (es decir, la que hab\'{\i}a antes de
promediar).

\item 
La \'unica expresi\'on desconocida que queda entonces,
ser\'a $S_n$ que podr\'a obtenerse integrando en:

$$\frac{\partial S_n}{\partial w_0} = \;\; {\rm funci\acute{o}n} \;\;
{\rm conocida} \;\; {\rm de} \;\;  w_0 \;\; {\rm y} \;\; J \; .$$

\item 
Sustituir $S_n$ en la ecuaci\'on completa para $E_n$.

\end{itemize}

$\;$\\
Una vez hecho esto, se puede continuar para el orden de perturbaci\'on $n+1$.

$\;$\\
Seg\'un podemos ver la determinaci\'on de la
energ\'{\i}a a un orden particular $n$ se determina s\'olo cuando se ha
hallado $S_{n-1}$, y $S_n$ puede determinarse s\'olo cuando se ha
hallado $E_n$.

$\;$\\
La teor\'{\i}a de perturbaciones independiente del tiempo, es muy similar al
esquema de perturbaciones de Rayleigh y Schr\"{o}dinger en la mec\'anica
ondulatoria.  
En la teor\'{\i}a ondulatoria, se conoce $E_n$ s\'olo si la funci\'on de onda
se conoce a un orden $n-1$. Y adem\'as, la funci\'on de onda de orden $n$ se
encuentra s\'olo cuando se ha calculado $E_n$. 





$\;$\\
\vspace{1.0in}

\begin{center}
{\Large \bf Bibliograf\'{\i}a}
\end{center}

\vspace{0.5in}

\begin{itemize}

\item
H. Goldstein, {\it Mec\'anica Cl\'asica},2a. ed. Versi\'on espa\~nola.  
Editorial Revert\'e S.A.,1992.\\

\item
R.A. Matzner and L.C. Shepley,  {\it Classical Mechanics},  Prentice-Hall, Inc.
U.S.A., 1991.\\

\item
R. Murray, {\it Mec\'anica Te\'orica}, Spiegel, 
Edici\'on en espa\~nol. McGraw-Hill S.A. de C.V.  M\'exico, 1976.\\

\item
L.D. Landau and E.M. Lifshitz, {\it Mechanics},  3rd. ed.  
Course of Theoretical Physics, Volume 1.  Pergamon Press, Ltd.  1976.\\

\end{itemize}


\newpage


\centerline{\Large 10. INVARIANTES ADIABATICOS}

\bigskip
\bigskip

\noindent
{\bf Pr\'ologo}:
Un invariante adiab\'atico es una funci\'on de los par\'ametros y de las
constantes del movimiento de un sistema, que permanece casi constante en el
l\'{\i}mite en el que los p\'arametros cambian infinitamente despacio en el
tiempo, aunque ellos puedan en \'ultima instancia cambiar por grandes
cantidades.

\bigskip
\bigskip

{\bf CONTENIDO}:

\bigskip

10.1 BREVE HISTORIA

\bigskip

10.1 GENERALIDADES

\newpage

{\bf 10.1 BREVE HISTORIA}\\

\noindent
La noci\'on de invarianza adiab\'atica se remonta a los primeros a\~nos
de la teor\'{\i}a
cu\'antica. Alrededor de 1910, quienes estudiaban
la emisi\'on y absorci\'on de radiaci\'on, notaron que los \'atomos
pod\'{\i}an existir en estados estables en los que su energ\'{\i}a
estaba fija. Einstein llam\'o la atenci\'on acerca de la cantidad
que permanec\'{\i}a casi invariante en un p\'endulo cuya longitud variaba
continua y lentamente (la cantidad $E/ \nu$).  \'El sugiri\'o
que podr\'{\i}an existir cantidades similares
asociadas con los sistemas at\'omicos, que determinaran la estabilidad de los
mismos cuando estas cantidades adquirieran ciertos valores. Tales
invariantes adiab\'aticas fueron luego encontradas por Ehrenfest
y su uso condujo a la primitiva teor\'{\i}a cu\'antica de Bohr y Sommerfeld.
Esta teor\'{\i}a trabajaba bien para los estados del hidr\'ogeno, pero
fracasaba cuando se le aplicaba a otros \'atomos.  M\'as tarde, en 1925-6,
surgi\'o una teor\'{\i}a cu\'antica altamente exitosa, debida a Schr\"odinger,
Heisenberg, Born y otros; esta teor\'{\i}a empleaba un enfoque diferente.

$\;$\\
\noindent
El tema de la invarianza adiab\'atica surgi\'o nuevamente despu\'es de
varias d\'ecadas, en el estudio de iones y electrones en movimiento
en el espacio.  Este tema era del inter\'es
de cient\'{\i}ficos escandinavos que estudiaban el fen\'omeno
de las auroras. Uno de ellos, H. Alfven,
mostr\'o en su libro {\it Electrodin\'amica C\'osmica}
que bajo condiciones apropiadas cierta combinaci\'on matem\'atica
de propiedades de los iones y electrones permanec\'{\i}an constantes a
primer orden. Aparentemente, Alfven no se percat\'o de que estaba
tratando con la misma clase de invariantes adiab\'aticas definidas
por Ehrenfest. Fueron L. Landau y E. Lifshitz quienes se\~nalaron la
relaci\'on que hab\'{\i}a con el tema de invariantes adiab\'aticas.

$\;$

{\bf 10.2 GENERALIDADES}\\

\noindent
Mostraremos primero, de una manera sencilla, cu\'al es la cantidad
invariante adiab\'atica en el caso del movimiento arm\'onico simple. El
m\'etodo a seguir consistir\'a en mostrar que, en el l\'{\i}mite de la
variaci\'on infinitamente lenta de los par\'ametros, el invariante
adiab\'atica del sistema unidimensional se aproxima a una cantidad que
se conserva exactamente en el correspondiente sistema bidimensional.\\

\noindent
Consideremos un sistema de un grado de libertad, inicialmente
conservativo y peri\'odico, que contenga un
par\'ametro $a$ que ser\'a inicialmente constante.
La variaci\'on lenta del par\'ametro no
alterar\'a la naturaleza peri\'odica del movimiento. Por una variaci\'on lenta
entendemos aqu\'ella en la que $a$ var\'{\i}a ligeramente durante un
per\'{\i}odo $\tau$ del movimiento:
\setcounter{equation}{0}\\
\begin{eqnarray}
\tau (da / dt) \ll a~.
\end{eqnarray}

$\;$\\
Pero, a\'un cuando las variaciones de $a$ sean peque\~nas en un
per\'{\i}odo cualquiera, al cabo de un tiempo suficientemente
largo las propiedades del movimiento pueden experimentar cambios grandes.

$\;$\\
Cuando el par\'ametro $a$ sea constante, el sistema vendr\'a descrito por
variables acci\'on \'angulo $(w_0, J_0)$ tales
que la Hamiltoniana ser\'a $H = H(J_0,a)$.  Supongamos
que la funci\'on generadora de la transformaci\'on $(q,p)
\rightarrow (w_0,J_0)$, es de la forma: $W^* (q, w_0,a)$.

$\;$\\
Cuando se deja variar $a$ con el tiempo, $(w_0, J_0)$ ser\'an a\'un
variables can\'onicas v\'alidas pero $W^*$ ser\'a funci\'on del tiempo a
trav\'es de $a$. Entonces $J_0$ no ser\'a constante y $w_0$ ya no ser\'a
m\'as funci\'on lineal del tiempo. La Hamiltoniana apropiada ser\'a ahora:

\begin{eqnarray}
K(w_0,J_0,a) &=& H(J_0,a) + \frac{\partial W^*}{\partial t} \nonumber \\
&=& H(J_0,a) + \dot a \frac{\partial W^*}{\partial a}~.
\end{eqnarray}

$\;$\\
El segundo miembro de la Hamiltoniana se puede ver como una perturbaci\'on y
la dependencia temporal de $J_0$ viene de:

\begin{eqnarray}
\dot{J_0} = - \frac{\partial K}{\partial w_0} = - \dot a \frac{\partial}
{\partial w_0} \left( \frac{\partial W^*}{\partial a} \right)~.
\end{eqnarray}

$\;$\\
Procediendo en una forma an\'aloga a como se procede en la teor\'{\i}a de
perturbaciones dependiente del tiempo,
buscamos la variaci\'on a primer orden en el valor medio de $\dot{J_0}$ a
lo largo del per\'{\i} odo del movimiento no perturbado. Como $a$ var\'{\i}a
lentamente, podemos
considerarla constante durante este intervalo y entonces podemos escribir:

\begin{eqnarray}
\langle \dot{J_0} \rangle = - \frac{1}{\tau} \int_{\tau} \dot{a}
\frac{\partial}{\partial w_0} \left( \frac{\partial W^*}{\partial a} \right) dt
= - \frac{\dot{a}}{\tau} \int_{\tau} \frac{\partial}{\partial w_0}
\left( \frac{\partial W^*}{\partial a} \right) dt + O(\dot{a}^2, \ddot{a}).
\end{eqnarray}

$\;$\\
Puede demostrarse que $W^*$ es una funci\'on peri\'odica en $w_0$, y por
consiguiente, tanto ella como su derivada respecto a $a$, pueden
escribirse como una serie de Fourier:

\begin{eqnarray}
\frac{\partial W^*}{\partial a} = \sum_k A_k e^{2 \pi ikw_0}.
\end{eqnarray}

$\;$\\
Sustituyendo $(81)$ en $(80)$:

\begin{eqnarray}
\langle \dot{J_0} \rangle = - \frac{\dot{a}}{\tau} \int_{\tau} \sum_k 2 \pi
ik A_k e^{2 \pi ikw_0} dt + O(\dot{a}^2, \ddot{a}).
\end{eqnarray}

$\;$\\
Como el integrando no tiene ning\'un t\'ermino constante, la
integral se anula. Luego,

\begin{eqnarray}
\langle \dot{J_0} \rangle = 0 + O(\dot{a}^2, \ddot{a}).
\end{eqnarray}

$\;$\\
Por tanto, $\langle \dot{J_0} \rangle$ no
tendr\'a variaci\'on secular de primer orden (o sea en $\dot{a}$), que
es una propiedad deseada de la invarianza adiab\'atica. De modo que el
t\'ermino {\it casi constante} en nuestra definici\'on de invariante
adiab\'atica, debe ser interpretada como {\it constante a primer orden}.

\vspace{0.1in}

\begin{center}
{\Large Bibliograf\'{\i}a complementaria}
\end{center}


\begin{itemize}





\item
L. Parker, {\it Adiabatic invariance in
simple harmonic motion}, Am. J. Phys. 39 (1971) pp. 24-27.\\

\item
A.E. Mayo, {\it Evidence for the adiabatic invariance of the black hole
horizon area}, Phys. Rev. D58 (1998) 104007 [gr-qc/9805047].

\end{itemize}


\newpage
\pagestyle{plain}

\begin{center}\section*{\Large 11. MECANICA DE SISTEMAS CONTINUOS}\end{center}

\bigskip

\noindent
{\bf Pr\'ologo}:
Todas las formulaciones de la mec\'{a}nica tratadas hasta ahora han estado
dirigidas al tratamiento de sistemas que tengan un n\'{u}mero de grados de
libertad finito o, como m\'{a}ximo, numerablemente infinito . Sin embargo,
existen ciertos problemas mec\'{a}nicos que entra\~{n}an sistemas continuos,
como por ejemplo el problema de un s\'{o}lido el\'{a}stico en vibraci\'{o}n.
En \'{e}l cada punto del s\'{o}lido continuo participa en las oscilaciones y
el movimiento total s\'{o}lo puede describirse especificando coordenadas de
posici\'{o}n de todos los puntos. No resulta dif\'{\i}cil modificar las
formulaciones anteriores de la mec\'{a}nica para poder tratar dichos
problemas. El m\'{e}todo m\'{a}s directo consiste en aproximar el sistema
continuo a uno que contenga part\'{\i}culas discretas y luego examinar como
cambian las ecuaciones que describen el movimiento, cuando nos aproximamos
al l\'{\i}mite continuo.

\bigskip
\bigskip

{\bf CONTENIDO}:

\bigskip
11.1 Formulaci\'on Lagrangiana: de discreto a continuo.

\bigskip
11.2 Formulaci\'on Lagrangiana para sistemas continuos.

\bigskip
11.3 Formulaci\'on Hamiltoniana, parentesis de Poisson.


\bigskip
11.4 Teorema de Noether.

\newpage

\section*{11.1 Formulaci\'{o}n Lagrangiana:
Transici\'{o}n de un sistema discreto a un sistema continuo.}

\noindent
Como uno de los casos mas sencillos en el que se puede pasar de un sistema
discreto a uno continuo consideremos el de una varilla el\'{a}stica
infinitamente larga que efect\'{u}a peque\~{n}as vibraciones longitudinales,
es decir, desplazamientos oscilantes de las part\'{\i}culas de la varilla
paralelos a su eje. Un sistema compuesto por part\'{\i}culas discretas que se
aproxime a la varilla continua es una cadena infinita de puntos materiales
iguales separados por distancias $a$ y unidos por resortes uniformes sin
masa de constante de rigides $k$ (ver la figura).

\vskip 1ex
\centerline{
\epsfxsize=220pt
\epsfbox{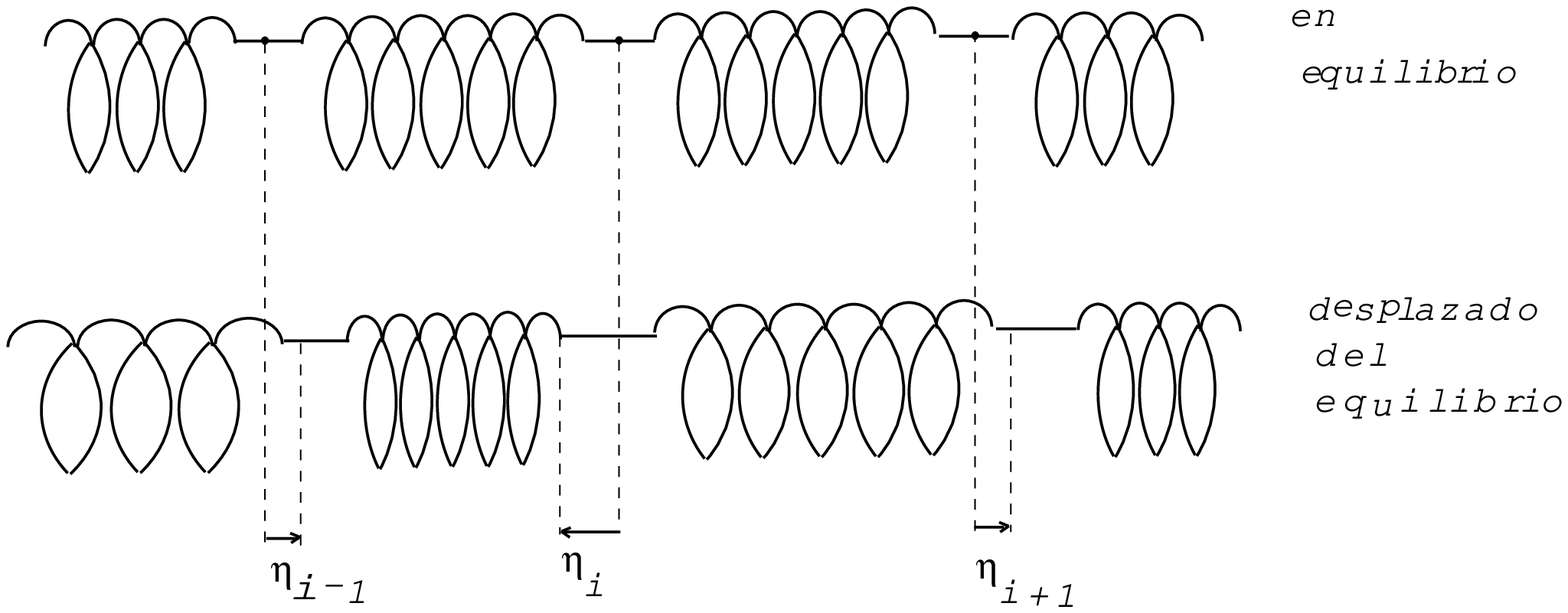}}
\vskip 2ex

\noindent
Supondremos que los puntos materiales s\'{o}lo pueden moverse a lo largo de
la direcci\'{o}n de la cadena . Podemos ver que el sistema discreto es una
extensi\'{o}n de la mol\'{e}cula poliat\'{o}mica lineal tratada en el
cap\'{\i}tulo 6 de Goldstein. Podremos, pues, obtener las ecuaciones que 
describen el
movimiento mediante las t\'{e}cnicas habituales para oscilaciones
peque\~{n}as. Representando por $\eta _i$ el desplazamiento de la
part\'{\i}cula $i$-\'{e}sima respecto a su posici\'{o}n de equilibrio,
la energ\'{\i}a cin\'{e}tica es
\setcounter{equation} {0}\\
\begin{equation}
T=\frac 12\sum_im\stackrel{.}{\eta }_i^2,  \label{1}
\end{equation}
donde $m$ es la masa de cada part\'{\i}cula. La energ\'{\i}a potencial
correspondiente es la suma de las energ\'{\i}as potenciales de cada resorte a
consecuencia de hallarse estirado o comprimido respecto a su longitud
natural:
\begin{equation}
V=\frac 12\sum_ik\left( \eta _{i+1}-\eta _i\right) ^2.  \label{2}
\end{equation}
De las ecuaciones (\ref{1}) y (\ref{2}) obtenemos la Lagrangiana del sistema
\begin{equation}
L=T-V=\frac 12\sum_i\left( m\stackrel{.}{\eta }_i^2-k\left( \eta _{i+1}-\eta
_i\right) ^2\right) ,  \label{3}
\end{equation}
que tambi\'{e}n puede escribirse de la forma 
\begin{equation}
L=\frac 12\sum_ia\left[ \frac ma\stackrel{.}{\eta }_i^2-ka\left( \frac{\eta
_{i+1}-\eta _i}a\right) ^2\right] =\sum_iaL_i,  \label{4}
\end{equation}
donde $a$ es la separaci\'{o}n de equilibrio entre puntos. Las ecuaciones de
movimiento de Lagrange para las coordenadas $\eta _i$ resultan ser 
\begin{equation}
\frac ma\stackrel{\cdot \cdot }{\eta }_i-ka\left( \frac{\eta _{i+1}-\eta _i}{%
a^2}\right) +ka\left( \frac{\eta _i-\eta _{i-1}}{a^2}\right) =0.  \label{5}
\end{equation}
La forma particular de $L$ de la ecuaci\'{o}n (\ref{4}) y las ecuaciones de
movimiento correspondientes las hemos elegido por ser comvenientes para
efectuar el paso al l\'{\i}mite a una varilla continua al tender a cero $a.$
Esta claro que $m/a$ se reduce a la masa por unidad de longitud $\mu $ del
sistema continuo, pero el valor l\'{\i}mite de $ka$ no resulta tan evidente.
Recordemos que en el caso de una varilla el\'{a}stica que cumpla la ley de
Hooke,el alrgamiento de la varilla por unidad de longitud es directamente
proporcional a la fuerza o tensi\'{o}n ejercida sobre ella, relaci\'{o}n que
podemos escribir en la forma 
\[
F=Y\xi , 
\]
donde $\xi $ es el alargamiento por unidad de longitud e $Y$ es el
m\'{o}dulo de Young. Ahora bien, el alargamiento de una longitud $a$ de un
sistema discreto, por unidad de longitud, ser\'{a} $\xi =\left( \eta
_{i+1}-\eta _i\right) /a.$ La fuerza necesaria para estirar el resorte esta
cantidad es 
\[
F=k\left( \eta _{i+1}-\eta _i\right) =ka\left( \frac{\eta _{i+1}-\eta _i}%
a\right) , 
\]
por lo que $\kappa a$ debe corresponder al m\'{o}dulo de Young de la varilla
continua. Al pasar del caso discreto al continuo, el indice entero $i$ que
identifica el punto material particular se convierte en la coordenada de
posici\'{o}n continua $x$; en vez de la variable $\eta _i$ tenemos $\eta
\left( x\right) .$ Adem\'{a}s, la cantidad 
\[
\frac{\eta _{i+1}-\eta _i}a=\frac{\eta \left( x+a\right) -\eta \left(
x\right) }a 
\]
que figura en $L_i$ tiende evidentemente al l\'{\i}mite 
\[
\frac{d\eta }{dx}, 
\]
cuando $a$ tiende a cero. Por \'{u}ltimo, la suma extendida a un n\'{u}mero
discreto de part\'{\i}culas se convierte en una integral extendida a $x,$ la
longitud de una varilla, y la Lagrangiana (\ref{4}) queda en la forma
\begin{equation}
L=\frac 12\int \left( \mu \stackrel{\cdot }{\eta }^2-Y\left( \frac{d\eta }{dx%
}\right) ^2\right) dx.  \label{6}
\end{equation}
En el l\'{\i}mite, cuando $a$ tiende a cero, los dos \'{u}ltimos t\'{e}rminos
de la ecuaci\'{o}n de movimiento (\ref{5}) resultan ser 
\[
{\rm Lim}_{a\rightarrow 0}
\;\frac{-Y}{a}\left\{ \left( \frac{d\eta }{dx}%
\right) _x-\left( \frac{d\eta }{dx}\right) _{x-a}\right\} , 
\]
tomando de nuevo el l\'{\i}mite cuando a tiende a cero la ecuaci\'{o}n define
claramente la segunda derivada de $\eta$. Por tanto, la ecuaci\'{o}n de
movimiento para la varilla el\'{a}stica ser\'{a} 
\begin{equation}
\mu \frac{d^2\eta }{dt^2}-Y\frac{d^2\eta }{dx^2}=0,  \label{7}
\end{equation}
que es la conocida ecuaci\'{o}n de onda en una dimensi\'{o}n con velocidad
de propagaci\'{o}n 
\begin{equation}
\upsilon =\sqrt{\frac Y\mu }.  \label{8}
\end{equation}
La ecuaci\'{o}n (\ref{8}) es la conocida f\'{o}rmula de la velocidad de
propagaci\'{o}n de las ondas el\'{a}sticas longitudes.

\noindent
Este sencillo ejemplo es suficiente para ilustrar las caracter\'{\i}sticas
principales de la transici\'{o}n de un sistema discreto a uno continuo. El
hecho mas importante que hemos de comprender es el papel que desempe\~{n}a
la coordenada $x.$ No se trata de una coordenada generalizada; s\'{o}lo hace
las veces de \'{\i}ndice continuo que sustituye al indice discreto $i$. Al
igual que cada valor de $x$ corresponde una coordenada generalizada $\eta
\left( x\right) .$ Como\thinspace $\eta $ depende tambi\'{e}n de la variable
continua $t,$ debemos tal vez escribir con mayor precisi\'{o}n que $\eta
\left( x.t\right)$. Indicando que $x$, al igual que t, puede considerarse
como par\'{a}metro que entra en la lagrangiana. Si el sistema continuo fuese
tridimensional y no unidimensional, como en este caso, las coordenadas
generalizadas se distinguir\'{\i}an mediante tres indices continuos $x,y,z$ y
se escribir\'{\i}an en la forma $\eta \left( x,y,z,t\right) .$ Notemos que
las cantidades $x,y,z,t$ son totalmente independientes unas de otras y
s\'{o}lo aparecen en $\eta $ como variables expl\'{\i}citas. Las derivadas de 
$\eta $ respecto a cualquiera de ellas podr\'{a}n, pues, escribirse siempre
en forma de derivadas totales sin ninguna ambig\"{u}edad. La ecuaci\'{o}n (%
\ref{6}) indica tambi\'{e}n que la lagrangiana aparece como integral para el
indice continuo $x;$ en el caso tridimensional, la lagrangiana tendr\'{\i}a
la forma 
\begin{equation}
L=\int \int \int{\cal L} dx dy dz,  \label{9}
\end{equation}
donde $\cal{L}$ se denomina densidad Lagrangiana. En el caso de vibraciones
longitudinales de la varilla continua, la densidad Lagrangiana es
\begin{equation}
{\cal L} =\frac{1}{2}\left\{ \mu \left( \frac{d\eta }{dt}\right)
^2-Y\left( \frac{d\eta }{dx}\right) ^2\right\} ,  \label{10}
\end{equation}
y corresponde al l\'{\i}mite continuo de la cantidad $L_i$ que aparece en la
ecuaci\'{o}n (\ref{4}). Es la densidad de Lagrangiana, m\'{a}s que la propia
Lagrangiana, la que utilizaremos para describir el movimiento del sistema.

\section*{11.2 Formulaci\'{o}n Lagrangiana para sistemas continuos}

Notemos en la ec. (\ref{9}) que la $\cal{L}$ para la varilla el\'{a}stica
depende de $\stackrel{\cdot }{\eta }=\partial \eta /\partial t$, la derivada
espacial de $\eta ,$ $\partial \eta /\partial x$; $x$ y $t$ desempe\~{n}an un
papel similar al de los par\'{a}metros de esta. Si adem\'{a}s de las
interacciones entre vecinos m\'{a}s pr\'{o}ximos hubiesen fuerzas locales,
$\cal{L}$ fuera funci\'{o}n de la misma $\eta$. En general $\cal{L}$ para
todo sistema continuo, puede ser funci\'{o}n expl\'{\i}cita de $x$ y $t$.
Por tanto, la densidad de Lagrangiana para todo sistema continuo debe
aparecer para todo sistema continuo unidimensional de la forma
\begin{equation}
{\cal L} = {\cal L} \left( \eta ,\frac{d\eta }{dx},\frac{d\eta }{dt}%
,x,t\right) .  \label{11}
\end{equation}
La Lagrangiana total siguiendo la forma de la ec.(\ref{9}) ser\'{a}
\[
L=\int {\cal L}dx,
\]
y el principio de Hamilton en el l\'{\i}mite del sistema continuo adopta la
forma 
\begin{equation}
\delta I=\delta \int_1^2\int {\cal L}dxdt=0.  \label{12}
\end{equation}
Del principio de Hamilton para el sistema continuo, deber\'{a} ser posible
deducir el l\'{\i}mite continuo de las ecuaciones de movimiento, para esto
como en la secci\'on 2-2 de Goldstein podemos obtener en el espacio $\eta $ un
camino variado
de integraci\'{o}n conveniente, eligiendo $\eta $ entre una familia de
funciones de $\eta $ dependiente de un par\'{a}metro: 
\begin{equation}
\eta \left( x,t;\alpha \right) =\eta \left( x,t;0\right) +\alpha \zeta
\left( x,t\right) .  \label{13}
\end{equation}
Donde $\eta \left( x,t;0\right) $ es la funci\'{o}n correcta que satisface
el principio de Hamilton y $\zeta $ es una funci\'{o}n cualquiera de buen
comportamiento que se anule en los puntos extremos en $t$ y en $x.$ Si
consideramos $I$ funci\'{o}n de $\alpha$, para que sea una extremal para la
derivada de $I$ respecto de $\alpha$ se anular\'{a} en $\alpha =0$. Ahora
por derivaci\'{o}n directa de $I$ tenemos 
\begin{equation}
\frac{dI}{da}=\int_{t_1}^{t_2}\int_{x_1}^{x_2}dxdt\left\{ \frac{\partial 
{\cal L}}{\partial \eta }\frac{\partial \eta }{\partial \alpha }+\frac{%
\partial {\cal L}}{\partial \frac{d\eta }{dt}}\frac \partial {\partial
\alpha }\left( \frac{d\eta }{dt}\right) +\left( \frac{\partial {\cal L}}{%
\partial \frac{d\eta }{dx}}\right) \frac \partial {\partial \alpha }\left( 
\frac{d\eta }{dx}\right) dt\right\} .  \label{14}
\end{equation}
Como la variaci\'{o}n de $\eta$, es decir $\alpha \zeta$, se anula en los
puntos extremos, integrando por partes seg\'{u}n $x$ y $t$ obtenemos las
relaciones 
\[
\int_{t_1}^{t_2}\frac{\partial {\cal L}}{\partial \frac{d\eta }{dt}}\frac
\partial {\partial \alpha }\left( \frac{d\eta }{dt}\right)
dt=-\int_{t_1}^{t_2}\frac d{dt}\left( \frac{\partial {\cal L}}{\partial 
\frac{d\eta }{dt}}\right) \frac{d\eta }{d\alpha }dt,
\]
y 
\[
\int_{x_1}^{x_2}\frac{\partial {\cal L}}{\partial \frac{d\eta }{dx}}\frac
\partial {\partial \alpha }\left( \frac{d\eta }{dx}\right)
dx=-\int_{x_1}^{x_2}\frac d{dx}\left( \frac{\partial {\cal L}}{\partial 
\frac{d\eta }{dx}}\right) \frac{d\eta }{d\alpha }dx.
\]
De aqu\'{\i} el principio de Hamilton se podr\'{a} escribir de la forma
\begin{equation}
\int_{t_1}^{t_2}\int_{x_1}^{x_2}dxdt\left\{ \frac{\partial {\cal L}}
{\partial \eta }-\frac d{dt}\left( \frac{\partial {\cal L}}{\partial
\frac{
d\eta }{dt}}\right) -\frac d{dx}\left( \frac{\partial {\cal L}}{\partial
\frac{d\eta }{dx}}\right) \right\} \left( \frac{\partial \eta }{\partial
\alpha }\right) _0=0~.  \label{15}
\end{equation}
Ahora debida a la naturaleza arbitraria del camino variado implica que la
expresi\'{o}n entre llaves es cero: 
\begin{equation}
\frac d{dt}\left( \frac{\partial {\cal L}}{\partial \frac{d\eta }{dt}}%
\right) +\frac d{dx}\left( \frac{\partial {\cal L}}{\partial \frac{d\eta }
{dx}}\right) -\frac{\partial {\cal L}}{\partial \eta }=0.  \label{16}
\end{equation}
La ecuaci\'{o}n anterior corresponde a la ecuaci\'{o}n correcta de
movimiento deducida del principio de Hamilton.

\noindent
En el caso concreto de las vibraciones longitudinales en una varilla
el\'{a}stica, la forma de la densidad de Lagrangiana ec. (\ref{10}) indica
que 
\[
\frac{\partial {\cal L}}{\partial \frac{d\eta }{dt}}=\mu \frac{d\eta }{dt}~
,\qquad \frac{\partial {\cal L}}{\partial \frac{d\eta }{dx}}=-Y\frac{d\eta }{
dx}~,\qquad \frac{\partial {\cal L}}{\partial \eta }=0.
\]
As\'{\i} pues, tal como quer\'{\i}amos, la ecuaci\'{o}n de Euler-Lagrange
(\ref{16}) se reduce a la ecuaci\'{o}n de movimiento (\ref{7}).

\noindent
La formulaci\'{o}n de Lagrange que acabamos de desarrollar corresponde a
sistemas continuos, evidentemente, se pude generalizar a sistemas bi- ,
tridimensionales y cuadridimensionales. Matem\'{a}ticamente, conviene pensar
en un espacio cuadridimensional de coordenadas $x_o=t,x_1=x,x_2=y,x_3=z.$

\noindent
Para un manejo matem\'{a}tico m\'{a}s f\'{a}cil introducimos la siguiente
notaci\'{o}n 
\begin{equation}
\eta _{\rho ,\nu }\equiv \frac{d\eta _\rho }{dx_\nu };\qquad \eta
_{,j}\equiv \frac{d\eta }{dx_j};\qquad \eta _{i,\mu \nu }\equiv \frac{%
d^2\eta _i}{dx_\mu dx_\nu }.  \label{17}
\end{equation}
Con esta notaci\'{o}n, y haciendo la extensi\'{o}n a un espacio
cuadridimensional, la forma general de la densidad de Lagrangiana (\ref{11})
toma la forma: 
\begin{equation}
\cal{L}={\cal L}\left( \eta _\rho ,\eta _{\rho ,\nu },x_{\nu} \right) .
\label{18}
\end{equation}
La Lagrangiana total es entonces una integral extendida al espacio
tridimensional: 
\begin{equation}
L=\int {\cal L}\left( dx_i\right) .  \label{19}
\end{equation}
Al principio de Hamilton corresponde una integral extendida a una regi\'{o}n
de un espacio cuadridimencional 
\begin{equation}
\delta I=\delta \int {\cal L}\left( dx_\mu \right) =0,  \label{20}
\end{equation}
donde la variaci\'{o}n de las $\eta _\rho$ se anulan en la superficie $S$
de contorno de la regi\'{o}n de integraci\'{o}n. La deducci\'{o}n de las
correspondientes ecuaciones de movimiento de Euler-Lagrange tiene lugar
simb\'{o}licamente como antes. Consideremos un sistema de funciones 
\[
\eta _\rho \left( x_\nu ;\alpha \right) =\eta _\rho \left( x_\nu \right)
+\alpha \zeta \left( x_\nu \right) 
\]
variadas de un solo par\'{a}metro que se reducen a $\eta _\rho \left( x_\nu
\right) $ cuando el par\'{a}metro $\alpha $ tiende a cero. La variaci\'{o}n
de $I$ equivale a hacer igual a cero la derivada de $I$ respecto a $\alpha $
es decir: 
\begin{equation}
\frac{dI}{d\alpha }=\int \left( \frac{\partial {\cal L}}{\partial \eta _\rho 
}\frac{\partial \eta _\rho }{\partial \alpha }+\frac{\partial {\cal L}}{%
\partial \eta _{\rho ,\nu }}\frac{\partial \eta _{\rho ,\nu }}{\partial
\alpha }\right) \left( dx_\mu \right) =0.  \label{21}
\end{equation}
Integrando por partes la ec. (\ref{21}), tenemos 
\[
\frac{dI}{d\alpha }=\int \left[ \frac{\partial  {\cal L}} {\partial \eta _\rho 
}-\frac d{dx_\nu }\left( \frac{\partial {\cal L}}{\partial \eta _{\rho ,\nu }%
}\right) \right] \frac{\partial \eta _\rho }{\partial \alpha }\left( dx_\mu
\right) +\int \left( dx_\mu \right) \frac d{dx_\nu }\left( \frac{\partial 
{\cal L}}{\partial \eta _{\rho ,\nu }}\frac{\partial \eta _{\rho ,\nu }}{%
\partial \alpha }\right) =0,
\]
y tomando el l\'{\i}mite cuando $\alpha $ tiende a cero la expresi\'{o}n
anterior se reduce a: 
\begin{equation}
\left( \frac{dI}{d\alpha }\right) _0=\int \left( dx_\mu \right) \left[ \frac{%
\partial {\cal L}}{\partial \eta _\rho }-\frac d{dx_\nu }\left( \frac{%
\partial {\cal L}}{\partial \eta _{\rho ,\nu }}\right) \right] \left( \frac{%
\partial \eta _\rho }{\partial \alpha }\right) _0=0.  \label{22}
\end{equation}
Ahora debido a la naturaleza arbitraria de la variaci\'{o}n de cada tipo de $%
\eta _\rho $ significa que la ecuaci\'{o}n (\ref{22}) cuando sea nulo cada
uno de los corchetes por separado esto es: 
\begin{equation}
\frac d{dx_\nu }\left( \frac{\partial {\cal L}}{\partial \eta _{\rho ,\nu }}%
\right) -\frac{\partial {\cal L}}{\partial \eta _\rho }=0.  \label{23}
\end{equation}
Las ecuaciones (\ref{23}) representan un sistema de ecuaciones en derivadas
parciales para las cantidades campo, que tiene tantas ecuaciones cuantos
valores diferentes de $\rho $ haya.

\bigskip

\noindent
\underline{{\LARGE Ejemplo:}}
Dada la densidad Lagrangiana para un campo ac\'{u}stico
\[
{\cal L}=\frac 12\left( \mu _0\stackrel{.}{\vec{\eta}}^2+2P_0\nabla
\cdot \vec{\eta} -\gamma P_0\left( \nabla \cdot \vec{\eta}\right)
^2\right) . 
\]
$\mu _0$ es la densidad m\'{a}sica de equilibrio y $P_0$ la
presi\'{o}n de equilibrio del gas. El primer t\'{e}rmino de ${\cal L}$ es
la densidad de energ\'{\i}a cin\'{e}tica, mientras que los t\'{e}rminos
restantes representan el cambio que sufre la energ\'{\i}a potencial del gas
por unidad de volumen a consecuencia del trabajo efectuado sobre el gas o
por el curso de las contracciones y expansiones que son la marca de las
vibraciones ac\'{u}sticas, $\gamma$ es el cociente entre los calores
molares a presi\'{o}n y a volumen constante obtener las ecuaciones de
movimiento.

\noindent
\underline{{\Large Soluci\'on:}}

\noindent
Con la notaci\'{o}n cuadridimensional, la densidad de Lagrangiana queda en
la forma

\begin{equation}
{\cal L}=\frac 12\left( \mu _0\eta _{i,0}\eta _{i,0}+2P_0\eta _{i,i}
-\gamma P_0\eta _{i,i}\eta _{j,j}\right)~.   \label{24}
\end{equation}
De la ecuaci\'{o}n (\ref{23}) se obtienen las siguientes ecuaciones de
movimiento 
\begin{equation}
\mu _0\eta _{j,00}-\gamma P_0\eta _{i,ij}=0,\qquad j=1,2,3.  \label{25}
\end{equation}
Volviendo a la notaci\'{o}n vectorial, las ecuaciones (\ref{25}) toman la
forma 
\begin{equation}
\mu _0\frac{d^2}{\vec{\eta}}{dt^2}-\gamma P_0\nabla \nabla \cdot
{\vec{\eta}}=0.  \label{26}
\end{equation}
Ahora utilizando el hecho de que en vibraciones de peque\~{n}a amplitud la
variaci\'{o}n relativa de la densidad del gas esta dado por la ecuaci\'{o}n 
\[
\sigma =-\nabla \cdot \vec{\eta}~.
\]
Ahora aplicando la divergencia y utilizando la ecuaci\'{o}n anterior
obtenemos 
\[
\nabla ^2\sigma -\frac{\mu _0}{\gamma P_0}\frac{d^2\sigma}{dt^2}=0
\]
la cual es una ecuaci\'{o}n de onda tridimensional, siendo 
\[
\upsilon =\sqrt{\frac{\gamma P_0}{\mu _0}}
\]
la velocidad del sonido en los gases.

\section*{11.3 Formulaci\'{o}n Hamiltoniana, par\'{e}ntesis de Poisson.}

\subsection*{11.3.1 Formulaci\'{o}n Hamiltoniana}

La formulaci\'{o}n de Hamilton para sistemas continuos se hace en forma
parecida como se hace para sistemas discretos. Para indicar el procedimiento
volvamos a la cadena de puntos materiales tratada anteriormente,
donde para cada $\eta _i$ hay una cantidad de movimiento can\'{o}nica

\begin{equation}
p_i=\frac{\partial L}{\partial \stackrel{.}{\eta }_i}=a\frac{\partial L_i}{%
\partial \stackrel{.}{\eta }_i}.  \label{27}
\end{equation}
La Hamiltoniana del sistema ser\'{a}, entonces
\begin{equation}
H\equiv p_i\stackrel{.}{\eta }_i-L=a\frac{\partial L_i}{\partial \stackrel{.%
}{\eta }_i}\stackrel{.}{\eta }_i-L,  \label{28}
\end{equation}
o sea 
\begin{equation}
H=a\left( \frac{\partial L_i}{\partial \stackrel{.}{\eta }_i}\stackrel{.}
{\eta }_i-L_i\right)~.   \label{29}
\end{equation}
Recordando que en l\'{\i}mite cuando $a$ tiende a cero, $L\rightarrow {\cal L}
$ y la suma de la ecuaci\'{o}n (\ref{29}) se convierte en una integral por
lo que el Hamiltoniano toma la forma:
\begin{equation}
H=\int dx\left( \frac{\partial {\cal L}}{\partial \stackrel{.}{\eta }}
\dot{\eta}-{\cal L}\right)~.   \label{30}
\end{equation}
Las cantidades de movimiento can\'{o}nicas individuales $p_i$, dadas por la
ecuaci\'{o}n (\ref{27}), se anulan en el l\'{\i}mite de la continuidad, pero
podemos definir una densidad de cantidad de movimiento $\pi $ que permanezca
finita: 
\begin{equation}
{\rm Lim}_{a\rightarrow 0}\frac{p_i}a\equiv \pi =\frac{
\partial {\cal L}}{\partial \stackrel{.}{\eta }}~.  \label{31}
\end{equation}
La ecuaci\'{o}n (\ref{30}) tiene la forma de integral espacial de una
densidad de Hamiltoniana ${\cal H}$ definida por
\begin{equation}
{\cal H}=\pi \stackrel{.}{\eta }-{\cal L}~.  \label{32}
\end{equation}
Aun cuando se pueda introducir as\'{\i} una formulaci\'{o}n de Hamilton de
manera directa para campos cl\'{a}sicos, d\'{e}monos cuenta de que el
procedimiento singulariza la variable tiempo a la que habr\'{a} que darle un
tratamiento especial. Contrasta, pues con el desarrollo que hemos dado a la
formulaci\'{o}n de Lagrange en el cual se trataban sim\'{e}tricamente las
variables independientes del tiempo y espaciales. Por esta raz\'{o}n, el
m\'{e}todo de Hamilton, se tratar\'{a} en forma un tanto distinta.

\noindent
La v\'{\i}a evidente para la generalizaci\'{o}n a un campo tridimensional
descrito por cantidades campo $\eta _\rho $ es la siguiente:

\noindent
Se define una cantidad de movimiento can\'{o}nica 
\begin{equation}
\pi _{_\rho }\left( x_{_\mu }\right) =\frac{\partial {\cal L}}{\partial 
\stackrel{.}{\eta }_{_\rho }}.  \label{33}
\end{equation}
Donde las cantidades $\eta _{_\rho }\left( x_i,t\right) ,\pi _{_\rho }\left(
x_i,t\right) $ juntas, definen el espacio f\'{a}sico de infinitas
dimensiones que describe el campo cl\'{a}sico y su desarrollo en el tiempo.

\noindent
An\'{a}logamente como en el sistema discreto podemos hallar un teorema de
conservaci\'{o}n para $\pi $ que sea algo parecido al correspondiente al de
cantidad de movimiento can\'{o}nico de sistemas discretos. Si una cantidad
de campo dada $\eta _\rho $ es c\'{\i}clica es decir que ${\cal L}$ no
contenga expl\'{\i}citamente a $\eta _\rho ,$ la ecuaci\'{o}n de campo de
Lagrange presenta el aspecto de enunciado de la existencia de una corriente
conservativa: 
\[
\frac d{dx_{_\mu }}\frac{\partial {\cal L}}{\partial \eta _{_{\rho ,\mu }}}=0
\]
o sea 
\begin{equation}
\frac{d\pi _\rho }{dt}-\frac d{dx_{_i}}\frac{\partial {\cal L}}{\partial
\eta _{_{\rho ,i}}}=0~.  \label{34}
\end{equation}
Se sigue que si es c\'{\i}clica $\eta _\rho ,$ existe una cantidad integral
que se conservativa 
\[
\Pi _\rho =\int dV\pi _{_\rho }\left( x_i,t\right) .
\]

\noindent
La generalizaci\'{o}n para la densidad de la ecuaci\'{o}n (\ref{32}) para
una densidad Hamiltoniana es
\begin{equation}
{\cal H}\left( \eta _{{\rho }},\eta _{_{\rho ,i}},\pi _{{ %
\rho }},x_{{ \mu }}\right) =\pi _{_\rho }\stackrel{.}{\eta }_{_\rho }-
\cal{L,}  \label{35}
\end{equation}
donde se supone que se puede eliminar la dependencia funcional
de $\stackrel{.}{\eta }_\rho$ por inversi\'{o}n de las ecuaciones de
definici\'{o}n (\ref{33}). De esta definici\'{o}n se deduce que
\begin{equation}
\frac{\partial \cal{H}}{\partial \pi _{{ \rho }}}=\stackrel{.}{%
\eta }_{_\rho }+\pi _{_\lambda }\frac{\partial \stackrel{.}{\eta }_{_\lambda
}}{\partial \pi _{_\rho }}-\frac{\partial {\cal L}}{\partial \stackrel{.}{%
\eta }_{_\lambda }}\frac{\partial \stackrel{.}{\eta }_{_\lambda }}{\partial
\pi _{_\rho }}=\stackrel{.}{\eta }_{_\rho }  \label{36}
\end{equation}
en virtud de la ecuaci\'{o}n (\ref{33}). An\'{a}logamente obtenemos
\begin{equation}
\frac{\partial \cal{H}}{\partial \eta _{{ \rho }}}=\pi _{_\lambda }
\frac{\partial \stackrel{.}{\eta }_{_\lambda }}{\partial \eta _{_\rho }}-
\frac{\partial {\cal L}}{\partial \stackrel{.}{\eta }_{_\lambda }}\frac{
\partial \stackrel{.}{\eta }_{_\lambda }}{\partial \eta _{_\rho }}-\frac{
\partial {\cal L}}{\partial \eta _{_\rho }}=-\frac{\partial {\cal L}}{
\partial \eta _{_\rho }}.  \label{37}
\end{equation}
Ahora utilizando las ecuaciones de Lagrange en la ec.(\ref{37}) queda como 
\begin{equation}
\frac{\partial \cal{H}}{\partial \eta _{{ \rho }}}=-\frac
d{dx_{_\mu }}\left( \frac{\partial {\cal L}}{\partial \eta _{_{\rho ,\mu }}}
\right) =-\stackrel{.}{\pi }_{_\rho }-\frac d{dx_{_i}}\left( \frac{\partial 
{\cal L}}{\partial \eta _{_{\rho ,i}}}\right) .  \label{38}
\end{equation}
Debido a la aparici\'{o}n de ${\cal L}$ aun no tenemos una forma \'{u}til.
Sin embargo, haciendo una deducci\'{o}n an\'{a}loga a la de los t\'{e}rminos 
$\frac{\partial {\cal H}}{\partial \pi _\rho }$ y $\frac{\partial
{\cal H}}{\partial \eta _\rho }$ para$\ \frac{\partial {\cal H}}{
\partial \eta _{\rho ,i}}$ tenemos 
\begin{equation}
\frac{\partial \cal{H}}{\partial \eta _{{ \rho ,i}}}=\pi
_{_\lambda }\frac{\partial \stackrel{.}{\eta }_{_\lambda }}{\partial \eta
_{_{\rho ,i}}}-\frac{\partial {\cal L}}{\partial \stackrel{.}{\eta }%
_{_\lambda }}\frac{\partial \stackrel{.}{\eta }_{_\lambda }}{\partial \eta
_{_{\rho ,i}}}-\frac{\partial {\cal L}}{\partial \eta _{_{\rho ,i}}}=-\frac{%
\partial {\cal L}}{\partial \eta _{_{\rho ,i}}}~.  \label{39}
\end{equation}
Por lo tanto sustituyendo (\ref{39}) en (\ref{38}) obtenemos 
\begin{equation}
\frac{\partial \cal{H}}{\partial \eta _{{ \rho }}}-\frac
d{dx_i}\left( \frac{\partial \cal{H}}{\partial \eta _{{ \rho ,i}}}%
\right) =-\stackrel{.}{\pi }_{_\rho}~.  \label{40}
\end{equation}

\noindent
Las ecuaciones (\ref{36}) y (\ref{40}) las podemos expresar con una
notaci\'{o}n m\'{a}s pr\'{o}xima a la de las ecuaciones de Hamilton para un
sistema\thinspace discreto introduciendo la noci\'{o}n de derivada funcional
definida en la forma 
\begin{equation}
\frac \delta {\delta \psi }=\frac \partial {\partial \psi }-\frac
d{dx_i}\frac \partial {\partial \psi _{,i}}.  \label{41}
\end{equation}
Como $\cal{H}$ no es funci\'{o}n de $\pi _{_{\rho ,i}}$ las ecuaciones (
\ref{36}) y (\ref{40}) se pueden escribir en la forma 
\begin{equation}
\stackrel{.}{\eta }_{_\rho }=\frac{\delta \cal{H}}{\delta \pi _{_\rho }}
,\qquad \stackrel{.}{\pi }_{_\rho }=-\frac{\delta \cal{H}}{\delta \eta _{
{ \rho }}}.  \label{42}
\end{equation}
Ahora con la misma notaci\'{o}n las ecuaciones de Lagrange (\ref{23}) toman
la forma 
\begin{equation}
\frac d{dt}\left( \frac{\partial {\cal L}}{\partial \stackrel{.}{\eta }
_{_\rho }}\right) -\frac{\delta {\cal L}}{\delta \eta _{{ \rho }}}=0.
\label{43}
\end{equation}

\noindent
Sin embargo, la ventaja casi \'{u}nica de la derivada funcional estriba en
la semejanza resultante con un sistema discreto. Por otra parte, sorprende
el tratamiento paralelo de las variables temporal y espaciales.

\subsection*{11.3.2 Par\'{e}ntesis de Poisson}

\noindent
Podemos obtener otras propiedades de $\cal{H}$ desarrollando la derivada
total respecto al tiempo de la ecuaci\'{o}n (\ref{35}), recordando que hay
que considerar que $\stackrel{.}{\eta }_{_\rho }$ es funci\'{o}n de $\eta
_{_\rho },\eta _{_{\rho ,j}},\pi _{_\rho }$ y $\pi _\mu $. Tenemos,
entonces que 
\[
\frac{d{\cal H}}{dt}=\stackrel{.}{\pi }_{_\rho }\stackrel{.}{\eta }
_{_\rho }+\pi _{_\rho }\frac{d\stackrel{.}{\eta }_{_\rho }}{dt}-\frac{
\partial {\cal L}}{\partial \eta _{{ \rho }}}\stackrel{.}{\eta }
_{_\rho }-\frac{\partial {\cal L}}{\partial \stackrel{.}{\eta }_{_\rho }}
\frac{d\stackrel{.}{\eta }_{_\rho }}{dt}-\frac{\partial {\cal L}}{\partial
\eta _{_{\rho ,i}}}\frac{d\eta _{_{\rho ,i}}}{dt}-\frac{\partial {\cal L}}{
\partial t}. 
\]

\noindent
En la expresi\'{o}n el segundo t\'{e}rmino y el cuarto se aniquilan debido a
la definici\'{o}n (\ref{33}), por lo que la derivada se simplifica quedando 
\begin{equation}
\frac{d{\cal H}}{dt}=\stackrel{.}{\pi }_{_\rho }\stackrel{.}{\eta }
_{_\rho }-\frac{\partial {\cal L}}{\partial \eta _{{ \rho }}}\stackrel{
.}{\eta }-\frac{\partial {\cal L}}{\partial \eta _{_{\rho ,i}}}\frac{d\eta
_{_{\rho ,i}}}{dt}-\frac{\partial {\cal L}}{\partial t}.  \label{44}
\end{equation}

\noindent
Por otra parte, considerando $\cal{H}$ funci\'{o}n de $\eta _{_\rho
},\eta _{_{\rho ,j}},\pi _{_\rho }$ y $\pi _\mu ,$ la derivada total
respecto al tiempo es 
\begin{equation}
\frac{d{\cal H}}{dt}=\stackrel{.}{\pi }_{_\rho }\frac{\partial {\cal H}
}{\partial \pi _{{ \rho }}}+\frac{\partial {\cal H}}{\partial \eta
_{{ \rho }}}\stackrel{.}{\eta }_{_\rho }+
\frac{\partial {\cal H}}{
\partial \eta _{{ \rho ,i}}}\frac{d\eta _{_{\rho ,i}}}{dt}+\frac{
\partial {\cal H}}{\partial t}~,  \label{45}
\end{equation}
donde la expresi\'{o}n se escribi\'{o} de tal manera que facilite la
comparaci\'{o}n con el segundo miembro de la ecuaci\'{o}n (\ref{44}), donde
usando las ecuaciones (\ref{36}), (\ref{37}) y (\ref{39}) obtenemos 
\begin{equation}
\frac{\partial {\cal H}}{\partial t}=-\frac{\partial {\cal L}}{\partial t},
\label{46}
\end{equation}
la cual es an\'{a}loga a la correspondiente para sistemas discretos.

\noindent
En cambio no se cumple que las derivadas total y parcial respecto al tiempo
no son, en general iguales. Utilizando las ecuaciones de movimiento de
hamilton (ec. (\ref{36}) y (\ref{40})) e intercambiando los ordenes de
derivaci\'{o}n, la ecuaci\'{o}n (\ref{45}) se puede escribir como 
\[
\frac{d{\cal H}}{dt}=\frac{\partial {\cal H}}{\partial \pi _{{
\rho }}}\frac d{dx_i}\left( \frac{\partial {\cal H}}{\partial \eta _{
{ \rho ,i}}}\right) +\frac{\partial {\cal H}}{\partial \eta _{
{ \rho ,i}}}\frac{d\stackrel{.}{\eta }_{_\rho }}{dx_i}+\frac{\partial 
{\cal H}}{\partial t}~.
\]
Ahora utilizando la ecuaci\'{o}n (\ref{46}) y combinando t\'{e}rminos
tenemos finalmente 
\begin{equation}
\frac{d{\cal H}}{dt}=\frac d{dx_i}\left( \stackrel{.}{\eta }_{_\rho }
\frac{\partial {\cal H}}{\partial \eta _{{ \rho ,i}}}\right) +\frac{
\partial {\cal H}}{\partial t},  \label{47}
\end{equation}
que es lo mas que podemos aproximarnos a la ecuaci\'{o}n correspondiente
para sistemas discretos.

\noindent
Por otra parte cuando ${\cal L}$ no contenga a $t$ expl\'{\i}citamente,
tampoco la contendr\'{a} $\cal{H}$ esto implica la existencia de una
corriente consservativa y por lo tanto la conservaci\'{o}n de una cantidad
integral, en este caso 
\begin{equation}
H=\int {\cal H}dV~.  \label{48}
\end{equation}
As\'{\i} pues, si ${\cal H}$ no es funci\'{o}n expl\'{\i}cita del tiempo,
la cantidad que se conserva no es ${\cal H},$ sino la cantidad integral $H$.

\noindent
La Hamiltoniana no es m\'{a}s que un ejemplo de funciones que son integrales
de volumen de densidades. Podemos formular directamente un formalismo
general para la derivada respecto al tiempo de dichas cantidades integrales.
Consideremos una cierta densidad $\cal{U}$ que sea funci\'{o}n de las
coordenadas del espacio f\'{a}sico $\left( \eta _\rho ,\pi _\rho \right)$,
de sus gradientes espaciales y posiblemente de $x_\mu$:
\begin{equation}
{\cal U}={\cal U}\left( \eta _{{ \rho }},\pi _{_\rho },\eta _{{ \rho
,i}},\pi _{_{\rho ,i}},x_\mu \right)~.   \label{49}
\end{equation}
La cantidad integral correspondiente es 
\begin{equation}
U\left( t\right) =\int {\cal U}dV  \label{50}
\end{equation}
donde la integral de volumen se extiende a todo el espacio limitado por la
superficie de contorno sobre la cual se anulan $\eta _\rho $ y $\pi _\rho .$
Derivando $U$ respecto al tiempo tenemos en general, 
\begin{equation}
\frac{dU}{dt}=\int \left\{ \frac{\partial \cal{U}}{\partial \eta _{%
{ \rho }}}\stackrel{.}{\eta }_{_\rho }+\frac{\partial \cal{U}}{%
\partial \eta _{{ \rho ,i}}}\stackrel{.}{\eta }_{{ \rho ,i}}+%
\frac{\partial \cal{U}}{\partial \pi _{{ \rho }}}\stackrel{.}{\pi }%
_{{ \rho }}+\frac{\partial {\cal U}}{\partial \pi _{{ \rho ,i}%
}}\stackrel{.}{\pi }_{{ \rho ,i}}+\frac{\partial {\cal U}}{\partial
t}\right\} dV~.  \label{51}
\end{equation}
Consideremos un t\'{e}rmino tal como 
\[
\int dV\frac{\partial {\cal U}}{\partial \eta _{{ \rho ,i}}}
\stackrel{.}{\eta }_{{ \rho ,i}}=\int dV\frac{\partial {\cal U}}{
\partial \eta _{{ \rho ,i}}}\frac{d\stackrel{.}{\eta }_{_\rho }}{dx_i}~.
\]
Ahora intengrando por partes, considerando que $\eta _\rho$ y las
derivadas se anulan en las superficies de contorno, tenemos 
\[
\int dV\frac{\partial {\cal U}}{\partial \eta _{{ \rho ,i}}}%
\stackrel{.}{\eta }_{{ \rho ,i}}=-\int dV\stackrel{.}{\eta }_{_\rho
}\frac d{dx_i}\left( \frac{\partial {\cal U}}{\partial \eta _{{
\rho ,i}}}\right) .
\]

\noindent
Para el t\'{e}rmino en $\stackrel{.}{\pi }_{\rho ,i}$ se hace un
procedimiento similar. Sustituyendo las expresiones obtenidas y agrupando
coeficientes de $\stackrel{.}{\eta }$ y de $\stackrel{.}{\pi }_\rho $
respectivamente, y usando la notaci\'{o}n de derivada funcional la
ecuaci\'{o}n (\ref{51}) se reduce a

\begin{equation}
\frac{dU}{dt}=\int dV\left\{ \frac{\delta \cal{U}}{\delta \eta _{{
\rho }}}\stackrel{.}{\eta }_{_\rho }+\frac{\delta \cal{U}}{\delta \pi _{
{ \rho }}}\stackrel{.}{\pi }_{{ \rho }}+\frac{\partial {\cal U
}}{\partial t}\right\}~.  \label{52}
\end{equation}
Por \'{u}ltimo, introduciendo las ecuaciones de movimiento can\'{o}nicas (
\ref{42}), tenemos 
\begin{equation}
\frac{dU}{dt}=\int dV\left\{ \frac{\delta \cal{U}}{\delta \eta _{{
\rho }}}\frac{\delta \cal{H}}{\delta \pi _{{ \rho }}}-\frac{\delta
\cal{H}}{\delta \eta _{{ \rho }}}\frac{\delta \cal{U}}{\delta
\pi _{{ \rho }}}\right\} +\int dV\frac{\partial {\cal U}}{\partial t
}~.  \label{53}
\end{equation}

\noindent
La primera integral del segundo miembro corresponde claramente a la forma de
corchete de Poisson. Si $\cal{U}$ y $\cal{W}$ son dos funciones de
densidad, estas consideraciones nos sugieren la definici\'{o}n del corchete
de Poisson de las cantidades integrales como 
\begin{equation}
\left[ U,W\right] =\int dV\left\{ \frac{\delta \cal{U}}{\delta \eta _{%
{ \rho }}}\frac{\delta \cal{W}}{\delta \pi _{{ \rho }}}-%
\frac{\delta \cal{W}}{\delta \eta _{{ \rho }}}\frac{\delta
\cal{U}}{\delta \pi _{{ \rho }}}\right\} .  \label{54}
\end{equation}
Definamos tambi\'{e}n la derivada parcial de $U$ respecto a $t,$ mediante la
siguiente expresi\'{o}n 
\begin{equation}
\frac{\partial U}{\partial t}=\int dV\frac{\partial {\cal U}}{\partial t}.
\label{55}
\end{equation}

\noindent
La ecuaci\'{o}n (\ref{53}) podr\'{a} entonces escribirse en la forma 
\begin{equation}
\frac{dU}{dt}=\left[ U,H\right] +\frac{\partial U}{\partial t},  \label{56}
\end{equation}
que corresponde precisamente, en esta notaci\'{o}n a la ecuaci\'{o}n para
sistemas discretos. Como por definici\'{o}n, el corchete de Poisson de $H$
consigo misma es nulo, la ecuaci\'{o}n (\ref{46}) se concretar\'{a} en 
\begin{equation}
\frac{dH}{dt}=\frac{\partial H}{\partial t}~,  \label{57}
\end{equation}
que es la forma integral de la ec. (\ref{47}).
As\'{\i} pues, el formalismo de corchetes de Poisson aparece como
consecuencia de la formulaci\'{o}n de Hamilton. Pero no podemos llevar a
cabo una descripci\'{o}n por corchetes de Poisson de la teor\'{\i}a de campos
en correspondencia paso a paso con la de los sistemas discretos.

\noindent
Sin embargo, hay una manera de tratar los campos cl\'{a}sicos que provee
casi todo lo de la formulaci\'{o}n de Hamilton y de corchetes de Poisson de
la Mec\'{a}nica para sistemas discretos. La idea fundamental de este
tratamiento es sustituir la variable espacial continua o el indice continuo
por un indice discreto numerable.

\noindent
El requisito de que $\eta $ se anule en los extremos es una condici\'{o}n de
contorno que se podr\'{\i}a realizar f\'{\i}sicamente colocando la varilla
entre dos paredes perfectamente rigidas. Entonces, la amplitud de
oscilaci\'{o}n se puede representar mediante una serie de Fourier: 
\begin{equation}
\eta \left( x\right) =\sum_{n=0}^\infty q_n\sin \frac{2\pi n\left(
x-x_1\right) }{2L}~.  \label{58}
\end{equation}
En vez del indice continuo $x$ tenemos el indice discreto $\eta .$ Podremos
utilizar esta representaci\'{o}n de $x$ solamente cuando $\eta \left(
x\right) $ sea una funci\'{o}n regular, cosa que sucede en la mayor\'{\i}a de
cantidades de campo f\'{\i}sicas.

\noindent
Supondremos que s\'{o}lo hay una cantidad campo real $\eta $ que se puede
desarrollar en serie de Fourier tridimensional de la forma 
\begin{equation}
\eta \left( \overrightarrow{r},t\right) =\frac 1{V^{1/2}}\sum_{k=0}q_k\left(
t\right) \exp \left( i\overrightarrow{k}\cdot \overrightarrow{r}\right) 
\label{59}
\end{equation}
Aqu\'{\i}, $\vec{k}$ es un vector de onda que solo puede tomar m\'{o}dulos y
direcciones discretos de manera que en una dimensi\'{o}n lineal dada
s\'{o}lo encaje un n\'{u}mero entero (o a veces semientero) de longitudes de
onda. Decimos que $\vec{k}$ tiene un espectro discreto. El subindice escalar 
$k$ representa una cierta ordenaci\'{o}n del sistema de indices enteros que se
utiliza para enumerar los valores discretos de $\vec{k}$; $V$ es el volumen del
sistema, el cual aparece en forma de factor de normalizaci\'{o}n.

\noindent
La ortogonalidad de las exponenciales en todo el volumen se puede enunciar
mediante la relaci\'{o}n 
\begin{equation}
\frac 1V\int e^{i\left( \vec{k}-\vec{k}^{^{\prime }}\right) .\vec{r}%
}dV=\delta _{k,k^{^{\prime }}}~.  \label{60}
\end{equation}
En realidad, los valores permitidos de $k$ son aquellos para los cuales se
satisface la condici\'{o}n (\ref{60}), y los coeficientes $q_k\left(
t\right) $ est\'{a}n dados por 
\begin{equation}
q_k\left( t\right) =\frac 1{V^{1/2}}\int e^{-i\overrightarrow{k}\cdot 
\overrightarrow{r}}\eta \left( \overrightarrow{r},t\right) dV~.  \label{61}
\end{equation}
De manera an\'{a}loga para la densidad de cantidad de movimiento
can\'{o}nica tenemos 
\begin{equation}
\pi \left( \overrightarrow{r},t\right) =\frac 1{V^{1/2}}\sum_kp_k\left(
t\right) e^{-i\overrightarrow{k}\cdot \overrightarrow{r}}  \label{62}
\end{equation}
con $p_k\left( t\right) $ definido como 
\begin{equation}
p_k\left( t\right) =\frac 1{V^{1/2}}\int e^{-i\overrightarrow{k}\cdot 
\overrightarrow{r}}\pi \left( \overrightarrow{r},t\right) dV~.  \label{63}
\end{equation}
Tanto $q_k$ como $p_k$ son cantidades integrales. Podemos, pues, buscar los
corchetes de Poisson de dichas cantidades. Como las exponenciales no
contienen las cantidades campo tenemos, por la ecuaci\'{o}n (\ref{54})
\begin{eqnarray*}
\left[ q_k,p_{k^{^{\prime }}}\right]  &=&\frac 1V\int dVe^{-i\overrightarrow{
k}\cdot \overrightarrow{r}}\left\{ \frac{\delta \eta }{\delta \eta }\frac{
\delta \pi }{\delta \pi }-\frac{\delta \pi }{\delta \eta }\frac{\delta \eta 
}{\delta \pi }\right\}  \\
&=&\frac 1V\int dVe^{-i\overrightarrow{k}\cdot \overrightarrow{r}}
\end{eqnarray*}
o sea, por la ecuaci\'{o}n (\ref{60}), 
\begin{equation}
\left[ q_k,p_{k^{^{\prime }}}\right] =\delta _{k,k^{^{\prime }}}~.  \label{64}
\end{equation}
De la definici\'{o}n de los corchetes de Poisson resulta evidente que 
\begin{equation}
\left[ q_k,q_{k^{^{\prime }}}\right] =\left[ p_k,p_{k^{^{\prime }}}\right]
=0~.
\label{65}
\end{equation}
La dependencia temporal de $q_k$ se hallar\'{a} a partir de
\[
\stackrel{.}{q}_k\left( t\right) =\left[ q_k,H\right] =\frac 1{V^{1/2}}\int
dVe^{-i\overrightarrow{k}\cdot \overrightarrow{r}}\left\{ \frac{\delta \eta 
}{\delta \eta }\frac{\delta \cal{H}}{\delta \pi }-\frac{\delta {\cal H
}}{\delta \eta }\frac{\delta \eta }{\delta \pi }\right\} 
\]
o sea 
\begin{equation}
\stackrel{.}{q}_k\left( t\right) =\frac 1{V^{1/2}}\int e^{-i\overrightarrow{k
}\cdot \overrightarrow{r}}\frac{\delta \cal{H}}{\delta \pi}~.  \label{66}
\end{equation}
Por otra parte, tenemos que 
\begin{equation}
\frac{\partial H}{\partial p_k}=\int dV\frac{\partial \cal{H}}{\partial
\pi }\frac{\partial \pi }{\partial p_k}  \label{67}
\end{equation}
por lo que obtenemos 
\begin{equation}
\frac{\partial \pi }{\partial p_k}=\frac 1{V^{1/2}}e^{-i\overrightarrow{k}
\cdot \overrightarrow{r}}~.  \label{68}
\end{equation}
Comparando las ecuaciones (\ref{67}) y (\ref{66}) tenemos 
\begin{equation}
\stackrel{.}{q}_k\left( t\right) =\frac{\partial H}{\partial p_k}~.  \label{69}
\end{equation}
De manera similar podemos obtener la ecuaci\'{o}n de movimiento para $p_k$ 
\begin{equation}
\stackrel{.}{p}_k=-\frac{\partial H}{\partial q_k}~.  \label{70}
\end{equation}
As\'{\i} las cantidades $p_k$ y $q_k,$ obedecen pues, las ecuaciones de
movimiento de Hamilton.

\section*{11.4 Teorema de Noether}

\noindent
Ya hemos visto m\'{u}ltiples veces que las propiedades de la Lagrangiana (o
de la Hamiltoniana) implican la existencia de cantidades conservativas.
As\'{\i}, si la Lagrangiana no contiene expl\'{\i}citamente una coordenada
particular de desplazamiento, se conserva la correspondiente cantidad de
movimiento can\'{o}nica. La ausencia de dependencia expl\'{\i}cita de la
coordenada significa que la Lagrangiana no queda afectada por una
transformaci\'{o}n que altere el valor de dicha coordenada; se dice que es
invariante o es sim\'{e}trica ante la transformaci\'{o}n dada.

\noindent
La simetr\'{\i}a ante una transformaci\'{o}n de coordenadas se refiere a los
efectos de una transformaci\'{o}n infinitesimal de la forma 
\begin{equation}
x_\mu \rightarrow x_\mu ^{^{\prime }}=x_\mu +\delta x_\mu ,  \label{71}
\end{equation}
donde la variaci\'{o}n $\delta x_\mu $ puede ser funci\'{o}n de las
dem\'{a}s $x_\nu$. El teorema de Noether considera tambi\'{e}n el efecto de
una transformaci\'{o}n de las propias cantidades campo, la cual podemos
escribir en la forma
\begin{equation}
\eta \left( x_\mu \right) \rightarrow \eta _\rho ^{^{\prime }}\left( x_\mu
^{^{\prime }}\right) =\eta _{_\rho }\left( x_{_\mu }\right) +\delta \eta
_{_\rho }\left( x_{_\mu }\right) .  \label{72}
\end{equation}
Aqu\'{\i} $\delta \eta _{_\rho }\left( x_{_\mu }\right) $ mide el efecto de
las variaciones de $x_{_\mu }$ y de $\eta _{_\rho }$ y puede ser funci\'{o}n
de las dem\'{a}s cantidades campo $\eta _{_\lambda }.$ La variaci\'{o}n de
una de las variables campo en un punto particular del espacio $x_{_\mu}$ es
una cantidad diferente $\overline{\delta }\eta _{_\rho }$:
\begin{equation}
\eta _\rho ^{^{\prime }}\left( x_\mu ^{^{\prime }}\right) =\eta _{_\rho
}\left( x_{_\mu }\right) +\overline{\delta }\eta _{_\rho }\left( x_{_\mu
}\right) .  \label{73}
\end{equation}
La descripci\'{o}n de las transformaciones en funci\'{o}n de varaiaciones
infinitesimales a partir de las cantidades no transformadas nos indica que
solo estamos tratando con transformaciones continuas. As\'{\i} la
simetr\'{\i}a ante la inversi\'{o}n en tres dimenciones no ser\'{a} una de
las simetr\'{\i}as a las que se pueda aplicar el teorema de Noether.
A consecuencia de
las transformaciones tanto de las coordenadas como de las cantidades campo,
la lagrangiana aparecer\'{a}, en general, como funci\'{o}n diferente de las
variables campo y de las coordenadas del espacio y tiempo:
\begin{equation}
{\cal L}\left( \eta _{_\rho }\left( x_{_\mu }\right) ,\eta _{_{\rho ,\nu
}}\left( x_{_\mu }\right) ,x_{_\mu }\right) \rightarrow {\cal L}^{^{\prime
}}\left( \eta _{_\rho }^{^{\prime }}\left( x_{_\mu }^{^{\prime }}\right)
,\eta _{_{\rho ,\nu }}^{^{\prime }}\left( x_{_\mu }^{^{\prime }}\right)
,x_{_\mu }^{^{\prime }}\right) .  \label{74}
\end{equation}

\noindent
La versi\'{o}n del teorema de Noether que vamos a presentar no constituye la
forma m\'{a}s general posible pero facilita la deducci\'{o}n sin restringir
de manera importante el \'{a}mbito de aplicaci\'{o}n del teorema ni la
utilidad de las conclusiones. Supondremos que se cumplen tres condiciones:

\begin{enumerate}
\item  El cuadriespacio es eucl\'{\i}deo. Este requisito, restringe el
espacio-tiempo relativista al espacio de Minkowski, que es complejo pero
euclideo.

\item  La densidad de lagrangiana presenta la misma forma funcional para las
cantidades transformadas que para las cantidades originales, es decir,
\begin{equation}
{\cal L}^{^{\prime }}\left( \eta _{_\rho }^{^{\prime }}\left( x_{_\mu
}^{^{\prime }}\right) ,\eta _{_{\rho ,\nu }}^{^{\prime }}\left( x_{_\mu
}^{^{\prime }}\right) ,x_{_\mu }^{^{\prime }}\right) ={\cal L}\left( \eta
_{_\rho }^{^{\prime }}\left( x_{_\mu }^{^{\prime }}\right) ,\eta _{_{\rho
,\nu }}^{^{\prime }}\left( x_{_\mu }^{^{\prime }}\right) ,x_{_\mu
}^{^{\prime }}\right) .  \label{75}
\end{equation}

\item  La magnitud de la integral de acci\'{o}n es invariante ante la
transformaci\'{o}n, es decir, 
\begin{equation}
I^{^{\prime }}\equiv \int_{\Omega ^{^{\prime }}}\left( dx_{_\mu }\right) 
{\cal L}^{^{\prime }}\left( \eta _{_\rho }^{^{\prime }}\left( x_{_\mu
}^{^{\prime }}\right) ,\eta _{_{\rho ,\nu }}^{^{\prime }}\left( x_{_\mu
}^{^{\prime }}\right) ,x_{_\mu }^{^{\prime }}\right) =\int_\Omega {\cal L}%
\left( \eta _{_\rho }\left( x_{_\mu }\right) ,\eta _{_{\rho ,\nu }}\left(
x_{_\mu }\right) ,x_{_\mu }\right) .  \label{76}
\end{equation}
\end{enumerate}

\noindent
La combinaci\'{o}n de las ecuaciones (\ref{75}) y (\ref{76}) nos da el
requisito
\begin{equation}
\int_{\Omega ^{^{\prime }}}\left( dx_{_\mu }\right) {\cal L}\left( \eta
_{_\rho }^{^{\prime }}\left( x_{_\mu }\right) ,\eta _{_{\rho ,\nu
}}^{^{\prime }}\left( x_{_\mu }\right) ,x_{_\mu }\right) -\int_\Omega {\cal L}
\left( \eta _{_\rho }\left( x_{_\mu }\right) ,\eta _{_{\rho ,\nu }}\left(
x_{_\mu }\right) ,x_{_\mu }\right) =0~.  \label{77}
\end{equation}
De la condici\'{o}n de invariancia, la ecuaci\'{o}n (\ref{77}) adopta la
forma 
\begin{eqnarray}
&&\int_{\Omega ^{^{\prime }}}dx_{_\mu }{\cal L}\left( \eta ^{^{\prime
}},x_{_\mu }\right) -\int_\Omega dx_{_\mu }{\cal L}\left( \eta ,x_{_\mu
}\right)   \label{78} \\
&=&\int_\Omega dx_{_\mu }\left[ {\cal L}\left( \eta ^{^{\prime }},x_{_\mu
}\right) -{\cal L}\left( \eta ,x_{_\mu }\right) \right] +\int_s{\cal L}%
\left( \eta \right) \delta x_{_\mu }dS_\mu =0~.  \nonumber
\end{eqnarray}
Aqu\'{\i} ${\cal L}\left( \eta ,x_{_\mu }\right) $ es una abreviatura de la
dependencia funcional total, $S$ es la superficie tridimensional de la regi%
\'{o}n $\Omega $ y $\delta x_{_\mu }$ es de hecho el vector diferencia entre
puntos de $S$ y los puntos correspondientes de la superficie transformada $%
S^{^{\prime }}.$ La \'{u}ltima integral se puede transformar mediante el
teorema de la divergencia cuadridimensional con lo que tendremos para la
condici\'{o}n de invarianza
\begin{equation}
0=\int_\Omega dx_{_\mu }\left\{ \left[ {\cal L}\left( \eta ^{^{\prime
}},x_{_\mu }\right) -{\cal L}\left( \eta ,x_{_\mu }\right) \right] +\frac
d{dx_{_\mu }}\left( {\cal L}\left( \eta ,x_{_\mu }\right) \delta x_{_\nu
}\right) \right\} .  \label{79}
\end{equation}
Ahora usando la ecuaci\'{o}n (\ref{73}), el t\'{e}rmino entre los corchetes
puede escribirse en primera aproximaci\'{o}n en la forma 
\[
{\cal L}\left( \eta _{_\rho }^{^{\prime }}\left( x_{_\mu }\right) ,\eta
_{_{\rho ,\nu }}^{^{\prime }}\left( x_{_\mu }\right) ,x_{_\mu }\right) -%
{\cal L}\left( \eta _{_\rho }\left( x_{_\mu }\right) ,\eta _{_{\rho ,\nu
}}\left( x_{_\mu }\right) ,x_{_\mu }\right) =\frac{\partial {\cal L}}{%
\partial \eta _{_\rho }}\overline{\delta }\eta _{_\rho }+\frac{\partial 
{\cal L}}{\partial \eta _{_{\rho ,\nu }}}\overline{\delta }\eta _{_{\rho
,\nu }}.
\]
Utilizando las ecuaciones de campo de Lagrange
\[
{\cal L}\left( \eta ^{\prime },x_{_\mu }\right) -{\cal L}\left( \eta
,x_{_\mu }\right) =\frac d{dx_\nu }\left( \frac{\partial {\cal L}}{\partial
\eta _{_{\rho ,\nu }}}\overline{\delta }\eta _{_\rho }\right) .
\]
Luego la condici\'{o}n de invarianza (\ref{79}) aparece en la forma
\begin{equation}
\int \left( dx_{_\mu }\right) \frac d{dx_\nu }\left\{ \frac{\partial {\cal L}%
}{\partial \eta _{_{\rho ,\nu }}}\overline{\delta }\eta _{_\rho }-{\cal L}%
\delta x_{_\nu }\right\} =0,  \label{80}
\end{equation}
que ya tiene la forma de una ecuaci\'{o}n de corriente conservativa.

\noindent
Sin embargo resulta \'{u}til desarrollar algo m\'{a}s la condici\'{o}n
especificando la forma de la transformaci\'{o}n infinitesimal en funci\'{o}n
de $R$ par\'{a}metros infinitesimales $\varepsilon _{r,}r=1,2,...,R$, tales
que las variaciones de $x_{_\mu }$ y $\eta _{_\rho }$ sean lineales en
los $\varepsilon _r$:
\begin{equation}
\delta x_{_\nu }=\varepsilon _rX_{r\nu },\qquad \delta \eta _{_\rho
}=\epsilon _r\Psi _{_{r\rho }}.  \label{81}
\end{equation}
Sustituyendo estas condiciones en la ecuaci\'{o}n (\ref{80}) obtenemos
\[
\int \epsilon _r\frac d{dx_\nu }\left\{ \left( \frac{\partial {\cal L}}{%
\partial \eta _{_{\rho ,\nu }}}\eta _{_{\rho ,\sigma }}-{\cal L}\delta
_{_{\nu \sigma }}\right) X_{r\sigma }-\frac{\partial {\cal L}}{\partial \eta
_{_{\rho ,\nu }}}\Psi _{_{r\rho }}\right\} \left( dx_{_\mu }\right) =0~.
\]
Como los par\'{a}metros $\varepsilon _r$ son arbitrarios, existen $r$
corrientes conservativas con teoremas de conservaci\'{o}n diferenciales:
\begin{equation}
\frac d{dx_\nu }\left\{ \left( \frac{\partial {\cal L}}{\partial \eta
_{_{\rho ,\nu }}}\eta _{_{\rho ,\sigma }}-{\cal L}\delta _{_{\nu \sigma
}}\right) X_{r\sigma }-\frac{\partial {\cal L}}{\partial \eta _{_{\rho ,\nu
}}}\Psi _{_{r\rho }}\right\} =0~.  \label{82}
\end{equation}
Las ecuaciones (\ref{82}) constituyen la principal conclusi\'{o}n del
teorema de Noether, el cual dice pues, que si el sistema tiene propiedades
de simetr\'{\i}a tales que se cumplan las condiciones (1) y (2) para
transformaciones del tipo indicado en las ecuaciones (\ref{81}),
existir\'{a}n $r$ cantidades conservativas.

\bigskip

\begin{center} BIBLIOGRAFIA COMPLEMENTARIA \end{center}

\bigskip

\noindent
R.D. Kamien, {\it Poisson bracket formulation of nematic polymer dynamics},
cond-mat/9906339 (1999)


\end{document}